\documentclass[a4paper,11pt,bookmarks=true]{article}
\pdfoutput=1 
\usepackage{jheppub} 
\usepackage[makeroom]{cancel}
\usepackage[T1]{fontenc} 
\usepackage{amsmath}
\usepackage{bm}
\usepackage{upgreek}
\usepackage{slashed}
\usepackage{tikz}
\newcommand{\subfig}[2]{%
\begin{tikzpicture}%
\node[rectangle] (image) at (0,0) {#2};
\node[anchor=south west] (label) at (image.south west) {(#1)};
\end{tikzpicture}%
}

\newcommand{\br}{{\bf r}}
\newcommand{\bk}{{\bf k}}

\newcommand{\bq}{{\bf q}}

\newcommand{\rhoh}{\hat{\bm{\uprho}}}

\renewcommand{\Re}{{{\mathfrak{Re}}}}

\title{\boldmath The medium-modified $g\to c\bar{c}$ splitting function in the BDMPS-Z formalism}

\author[a]{Maximilian Attems,}
\author[a]{Jasmine Brewer,}
\author[b]{Gian Michele Innocenti,}
\author[a]{Aleksas Mazeliauskas,}
\author[a]{Sohyun Park,}
\author[a]{Wilke van der Schee}
\author[a]{and Urs Achim Wiedemann}

\affiliation[a]{Theoretical Physics Department, CERN, CH-1211 Geneva 23, Switzerland}
\affiliation[b]{Experimental Physics Department, CERN, CH-1211 Geneva 23, Switzerland}
\emailAdd{maximilian.attems@cern.ch}
\emailAdd{jasmine.brewer@cern.ch}
\emailAdd{gian.michele.innocenti@cern.ch}
\emailAdd{aleksas.mazeliauskas@cern.ch}
\emailAdd{sohyun.park@cern.ch}
\emailAdd{wilke.van.der.schee@cern.ch}
\emailAdd{urs.wiedemann@cern.ch}

\abstract{
The formalism of Baier-Dokshitzer-Mueller-Peign\'e-Schiff and Zakharov  determines the modifications of parton splittings in the QCD plasma that arise from medium-induced gluon radiation. Here, we study medium-modifications of the gluon splitting into a quark--anti-quark pair in this BDMPS-Z formalism. We derive a compact path-integral formulation that resums effects from an arbitrary number of interactions with the medium to leading order in the $1/N_c^2$ expansion. 
Analyses in the $N=1$ opacity and the saddle point approximations reveal two phenomena: a medium-induced momentum broadening of the relative quark- anti-quark pair momentum that increases the invariant mass of quark–anti-quark pairs, and a medium-enhanced production of such pairs. We note that both effects are numerically sizeable if the average momentum transfer from the medium is comparable to the quark mass. In ultra-relativistic heavy-ion collisions, this condition is satisfied for charm quarks. We therefore focus our numerical analysis on the medium modification of $g\to c\bar{c}$, although our derivation applies equally well to $g\to b\bar{b}$ and to gluons splitting into light-flavoured quark--anti-quark pairs. 
}

\begin{document} 
\preprint{CERN-TH-2022-045}
\maketitle
\flushbottom

\section{Introduction}
\label{sec:intro}

The non-Abelian gauge theory of strong interactions---Quantum Chromodynamics (QCD)---is studied extensively in high-energy particle colliders. The high-density nuclear matter created in ultra-relativistic heavy-ion collisions provides unique access to the emergent phenomena associated with many-body QCD interactions. For example, energetic quarks and gluons (partons) traversing the quark-gluon plasma (QGP) experience energy loss due to medium-induced scatterings and radiation. This leads to the experimentally-observed suppression of high transverse momentum hadron and jet spectra in heavy-ion collisions~\cite{Connors:2017ptx}.

Motivated by pioneering works of Bjorken~\cite{Bjorken:1982tu}, and of Gyulassy, Pluemer and Wang~\cite{Wang:1994fx,Gyulassy:1993hr} 
the first QCD formulation of the modification of parton splittings in a dense QCD medium was derived by Baier, Dokshitzer, Mueller, Peign\'e, Schifff (BDMPS)~\cite{Baier:1996kr,Baier:1996sk}, and Zakharov (Z)~\cite{Zakharov:1996fv,Zakharov:1997uu}. This BDMPS-Z formalism
resums in a close-to-eikonal approximation interactions with the medium in the calculation of leading order (LO) $q\to qg$ and $g\to gg$ splitting functions. BDMPS-Z is a central part of the jet quenching phenomenology in ultra-relativistic heavy ion collisions at RHIC and at the LHC~\cite{Cao:2020wlm,Qin:2015srf}.  It has been generalised 
to rapidly expanding media~\cite{Baier:1998yf,Salgado:2003gb}, to a differential description of the angular dependence~\cite{Wiedemann:2000za,Wiedemann:2000tf,Baier:2001qw,Zakharov:1999zk}, and to gluon radiation from massive quarks~\cite{Dokshitzer:2001zm}. Its main parametric properties, such as a quadratic growth of the radiated gluon energy with in-medium path length and a characteristic transverse momentum broadening, were corroborated independently in several related approaches~\cite{Gyulassy:2000er,Wang:2001ifa}. 

In the last decade, the BDMPS-Z formalism has been extended further to determine double differential medium-induced gluon radiation beyond the soft approximation~\cite{Blaizot:2012fh,Apolinario:2014csa}. Numerical techniques for evaluating medium-induced gluon emission have been developed in several improved 
approximation schemes~\cite{Mehtar-Tani:2019tvy,Feal:2019xfl,Andres:2020vxs,Andres:2020kfg,Barata:2021wuf,Schlichting:2021idr,Isaksen:2022pkj}, and for the more 
complicated medium-averages that arise in some double-differential distributions~\cite{Isaksen:2020npj}. In addition, there has been significant theoretical effort to improve over 
the BDMPS-Z formulation. Radiative corrections to the BDMPS-Z formulation were understood to lead to double logarithmic enhancements with in-medium path length that can be absorbed in a renormalisation of the quenching parameter~\cite{Liou:2013qya,Blaizot:2014bha,Wu:2014nca}. The study of medium-induced gluon radiation from a QCD antenna~\cite{Armesto:2011ir,Casalderrey-Solana:2011ule,Mehtar-Tani:2012mfa,Barata:2021byj} was a first step toward understanding the destructive interference 
patterns arising in the case of multiple medium-induced parton branchings. The problem of overlapping formation times in sequential bremsstrahlung gluons has motivated a complete analysis of medium-induced QCD bremsstrahlung beyond leading order in $\alpha_s$~\cite{Arnold:2015qya,Arnold:2020uzm}. In addition, substantial work has been aimed at extending formulations of medium-induced gluon radiation to multi-parton final states
using dedicated Monte Carlo simulation tools~\cite{Zapp:2008af,Zapp:2012ak,Zapp:2013vla,Armesto:2009fj,Schenke:2009gb,Caucal:2018ofz,Putschke:2019yrg,Caucal:2018dla,Caucal:2019uvr}.

The medium modification of quark-anti-quark pair production  $g\to q\bar{q}$ in the BDMPS-Z formalism has received less attention so far. For the 
transverse momentum integrated splitting probability, the generalisation of Zakharov's path 
integral formalism~\cite{Zakharov:1996fv,Zakharov:1997uu} to $g\to q\bar{q}$ was given in~\cite{Caron-Huot:2010qjx}. To first order in opacity, an explicit expression for the
transverse momentum differential $g\to q\bar{q}$ distribution was derived in~\cite{Kang:2016ofv,Sievert:2019cwq} and used for the calculation of the nuclear modification factor of 
heavy-flavoured mesons~\cite{Kang:2016ofv,Sievert:2019cwq,Ke:2022gkq}. In the present work, we derive a transverse momentum differential expression for the $g\to q\bar{q}$ splitting
function in the BDMPS-Z path integral formalism and we provide a detailed study of its kinematic dependencies and their physical origin. We note that in the related problem of the medium-modified photon splitting function $\gamma \to q\, \bar{q}$, non-trivial target averages over four Wilson lines arise even to leading order in the $1/N_c^2$ expansion~\cite{Dominguez:2019ges}. Curiously, the splitting  $g\to q\bar{q}$ allows for a technically simpler formulation since it involves target averages over 
two Wilson lines only to leading $\mathcal{O} \left( \tfrac{1}{N_c^2} \right)$. Our calculations will be restricted to this leading order in the $1/N_c^2$ expansion that allows for a simpler description. 

In contrast to the $g \to gg$ and $q \to qg$ splitting functions,  $g \to q\bar{q}$ does not have a soft singularity. We thus calculate the medium modification of a formally-subleading contribution to the vacuum parton shower.
In general, heavy quark production at collider energies is described perturbatively~\cite{Cacciari:2012ny,Mangano:1991jk}. It is dominated by back-to-back topologies in which $q\bar{q}$-pairs recoil against each other with invariant masses of order of the partonic centre of mass, $Q^2 \sim {\cal O}(\hat{s})$. In this topology, the dominant medium modification of heavy-flavoured hadron spectra and flavour-tagged jets results from the medium-modified $q\to q g$ splitting functions determined in~\cite{Dokshitzer:2001zm}.
However, there is also the kinematic region $Q^2 \ll \hat{s}$ in which $q\bar{q}$-pairs are produced not back-to-back but almost collinear. In this region, heavy quark production factorises into a cross section for a final state gluon times a $g\to q\bar{q}$  splitting function~\cite{Ellis:1986ef}. In this collinear topology, it is the $g\to q\bar{q}$ splitting function that can be medium-modified. This contribution is phase space suppressed by 
$\mathcal{O}\left( \tfrac{Q^2}{\hat{s}}\right)$ compared to back-to-back topologies, but it is 
the leading production mechanism of  heavy flavour quark--anti-quark \emph{pairs} inside jets. 
Heavy-flavoured quarks are stable over the lifetime of the QGP and are therefore widely used probes of the medium~\cite{Andronic:2015wma}.
Dedicated adaptations \cite{Ilten:2017rbd} of modern jet finding algorithms~\cite{Cacciari:2011ma,Larkoski:2014wba} have demonstrated how to identify close-to-collinear $g\to c\bar{c}$  and $g \to b\bar{b}$  splitting topologies in hadronic collisions and how to relate them to the kinematics of the leading-order Altarelli-Parisi splitting functions.  
It is conceivable that such techniques can be adapted to disentangle medium-modifications of the collinear $g\to q\bar{q}$ production channel in the high-multiplicity environment of nucleus-nucleus collisions. The improved heavy flavour capabilities and higher integrated luminosities
of future heavy ion experiments at RHIC~\cite{Aidala:2012nz} and at the LHC~\cite{Citron:2018lsq,Adamova:2019vkf} may give access to such observables.
The present work lays the theoretical ground for the phenomenological studies we have started in a separate work~\cite{Attems:2022otp}.

Our work will clarify the size of the medium modification of the $g\to q\bar{q}$ splitting function. In the BDMPS-Z formalism, the properties of the QCD plasma are characterised  by a single parameter $\hat{q}$. This so-called quenching parameter measures the
average squared momentum transferred per unit path-length from the medium to a high-energy parton, and it can be determined for different microscopic model descriptions of the QCD plasma~\cite{Armesto:2011ht,JET:2013cls}. In quantum field theory,
the quenching parameter is a non-perturbative quantity defined in terms of the thermal expectation value of a null Wilson loop. This quantity was calculated first in the gravity duals of strongly coupled non-Abelian gauge theories~\cite{Liu:2006ug,Casalderrey-Solana:2006fio,DEramo:2010wup}. 
Following a proposal of Caron-Huot~\cite{Caron-Huot:2008zna}, it can be mapped to a problem in a dimensionally-reduced 3D effective theory for QCD that allows $\hat{q}$ to be calculated with lattice techniques~\cite{Panero:2013pla,Moore:2021jwe}. In the phenomenology of ultra-relativistic nucleus-nucleus collisions, much work has gone into understanding the measured suppression patterns of high-$p_T$ hadron spectra in terms of a QCD factorised formulation of single inclusive hadron spectra supplemented by quenching effects~\cite{Chien:2015vja,Bianchi:2017wpt,Andres:2016iys,Noronha-Hostler:2016eow,Casalderrey-Solana:2014bpa,Zigic:2018ovr,Andres:2019eus,Huss:2020whe,JETSCAPE:2021ehl}.
While individual phenomenological extractions of $\hat{q}$~\cite{JET:2013cls,Andres:2016iys,Andres:2019eus,Huss:2020whe,JETSCAPE:2021ehl} differ in the preferred value of $\hat{q}/T^3$, for the present work we will use that momentum transfers from the medium to the parton can reach and exceed the charm mass scale
\footnote{Following the convention of \cite{Arnold:2008iy}, the quenching parameter for momentum broadening of quarks is proportional to the fundamental
Casimir, $\hat{q} = C_F\, \hat{\bar{q}}$.  }
\begin{equation}
\langle {\bf q}^2\rangle_\text{med} =\int_{\tau_i}^{\tau_f} d\tau\, \hat{q}(\tau) \sim \mathcal{O}(m_c^2)\, ,
\label{eq1.1}
\end{equation}
where  the partonic trajectory is from initial time $\tau_i$ to final time $\tau_f$.  For the $g\to q\bar{q}$ splitting functions calculated in this manuscript, we show that medium corrections are 
\begin{equation}
P^{\text{med}}_{g\to q\bar{q}} \sim \mathcal{O}\left(\tfrac{\langle {\bf q}^2\rangle_\text{med}}{Q^2}\right) \, .
\end{equation} 
Splitting functions are the building blocks of parton showers. Therefore, we will put the size of these corrections in context by comparing them to other relevant terms in the vacuum splitting function.

We recall that 
in the collinear limit, the QCD differential cross section $d\sigma_{n+1}$ for any process with an additional parton in the final state is given by multiplying the $n$-parton final state with a process-independent factor
\begin{equation}
d\sigma_{n+1} = \sigma_n \frac{dQ^2}{Q^2} dz \frac{\alpha_s}{2\pi} P^\text{vac}(z)\, .
\label{eq1.2}
\end{equation}
Here, $P^\text{vac}(z)$ denotes the $1\to2$ splitting function in the absence of a medium for a daughter with momentum fraction $z$. As $Q^2$ can range 
from the hard scale $Q_h^2 \sim \hat{s}$ that governs $\sigma_n(\hat{s})$ to much smaller scales $Q_l^2$ (down to the non-perturbative scale), the factor $\mathcal{O}\left( \alpha_s \log Q_h^2/Q_l^2\right)$ in \eqref{eq1.2} requires resummation. In essence, Monte Carlo (MC) parton showers achieve this resummation by associating to each of the $n$ partons leaving $\sigma_n$ a Sudakov form 
factor~\cite{Platzer:2011dq,Lonnblad:2012hz,Kleiss:2016esx}
\begin{equation}
	S^{\text{vac}}_{q/g}(Q^2_h,Q^2_l) = \exp \left[ - \frac{\alpha_s}{2\pi}  \int_{Q^2_l}^{Q^2_h} \frac{dQ^2}{Q^2}   \int_{z_\text{min}}^{z_\text{max}} dz  
	\sum_{\text{channel}\, i} P^{\text{vac}}_{q/g \to i} \right]\,
	\label{eq1.3}
\end{equation}
that can be interpreted as the probability that a quark or gluon emits no resolvable radiation between the scale $Q_h^2$ and  $Q_l^2$. With the probability that a resolvable radiation is found, the parton shower decides on the scale $Q_h^2 < Q^2 < Q_l^2$ and on the kinematics of the resulting branching. The procedure is iterated by evaluating the no-splitting probability \eqref{eq1.3} for all splittees, until no further resolvable radiation is found.

In this way, the vacuum parton shower determines \emph{long-distance} contributions to the high-momentum transfer process $\sigma_n(\hat{s})$. As $Q^2$ becomes smaller, the associated distances become larger. In ultra-relativistic nucleus-nucleus collisions, the $\sigma_n(\hat{s})$ are embedded in QCD plasma. 
For sufficiently low $Q^2$, these distances can then become large enough for components of the medium to interact with the splitting process.  It is this situation that the calculation of medium-modified splitting functions addresses.  One may think schematically of a medium-modified splitting as a correction of the vacuum splitting functions $P^{\text{vac}} \to  P^{\text{vac}} + P^{\text{med}} $ that affects the no-splitting probability \eqref{eq1.3}. Expression  \eqref{eq1.3} is valid to leading logarithmic accuracy.
The literature does not provide firm guidance on how to include power-law corrections to parton showers of leading logarithmic accuracy. However, for the mass-term $2\,m_c^2/Q^2$, which is a subleading power-law correction to the vacuum splitting function, the phenomenological practice~\cite{Ellis:1996mzs,Bahr:2008pv,Hoche:2015sya} is to include it in \eqref{eq1.3}.
\emph{The correction $P^{\text{\rm med}}_{g\to c\bar{c}} \sim \mathcal{O}\left(\tfrac{\langle {\bf q}^2\rangle_\text{\rm med}}{Q^2}\right) $ that we compute in this work is parametrically and numerically comparable to this mass correction. This motivates us to include it on the same footing as the mass term in the vacuum parton shower.}

The paper is organised as follows. Section~\ref{sec2} states our main result  \eqref{eq2.2}, a path integral formula for the medium-modified splitting function $P^{\text{med}}_{g\to c\bar{c}}$. The derivation of this result is deferred to section~\ref{sec7}. In sections \ref{sec3} and \ref{sec4}, we first exhibit the main phenomena described by 
$P^{\text{med}}_{g\to c\bar{c}}$
in a number of physical approximations and interesting limits.  In particular, we compute $P^{\text{med}}_{g\to c\bar{c}}$ in the opacity expansion and multiple soft scattering approximations for static and time-dependent expanding media. 
We then discuss in section~\ref{sec6} on how $P^{\text{med}}_{g\to c\bar{c}}$ affects branching probabilities, and we shortly comment in section~\ref{sec8b} on $g\to b\bar{b}$ and on the gluon splitting into light-flavoured quark--anti-quark pairs.

\clearpage
\section{The medium-modified splitting function in the BDMPS-Z formalism }
\label{sec2}

In the vacuum, the $g\to c\bar{c}$ splitting function to leading order in $\alpha_s$ reads~\cite{Ellis:1996mzs}
\begin{equation}
  P^\text{vac}_{g \to c\, \bar{c}}(z, Q^2) = \frac{1}{2} \left(z^2 + (1-z)^2 + 2\frac{m_c^2}{Q^2}\right)\, ,
  \label{eq2.1}
 \end{equation}
 where we use $m_c = 1.27$ GeV for the charm mass~\cite{ParticleDataGroup:2020ssz}. 
 In this section, we present our main result for the medium-modification of this splitting function and discuss its relation to the vacuum contribution~\eqref{eq2.1}.

\subsection{The path integral formula for \texorpdfstring{$\left(\frac{1}{Q^2}\, P_{g \to c\, \bar{c}} \right)^{\rm med}$}{Pmed} }
\label{sec2.1}
\begin{figure}
    \centering
      \includegraphics[width=\textwidth]{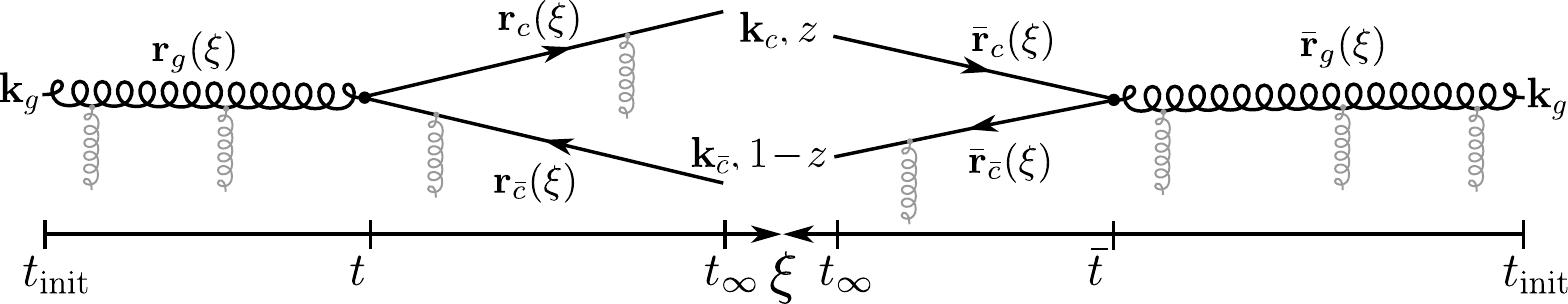}
    \caption{ Configuration space picture of the squared amplitude for the splitting of a gluon of initial transverse momentum $\bk_g$ into a $c\bar{c}$-pair carrying final 
    transverse momenta $\bk_c$ and $\bk_{\bar{c}}$ and longitudinal momentum fractions $z$ and $(1-z)$, respectively. The splitting occurs at longitudinal positions $t$ ($\bar{t}$) in amplitude (complex conjugate amplitude). Taking twice the real part of the contributions for $\bar{t} > t$ accounts for contributions from $\bar{t} < t$. Transverse positions of partons are evolved as a function of time $\xi$ in amplitude and complex conjugate amplitude. After integration over the average pair momentum $\frac{1}{2}(\bk_c + \bk_{\bar{c}})$, the resulting 
    medium-modified $g\to c\bar{c}$-splitting function takes the form \eqref{eq2.2}. See section~\ref{sec7} for the full derivation. }
    \label{fig1}
\end{figure}

We have considered the leading order $g\to c \bar{c}$ splitting diagram of Fig.~\ref{fig1} embedded in a QCD plasma. Within the BDMPS-Z formalism, we have formulated the resulting medium-modifications in time-ordered perturbation theory in the close-to-eikonal limit. In this case, the transverse momenta exchanged with the medium are small compared to the longitudinal momentum of parent and daughter partons. This leads to the following path integral formulation for the total in-medium splitting function,
\begin{align}
&	\left(\frac{1}{Q^2}\, P_{g \to c\, \bar{c}} \right)^{\text{tot}} \equiv \left(\frac{1}{Q^2}\, P_{g \to c\, \bar{c}} \right)^{\text{vac}} 
		+ \left(\frac{1}{Q^2}\, P_{g \to c\, \bar{c}} \right)^{\rm med}
		 		 \nonumber \\
		&\qquad= 2\, \mathfrak{Re}\, \frac{1}{8\, E_g^2}\, \int_{t_\text{init}}^{t_\infty} dt \int_t^{t_\infty} d\bar{t}\, 
    \exp\left[ i\frac{m_c^2}{2E_g z(1-z)} (t-\bar{t}) - \epsilon |t| - \epsilon |\bar{t}| \right]\, \int d{\bf r}_\text{out}
		\nonumber \\
		 &\qquad\qquad \times 
     \exp\left[ - \frac{1}{2} \int_{\bar{t}}^\infty d\xi\, n(\xi)\, \sigma_3({\bf r}_\text{out},z )\right]\, 
		  \exp\left[{-i\, \bm{\upkappa} \cdot{\bf r}_\text{out}}\right]
		  \nonumber \\		  		
		&\qquad\qquad \times \left[ \left( m_\text{c}^2 + \frac{\partial}{\partial {\bf r}_\text{in}}\cdot \frac{\partial}{\partial {\bf r}_\text{out}}
		  \right) \frac{z^2 + (1-z)^2}{z(1-z)}  + 2 m_c^2  \right] \, {\cal K}\left[{\bf r}_\text{in}=0,t;{\bf r}_\text{out},\bar{t} \vert \mu \right]\, .
		  \label{eq2.2} 
\end{align}
This splitting function depends on $Q^2$ (equivalently, $\kappa^2$) and on $z$, 
 and as shown in section~\ref{sec2.2} it reduces to the vacuum splitting function \eqref{eq2.1} in the vacuum limit $n(\xi)\to0$.
The derivation of \eqref{eq2.2} is discussed in section~\ref{sec7}. Here we determine its physical consequences. 
In this way, the reader who is not interested in technical details will still find in the following sections a significant number of non-trivial cross checks of the validity of \eqref{eq2.2}. 

The physics entering \eqref{eq2.2} can be summarised as follows:
The internal longitudinal integration variables $t$, $\bar t$ in \eqref{eq2.2} can be viewed as denoting the times at which the $g\to c\bar{c}$ splitting occurs in amplitude and complex conjugate amplitude, see Fig.~\ref{fig1}.  The transverse integration variable ${\bf r}_\text{out}$ in \eqref{eq2.2} is related to the relative transverse distance between charm- and anti-charm quark in the amplitude at the time $\bar{t}$ at which splitting occurs in the complex conjugate amplitude\footnote{We denote two dimensional variables by bold roman type symbols, e.g., $\br_\text{out}$, $\bm{\upkappa}$.}. 
We denote by $z$, $\bk_c$ ($1-z$, $\bk_{\bar{c}}$) the longitudinal momentum fraction and transverse momentum of the charm (anti-charm) quark, respectively, and the gluon energy by $E_g$. Equation~\eqref{eq2.2} is differential  in the transverse
momentum $\bm{\upkappa}$ where 
\begin{equation}
	\bm{\upkappa} = \frac{1}{2} \left( \bk_c - \bk_{\bar{c}}\right) =   \bk_c \quad \hbox{in the transverse pair rest frame where}\quad \bk_c + \bk_{\bar{c}} =0\, .
	\label{eq2.3}
\end{equation}
 A frame-independent definition is given in \eqref{eq77.43}.

Each parton propagating through the medium sees a medium-induced transverse colour field strength which is parameterised 
by a longitudinal density $n(\xi)$ of coloured scattering centres times a \emph{dipole cross section} 
\begin{equation}
	\sigma({\bf r}) = \int \frac{d{\bf q}}{(2\pi)^2}\, \vert a({\bf q}) \vert^2\, \left( 1 - e^{-i\, {\bf q}\cdot {\bf r}} \right)\, .
	\label{eq2.4}
\end{equation}
Here, $\sigma({\bf r})$ characterises the ${\bf q}$-differential elastic cross section $\propto \vert a({\bf q}) \vert^2$  of a medium scattering centre that interacts with a projectile parton in the amplitude and with the same (or another) parton in the complex conjugate amplitude, with ${\bf q}$ the momentum transferred from the medium to the parton during a single scattering.
Technically, $\sigma({\bf r})$  arises from comparing the transverse positions of the two partons in the amplitude and complex conjugate amplitude, so the cross section \eqref{eq2.4} should not be regarded as the cross section of a  $c\bar{c}$-dipole but as a measure of the characteristic momentum transferred per scattering from the medium to any parton. 

The interaction between the projectile parton and the medium depends on the colour representation of the parton and the longitudinal momentum fraction $z$ carried by it. This is encoded in the combination of dipole cross sections~\cite{Zakharov:1999zk,Caron-Huot:2010qjx} 
\begin{equation}
	\sigma_3({\bf r},z) \equiv	- \frac{1}{2 N_c}\sigma({\bf r})+\frac{N_c}{2}	\sigma(z{\bf r})+\frac{N_c}{2}	\sigma((1-z){\bf r})\, .
		\label{eq2.5}
\end{equation}
We write this expression including a subleading term in the ${1}/{N_c^2}$ expansion. As we shall show in sections~\ref{sec2.3} and~\ref{sec7.6}, the inclusion of this term ensures that the $\bm{\upkappa}$-integrated version of \eqref{eq2.2} is correct to all orders in $1/N_c^2$. However, the $\bm{\upkappa}$-differential distribution \eqref{eq2.2} is correct only to leading order in $1/N_c^2$ expansion. 

The path-integral ${\cal K}$ in \eqref{eq2.2} is defined in terms of the product $n(\xi)\, \sigma_3({\bf r},z)$, 
\begin{equation}
{\cal K}\big[{\bf r}_\text{in}=0,t;{\bf r}_\text{out},\bar{t} \vert \mu\big] = \int_{{\bf r}(t)={\bf r}_\text{in}}^{{\bf r}(\bar{t})={\bf r}_\text{out}}
{\cal D}{\bf r}\, \exp\left[ i \int_t^{\bar t} d\xi\, \left(  \frac{\mu}{2}  \dot{\bf r}^2 - \frac{n(\xi)\, \sigma_3({\bf r},z)}{2\, i} \right) \right]\, .
	\label{eq2.6}
\end{equation}

In this close-to-eikonal formalism, longitudinal position is identified with time. A gluon produced within the medium is a gluon produced at a finite initial time which we set to  $t_\text{init} = 0$. If the gluon were impinging on the target from the distant past, we would set $t_\text{init} = -\infty$, see section~\ref{sec3.4.2}. 
The longitudinal integrations in \eqref{eq2.2} go from $t_\text{init}$ to $t_\infty$ and they involve an $\epsilon$-regularisation. One needs to integrate \eqref{eq2.2} for 
 $t_\infty \to \infty$, then take the real part and  then remove the regulator $\epsilon \to 0$, since these operations do not commute.

The path integral \eqref{eq2.6} evolves  a dipole of initial transverse separation ${\bf r}_\text{in} = 0$ at time $t$ in a purely imaginary potential $\tfrac{n(\xi)\, \sigma_3({\bf r},z)}{2\, i}$  according to a standard light-cone Hamiltonian with  kinetic term 
$ \frac{\mu}{2}  \dot{\bf r}^2 $. Here, the `mass' 
\begin{equation}
	\mu \equiv E_g\, z\, (1-z)
	\label{eq2.7}
\end{equation}
controls the transverse growth of the dipole size in the plane orthogonal to the longitudinal direction of propagation.
For times $\xi > \bar{t}$, the dipole is fully formed and its legs travel on eikonal trajectories with fixed separation ${\bf r}_\text{out}$ from $\bar{t}$ to infinitely late times. 

The formula \eqref{eq2.2} relates this configuration space picture of an evolving dipole to information about the momentum transferred from the medium to the splitting vertex. In particular, we will demonstrate in the following subsection that the brackets $\left[ \dots \right]$ in the last line of \eqref{eq2.2} are a configuration space version of the corresponding terms in the vacuum expression \eqref{eq2.1}.

\subsection{Consistency with the vacuum splitting function}
\label{sec2.2}
In the absence of any medium, $n(\xi) = 0$, 
the path integral ${\cal K}$ in \eqref{eq2.6} reduces to the free propagator
\begin{equation}
  {\cal K}_0\big[{\bf r}_\text{in},t;{\bf r}_\text{out},\bar{t} \vert \mu \big] = 
	\frac{\mu}{2\pi \, i\, \left(\bar{t}-t \right)}\,
	\exp\left[ \frac{i\, \mu}{2}  \frac{\left( {\bf r}_\text{out} - {\bf r}_\text{in} \right)^2}{\left(\bar{t}-t \right)} \right]\,
	\label{eq2.8}
\end{equation}
and the absorption factor in the next-to-last line of \eqref{eq2.2} becomes unity. Evolving the phase factor $\exp\left[{-i\,\bm{\upkappa} \cdot{\bf r}_\text{out}}\right]$ in \eqref{eq2.2} with 
the  propagator \eqref{eq2.8} yields 
 \begin{equation}
   \int d{\bf r}_\text{out}\, {\cal K}_0\big[{\bf r}_\text{in},t;{\bf r}_\text{out},\bar{t} \vert \mu \big] \,
	\exp\left[{-i\, \bm{\upkappa} \cdot{\bf r}_\text{out}}\right] 
	= \exp\left[ - i \frac{\bm{\upkappa}^2}{2\mu} \left( \bar{t} - t \right) \right]\, 
		\exp\left[ {-i\, \bm{\upkappa} \cdot{\bf r}_\text{in}} \right]\, .
			\label{eq2.9}
 \end{equation}
Inserting these expressions into \eqref{eq2.2}, taking the derivatives $\frac{\partial}{\partial {\bf r}_\text{in}}\cdot\frac{\partial}{\partial {\bf r}_\text{out}}$ with the help of partial
integration and setting ${\bf r}_\text{in} = 0$, we find 
\begin{eqnarray}
	\left(\frac{1}{Q^2}\, P_{g \to c\, \bar{c}} \right)^{\rm vac} &=&  2\, \mathfrak{Re}\, \frac{1}{{8} E_g^2}\, \int_{t_\text{init}}^{t_\infty} dt \int_t^{t_\infty} d\bar{t}\, 
		\exp\left[ i \frac{m_c^2+\bm{\upkappa}^2}{2 \mu} (t-\bar{t}) - \epsilon |t| - \epsilon |\bar{t}| \right]\, 
		\nonumber \\
		&& \times \left[ \left( m_c^2 + \bm{\upkappa}^2
		  \right) \frac{z^2 + (1-z)^2}{z(1-z)}  + 2 m_c^2  \right] \, .
		  \label{eq2.10}
\end{eqnarray}
The longitudinal phase of \eqref{eq2.9} is absorbed in a phase factor $\Gamma_0$. In the limit of large gluon energy, $\Gamma_0$ can be written in terms of the invariant mass $Q^2$ of the $c\bar{c}$ pair
\begin{equation}
	\Gamma_0 \equiv \frac{m_c^2+\bm{\upkappa}^2}{2 \mu} = \frac{Q^2}{2 E_g} \, ,
	\label{eq2.11}
\end{equation}
since in the close-to-eikonal approximation $Q^2$ is given by
\begin{equation}
	Q^2 = \frac{m_c^2+\bm{\upkappa}^2}{z (1-z)}\, .
	\label{eq2.12}
\end{equation}
Doing the longitudinal integrations in \eqref{eq2.10} 
\begin{eqnarray}
	\lim_{\epsilon\to 0} \lim_{t_\infty \to \infty} 2\, \mathfrak{Re}\,  \int_{t_\text{init}}^{t_\infty} dt \int_t^{t_\infty} d\bar{t}\, 
		\exp\left[ i \Gamma_0\, (t-\bar{t}) - \epsilon |t| - \epsilon |\bar{t}| \right] = \frac{1}{\Gamma_0^2} = \frac{4 E_g^2}{Q^4}\,
		  \label{eq2.13}
\end{eqnarray}
one finds
\begin{equation}
\left(\frac{1}{Q^2}\, P_{g \to c\, \bar{c}} \right)^{\rm vac} 
 = \frac{1}{ 2Q^4} \left[ \left( m_c^2 + \bm{\upkappa}^2 \right) \frac{z^2 + (1-z)^2}{z(1-z)}  + 2 m_c^2 \right]\, .
	\label{eq2.14}
\end{equation}
In the close-to-eikonal approximation, \eqref{eq2.14} is the standard vacuum splitting function \eqref{eq2.1}. This is a first consistency check of \eqref{eq2.2}. 
The overall normalisation that we had left unspecified in the derivation in section~\ref{sec7}
has been fixed in \eqref{eq2.2} such that the 
prefactor of \eqref{eq2.14} agrees with the standard vacuum expression~\eqref{eq2.1}.

Equation~\eqref{eq2.13} illustrates how the longitudinal integrals in \eqref{eq2.2} yield two inverse powers of $Q^2$. We anticipate that in the presence of 
a medium, these factors $\tfrac{1}{Q^4}$ are shifted in transverse momentum. This is the technical reason for why it is convenient to define the total in-medium splitting function \eqref{eq2.2} multiplied by $\tfrac{1}{Q^2}$.

\subsection{Consistency with \texorpdfstring{$\bm{\upkappa}^2$}{kappasquared}-integrated in-medium splitting functions}
\label{sec2.3}
Expression \eqref{eq2.2} is double differential in $z$ and ${\bm \upkappa}$.
According to \eqref{eq1.2} the distribution function of a single medium-induced emission is proportional to the splitting function 
\begin{equation}
\frac{dN^\text{med}_{g\to c\bar{c}}}{dzdQ^2} = \frac{\alpha_s}{2\pi} \left(\frac{1}{ Q^2 }\, P_{g \to c\,\bar{c}} \right)^{\text{med}}
\end{equation}
Integrating \eqref{eq2.2} over virtuality $dQ^2 = \frac{d {\bm \upkappa}^2}{z(1-z)}$ we find
\begin{align}
\frac{dN^\text{med}_{g\to c\bar{c}}}{dz}
		&=\frac{\alpha_s}{2E_g^2 z(1-z)}\, \mathfrak{Re}\,\, \int_{t_\text{init}}^{t_\infty} dt \int_t^{t_\infty} d\bar{t}\, 
    \exp\left[ i\frac{m_c^2}{2E_g z(1-z)} (t-\bar{t}) - \epsilon |t| - \epsilon |\bar{t}| \right]\,
		\nonumber \\
		&\times \left[ \left( m_\text{c}^2 + \frac{\partial}{\partial {\bf r}_\text{in}}\cdot \frac{\partial}{\partial {\bf r}_\text{out}}
		  \right) \frac{z^2 + (1-z)^2}{z(1-z)}  + 2 m_c^2  \right] \, 
		  \nonumber \\
		  & \times \left( {\cal K}\big[{\bf r}_\text{in}=0,t;{\bf r}_\text{out}=0,\bar{t} \vert \mu \big] -
		   {\cal K}_0\big[{\bf r}_\text{in}=0,t;{\bf r}_\text{out}=0,\bar{t} \vert \mu \big] \right)	 \, .
		  \label{eq3.30} 
\end{align}
This expression is consistent with the result in Ref.~\cite{Caron-Huot:2010qjx} for the medium-modified rate of $g \to c\bar{c}$ splitting  (with a factor of $2$ difference due to the identification of $z$ and $1-z$ splittings in \cite{Caron-Huot:2010qjx}).
We emphasise that this 
${\bm \upkappa}$-integrated rate is \emph{positive for any z}. 
Non-perturbative computations supporting this statement can be found, e.g., in Ref.~\cite{Moore:2021jwe}.

\section{First order opacity expansion}
\label{sec3}
In this section, we derive the medium-modification of the vacuum splitting function \eqref{eq2.1} to first order in opacity. Massive splitting functions  to first order in opacity and all orders in $1/N_c^2$ have been derived in \cite{Kang:2016ofv}.
The total in-medium splitting function \eqref{eq2.2} depends on momenta measured in the transverse centre-of-mass frame of the $c\bar{c}$-pair. If the parent gluon receives a momentum transfer from the medium \emph{before} splitting into a $c\bar{c}$-pair, then its direction of propagation (which \eqref{eq2.2} is not sensitive to) is modified, but its invariant mass is not. One thus expects that the splitting function \eqref{eq2.2} is sensitive to the \emph{formation time} (i.e., the longitudinal distance) at which the $c\bar{c}$ splitting occurs since only scattering centres located \emph{after} that distance will modify \eqref{eq2.2}. The following calculations will give a physical meaning to the notions `{before}', `{after}' and `{formation time}' used in this paragraph.

\subsection{Structure of the opacity expansion }
\label{sec3.1}
The opacity expansion is an expansion of the integrand of \eqref{eq2.2} in powers of $n(\xi)\, \sigma_3({\bf r},z)$. 
The $N$-th order expansion of the integral \eqref{eq2.2} is of the parametric size 
$\sim \left( L\, n\, \sigma_{\text{el}}\right)^N$ where $\sigma_\text{el}$ denotes the elastic cross section associated to a single scattering centre
 \begin{equation}
 \sigma_\text{el} \equiv \int \frac{d{\bf q}}{(2\pi)^2}\, \vert a({\bf q}) \vert^2\, .
 \label{eq3.1}
 \end{equation}
 This is called an `opacity expansion' because the dimensionless product $L\, n\, \sigma_{\text{el}}$ characterises how opaque a medium with scattering centres of density $n$ and elastic cross section $\sigma_{\text{el}}$ is to a projectile.
 
The $N$-th order opacity contribution to the full propagator ${\cal K}$ in \eqref{eq2.6} can be obtained by reiterating $N$ times the recursion relation 
\begin{eqnarray}
  &&{\cal K}\big[{\bf r}_\text{in},t;{\bf r}_\text{out},\bar{t} \vert \mu  \big] = {\cal K}_0\big[{\bf r}_\text{in},t;{\bf r}_\text{out},\bar{t}  \vert \mu \big] 
	\nonumber \\
	&& \qquad \qquad \qquad	- \frac{1}{2} \int_t^{\bar{t}} d\xi\, n(\xi) \int d{\bm{\uprho}}	\, 
  {\cal K}_0\big[{\bf r}_\text{in},t;{\bm{\uprho}},\xi  \vert \mu \big] \, \sigma_3({\bm{\uprho}},z)\, {\cal K}\big[{\bm{\uprho}},\xi ;{\bf r}_\text{out},\bar{t}  \vert \mu \big]\, .
	\label{eq3.2}
\end{eqnarray}
One can check that this implements the path ordering of \eqref{eq2.6}. The propagation of the phase factor $\exp\left[{-i\, \bm{\upkappa}\cdot {\bf r}_\text{out}}\right]$ in \eqref{eq2.2} with the perturbatively-expanded path integral proceeds in close analogy to the free evolution \eqref{eq2.9}. For instance, the propagation of the phase to first order in  opacity yields 
\begin{eqnarray}
	&&- \frac{1}{2} \int_t^{\bar{t}} d\xi\, n(\xi) \int d{\bm{\uprho}}	\, d{\bf r}_\text{out}
  {\cal K}_0\big[{\bf r}_\text{in},t;{\bm{\uprho}},\xi| \mu \big] \, \sigma_3({\bm{\uprho}},z)\, {\cal K}_0\big[{\bm{\uprho}},\xi ;{\bf r}_\text{out},\bar{t}| \mu \big] \, 
	e^{-i\,  \bm{\upkappa}\cdot{\bf r}_\text{out}} 		\label{eq3.3}\\
	&& = - \frac{1}{2} \int_t^{\bar{t}} d\xi\, n(\xi)  
	\int_{a_3({\bf q},z) }
	\int d{\bm{\uprho}}	\, d{\bf r}_\text{out}
  {\cal K}_0\big[{\bf r}_\text{in},t;{\bm{\uprho}},\xi|\mu \big] \, \left( 1 - e^{-i\, {\bf q}\cdot{\bm{\uprho}}} \right)\, {\cal K}_0\big[{\bm{\uprho}},\xi ;{\bf r}_\text{out},\bar{t}| \mu \big] \, 
	e^{-i\,  \bm{\upkappa}\cdot{\bf r}_\text{out}} \nonumber \\
	&&= - \frac{1}{2} \int_t^{\bar{t}} d\xi\, n(\xi)  
	\int_{a_3({\bf q},z) }
	\left( e^{ - i \tfrac{ \bm{\upkappa}^2}{2\mu} \left( \bar{t} - t \right) }\, e^{-i\,  \bm{\upkappa}\cdot{\bf r}_\text{in}} - 
	e^{ - i \tfrac{ \bm{\upkappa}^2}{2\mu} \left( \bar{t} - \xi \right) }\, e^{ - i \tfrac{\left(  \bm{\upkappa}+ {\bf q} \right)^2}{2\mu} \left( \xi - t \right) }\, 
		e^{-i\, \left({\bf q}+  \bm{\upkappa}\right)\cdot{\bf r}_\text{in}} \right)\, .
		\nonumber 
\end{eqnarray}
Here we have introduced the notational short-hand  $\int_{a_3({\bf q},z) }\dots  \equiv \int \frac{d{\bf q}}{(2\pi)^2}\, \vert a_3({\bf q},z) \vert^2 \dots$ for the integral over the \emph{effective elastic scattering cross section}\footnote{The term $\vert a_3\left({{\bf q}},z\right)\vert^2$ always appears in an integral over ${\bf q}$, so the resulting expression will be free of soft singularities in the limits $z \to 0, 1$. }
\begin{equation}
  \vert a_3\left({{\bf q}},z\right)\vert^2\equiv -\frac{1}{2 N_c}\vert a({\bf q}) \vert^2+\frac{N_c}{2 z^2}\vert a\left({{\bf q}}/{z}\right)\vert^2 +\frac{N_c}{2 (1-z)^2}\vert a\left({{\bf q}}/{(1-z)}\right)\vert^2.
  \label{eq3.4}
 \end{equation}
 which we obtained from \eqref{eq2.4} and \eqref{eq2.5} by rescaling the integration variable $\bf{q}$.
According to \eqref{eq3.3}, the original phase is freely back-propagated from $\bar{t}$ to $\xi$ and then multiplied with the factor
$ \left( 1 - e^{-i\, {\bf q}\cdot{\bm{\uprho}}} \right)$ associated to the dipole cross section of the scattering centre as in \eqref{eq2.4}. 
This factor gives two contributions, with the first only depending on ${\bf q}$ through $a_3\left({{\bf q}},z\right)$. This term is freely back-propagated up to $t$ and thus reduces the vacuum-like contribution by a prefactor  $- \tfrac{1}{2} \sigma_\text{el,3} \int_t^{\bar{t}} d\xi\, n(\xi)\,  $, where $\sigma_\text{el,3} = \int \frac{d{\bf q}}{(2\pi)^2}\, \vert a_3({\bf q},z) \vert^2 $.
After integrating over ${\bf \rho}$, the second contribution changes the transverse phase at $\xi$ from $ \bm{\upkappa} \to  \bm{\upkappa} + {\bf q}$ and thus yields a term with medium-modified longitudinal phase $ e^{ - i \tfrac{\left(  \bm{\upkappa}+ {\bf q} \right)^2}{2\mu} \left( \xi - t \right) }$ that is multiplied with the probability that a scattering with momentum transfer ${\bf q}$ occurs. The combination of both contributions conserves probability: a scattering that enhances the vacuum distribution at ${\bm \upkappa}$ also depletes the distribution by the same amount at ${\bm \upkappa}+{\bf q}$.
The opacity expansion of the path integral formula \eqref{eq2.2} is based on expanding both the full propagator ${\cal K}$ and the absorption factor  in the next-to-last line of \eqref{eq2.2} consistently to given order in $L \,n\,\sigma_\text{el}$.

\subsection{Emergence of a testable formation time to first order in opacity}
\label{sec3.2}
We consider a gluon initiated at time $t_\text{init} = 0$ that propagates along the longitudinal direction and that splits eventually. We position a single scattering centre along the path of this gluon and ask to what extent the scattering-induced modification of the gluon splitting \eqref{eq2.2} depends on the distance between $t_\text{init} = 0$ and the position of that scattering centre. Technically, we realise this situation by evaluating \eqref{eq2.2}  to first order in opacity for a static medium with 
\begin{equation}
	n(\xi) = \left\{
                \begin{array}{ll}
                  n_0, \qquad  \xi \leq L\, ,\\
                  0, \, \, \, \qquad  \xi > L\, .
                \end{array}
              \right.
              \label{eq3.5}
\end{equation}
The result will be first order in $\left( L\, n_0\, \sigma_\text{el} \right)$. By keeping $L\, n_0$ fixed but varying the total in-medium path length $L$, we can move the average position of the scattering centre to arbitrarily late or early times.

The calculation is a straightforward application of section~\ref{sec3.1}. Deferring technical details to Appendix \ref{appb}, we write the final result 
for the medium-modification of the splitting function \eqref{eq2.2} to order  $N=1$  in opacity as 
\begin{eqnarray}
&&	\left(\frac{1}{Q^2}\, P_{g \to c\, \bar{c}} \right)^{\rm med}_{N=1}
	= \frac{1}{2} n_0\, L \int \frac{d{\bf q}}{(2\pi)^2} \vert a_3({\bf q},z) \vert^2\, 
	\left(  1 - \frac{1}{L\Gamma_1} \sin\left[ L\Gamma_1\right]   \right)
	\nonumber \\
	&&\qquad\qquad\qquad \times \left[	 \left(\frac{1}{Q^2}\, P_{g \to c\, \bar{c}} \right)^{\rm vac}_{\bm{\upkappa} \to \bm{\upkappa} + {\bf q}}  -	 
  \left(\frac{1}{Q^2}\, P_{g \to c\, \bar{c}} \right)^{\rm vac}  \right.
	\nonumber \\
		&&\qquad\qquad\qquad  + \left.	\left( \frac{1}{Q_1^2} - \frac{1}{Q^2}  \right)^2
		\frac{ m_c^2}{2z(1-z)}
	+ \left( \frac{\left(\bm{\upkappa}
		   + {\bf q} \right)}{Q_1^2} - \frac{\bm{\upkappa}}{Q^2}  \right)^2  \frac{z^2 + (1-z)^2}{2z(1-z)} \right]\,.
		   \label{eq3.6}
\end{eqnarray}
We have checked that \eqref{eq3.6} is consistent with the leading $\mathcal{O}(\frac{1}{N_c^2})$ limit of (2.52) in \cite{Kang:2016ofv}.
Here, we denote by $\left(\tfrac{1}{Q^2}\, P_{g \to c\, \bar{c}} \right)^{\rm vac}_{\bm{\upkappa} \to \bm{\upkappa} + {\bf q}}$ a vacuum 
splitting function of the form \eqref{eq2.1} in which all transverse momenta are shifted. 
In close analogy to \eqref{eq2.11}, we also
introduce a phase factor
\begin{equation}
	\Gamma_1 \equiv \frac{m_c^2+ \left(\bm{\upkappa} + {\bf q}\right)^2}{2 \mu} = \frac{Q_1^2}{2 E_g} \, ,
	\label{eq3.7}
\end{equation}
where $Q_1^2$ denotes a virtuality with shifted transverse momentum. As $\bm{\upkappa}$ is the transverse momentum in the final state, $\bm{\upkappa} + {\bf q}$ is the initial  momentum \emph{prior} to scattering.\footnote{The reader whose intuition prefers to associate $\bm{\upkappa} - {\bf q}$ with the transverse momentum \emph{prior} to scattering may notice that a change ${\bf q} \to - {\bf q}$ of the integration variable in \eqref{eq3.6} would change trivially all arguments $\bm{\upkappa} + {\bf q} \to \bm{\upkappa}- {\bf q}$. The relative sign in $\bm{\upkappa} \pm {\bf q}$ is unimportant since the integral in \eqref{eq3.6}  goes over all azimuthal orientations of ${\bf q}$ and since $\vert a_3({\bf q},z)\vert^2$ is a scalar.} The virtuality $Q_1^2$ can therefore be interpreted as the virtuality of the vacuum splitting $g\to c\bar{c}$ \emph{prior} to scattering, i.e., 
$Q_1^2$ is the virtuality of the incoming parent gluon prior to interaction with the medium.
The interference term in the first 
line of \eqref{eq3.6} interpolates between simple limiting cases
\begin{equation}
1 - \frac{1}{L\Gamma_1}\sin\left[L \Gamma_1 \right]  \longrightarrow \left\{
                \begin{array}{ll}
                  1, \qquad  L \gg  \Gamma_1^{-1}\, ,\\
                  0, \qquad  L \ll  \Gamma_1^{-1}\, .
                \end{array}
              \right.
              \label{eq3.8}
\end{equation}
 This motivates associating $\Gamma^{-1}_1$ with the vacuum formation time of the splitting,
\begin{equation}
 \tau_{g\to c\bar{c}} \equiv	\frac{1}{\Gamma_1} \, .
	\label{eq3.9}
\end{equation}
If the formation time $\tau_{g\to c\bar{c}}$ becomes much larger than the average longitudinal position of the scattering centre, then the modification of the $g\to c\bar{c}$ splitting is strongly suppressed, i.e.,
\begin{equation}
	\left(\frac{1}{Q^2}\, P_{g \to c\, \bar{c}} \right)^{\rm med}_{N=1} 
	\xrightarrow{{ \tau_{g\to c\bar{c}}} \gg L } 0\, .
	 \label{eq3.10}
\end{equation}
This is consistent with the probabilistic interpretation that the $g\to c\bar{c}$ splitting did not occur prior to $L$. 
\emph{In this sense, the formation time \eqref{eq3.9} sets an observable\footnote{Strictly speaking, we establish here only that the formation time can be inferred from knowledge of final state \emph{parton} momenta. To what extent this information can be extracted unambiguously from experimentally accessible hadronic distributions remains to be studied in future work.} minimal length scale below which
medium-modifications of the $c\bar{c}$ final state cannot occur. }

\begin{figure}[t]
   \centering
\subfig{a}{\includegraphics[width=0.45\linewidth]{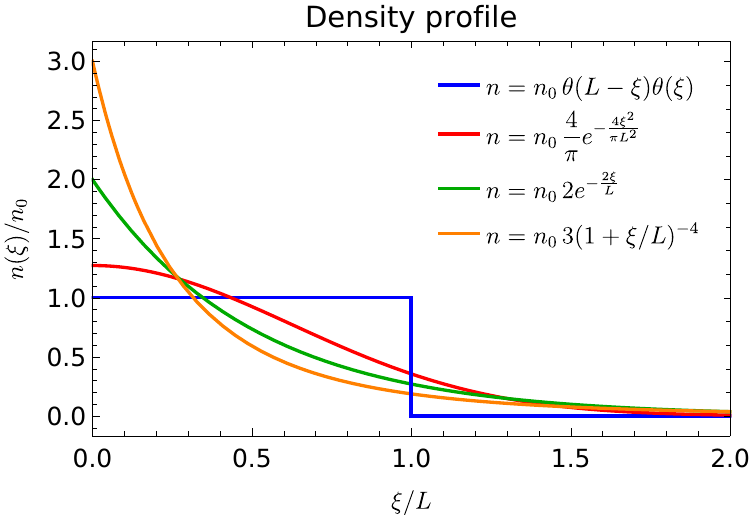}}
\subfig{b}{\includegraphics[width=0.45\linewidth]{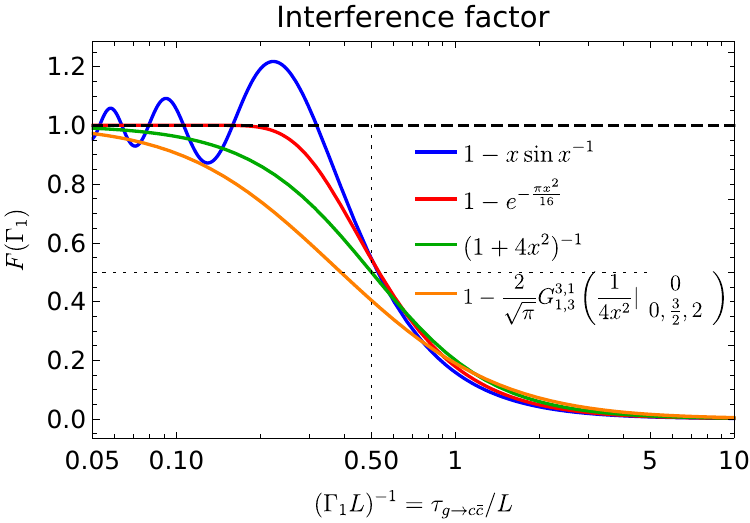}}
\caption{ a) Different density profiles that can be related to the same equivalent static medium of length $L$ and density $\bar{n}_0$ since they 
satisfy the same line integrals \eqref{eq3.12} and \eqref{eq3.15}. b) The interference factors \eqref{eq3.13} for these density profiles interpolate between
totally coherent and incoherent limits on a similar scale $\Gamma_1 L \approx 2$. $G_{p,q}^{m,n}$ denotes the Meijer G-function. 
\label{fig:interference_factor}}
 \end{figure}

The above line of argument is parametric. 
Numerically, the interference term in \eqref{eq3.6} is shown in blue in Fig.~\ref{fig:interference_factor}(b).
One finds that the interference factor reaches the mid-point  between the totally coherent and incoherent limiting cases ($F(\Gamma_1)\approx 0.5$) when $\tau_{g\to c\bar{c}}/L=1/2$, i.e., when the formation time is equal to the average position of the scattering centre $L/2$. 

\subsection{The \texorpdfstring{$g\to c\bar{c}$}{g->ccbar} splitting in an expanding medium}
\label{sec3.3}
In ultra-relativistic nucleus-nucleus collisions,  parton showers initiated in high momentum-transfer processes are embedded in a rapidly-expanding medium. 
To gain insight into the effects of an expanding medium, we consider here the medium-modified splitting function in the $N=1$ opacity expansion 
for a general time-dependent density profile $n(\xi)$ (see Appendix~\ref{appb} for details)
 \begin{align}
   \left(\frac{1}{Q^2}\, P_{g \to c\, \bar{c}} \right)^{\text{med}}_{N=1}=&\frac{1}{2}\int \frac{d{\bf q}}{(2\pi)^2}\, \vert a_3({\bf q},z) \vert^2 \int_{0}^{\infty} d\xi n(\xi)\left\{1-\cos[\Gamma_1 \xi]\right\}
\nonumber\\
&\times\Bigg[\left(\frac{1}{Q^2}\, P_{g \to c\, \bar{c}} \right)^{\text{vac}}_{{\bm{\upkappa}\to \bm{\upkappa}+{\bf q}}} - \left(\frac{1}{Q^2}\, P_{g \to c\, \bar{c}} \right)^{\text{vac}} \nonumber\\
    &+\frac{m_c^2}{2z(1-z)} \left( \frac{1}{Q_1^2} - \frac{1}{Q^2} \right)^2 + \frac{z^2 + (1-z)^2}{2z(1-z)}\left(\frac{\bm{\upkappa}+{\bf q}}{Q_1^2} - \frac{\bm{\upkappa}}{Q^2}\right)^2\Bigg].
    \label{eq3.11}
\end{align}

 We ask to what extent this result, obtained for an arbitrary expanding medium of density $n(\xi)$, can be related to the result for an \emph{equivalent static medium}  of density $\bar{n}_0$ and length $L$ that we define in terms of the line integral
 \begin{equation}
 	\bar{n}_0 L \equiv \int^\infty_{0} d\xi\, n(\xi)\, .
	\label{eq3.12}
 \end{equation} 
The $\xi$ integral in  \eqref{eq3.11} can be written as a product of $\bar{n}_0 L$ and the  interference factor 
 \begin{equation}
  F(\Gamma_1) \equiv  1 - \frac{\int_{0}^\infty d\xi \,n(\xi)\cos[ \Gamma_1 \xi] }{\int^\infty_{0} d\xi\, n(\xi)}\, .
  \label{eq3.13}
\end{equation}
For the static medium of constant density~\eqref{eq3.5}, the interference factor  $F(\Gamma_1)$ reduces to \eqref{eq3.8}. For the general, 
time-dependent case, the interference factor  $F(\Gamma_1) $ vanishes for $\Gamma_1\to 0$  (equivalently, $\tau_{g\to c\bar{c}}\to\infty$) as $\mathcal{O}(\Gamma_1^2)$. To compare interference factors of different density profiles, we choose to match the integral of $F(\Gamma_1)$ with respect to the formation time $\tau_{g\to c\bar{c}} = \Gamma_1^{-1}$, i.e.,
\begin{equation}
    \int_0^\infty d\Gamma_1^{-1} F(\Gamma_1) = \frac{\pi}{2}\frac{\int_{0}^\infty d\xi\, n(\xi) \xi}{\bar{n}_0 L}
    \label{eq3.14}.
\end{equation}

This suggests to fix the density $\bar{n}_0$ and the in-medium path-length $L$ of an \emph{equivalent static medium} in terms of
the line integral \eqref{eq3.12} and the linearly weighted line-integral appearing in  \eqref{eq3.14},
\begin{equation}
	\bar{n}_0 L^2 \equiv  2 \int_{0}^\infty d\xi\, n(\xi)\, \xi\, .
	     \label{eq3.15}
\end{equation}
In Fig.~\ref{fig:interference_factor}, we make use of \eqref{eq3.12} and \eqref{eq3.15} to compare different time-dependent density profiles to an equivalent static scenario. Specifically, the area under the curves is the same for different profiles shown in Fig.~\ref{fig:interference_factor}(a) and the corresponding interference factors are shown in Fig.~\ref{fig:interference_factor}(b) (note the log scale). The interference factor \eqref{eq3.13} 
is then independent of  $\bar{n}_0$ and it depends only on the 
dimensionless quantity $\left( \Gamma_1 L \right)^{-1} = \tau_{g\to c\bar{c}}/L$. Remarkably, while details of the shape depend on the density profile, the interference factor $F(\Gamma_1)$ 
interpolates in all cases between the
totally coherent and incoherent limiting cases on approximately the same scale $\Gamma_1^{-1} =L/2$. This reinforces the interpretation of a formation time given to 
$\tau_{g\to c\bar{c}} $ in \eqref{eq3.9}.

Where numerical accuracy would require it, techniques exist to evaluate \eqref{eq2.2} for expanding scenarios. However, the relations
established here indicate that quantitative guidance for the results expected for expanding scenarios can be gained from studying equivalent
static media. In particular,  in the incoherent limit, results for $\left(\frac{1}{Q^2}\, P_{g \to c\, \bar{c}} \right)^{\rm med}$ in an expanding scenario can be 
related to an equivalent static scenario satisfying \eqref{eq3.12}. The scale at which coherence effects start reducing the medium-modification is then set by 
$L\, \Gamma_1 \approx 2$ with $L$ being defined via \eqref{eq3.12} and \eqref{eq3.15}.
We anticipate that these observations can be generalised beyond the $N=1$ opacity expansion, see section~\ref{sec4}.

\subsection{Incoherent limit of the \texorpdfstring{$N=1$}{N=1} opacity expansion}
\label{sec3.4}
\subsubsection{Gluons produced at \texorpdfstring{$t_\text{init}=0$}{yinit=0} inside the medium.}
\label{sec3.4.1}
To zeroth order in opacity, we found in section~\ref{sec2.2}
\begin{eqnarray}
	\left(\frac{1}{Q^2}\, P_{g \to c\, \bar{c}} \right)^{\text{tot}}_{N=0} = \left(\frac{1}{Q^2}\, P_{g \to c\, \bar{c}} \right)^{\text{vac}}\, .
		  \label{eq3.16}
\end{eqnarray}
In this sense the gluon initialised at $t_\text{init} = 0$ is off shell since it splits in the absence of medium-induced scattering.

Equation \eqref{eq3.6} gives the correction to \eqref{eq3.16} that arises from a single scattering centre uniformly distributed between $0$ and $L$. In the incoherent limit $L \gg \tau_{g\to c\bar{c}}$ with $n_0\, L$ constant, the interference factor \eqref{eq3.8} is unity. This corresponds to placing the scattering centre far away from the production point $t_\text{init}=0$ of the virtual gluon. In this case, the physics occurring at $t_\text{init}=0$ does not interfere with the physics at $L$, and the entire process can be described as the probabilistic product of what happens around $t_\text{init}=0$ and around $L$. In the present case, this 
 \emph{probabilistic interpretation} is in terms of two mechanisms:
\begin{enumerate}
\item[A.] {\bf Broadening:} \emph{Probability-conserving momentum redistribution due to medium effects.}\\
The probabilistic interpretation of this mechanism is that a $g\to c\bar{c}$ vacuum splitting around $t_\text{init}=0$ is followed by a redistribution of $c\bar{c}$-pairs in transverse momentum due to elastic scattering at a distant scattering centre.
\item[B.] {\bf Medium-induced $c\bar{c}$-radiation:}  \emph{\it  Enhanced production of $c\bar{c}$-pairs due to scattering.}\\
Gluons that do not split close to $t_\text{init}=0$ may split due to the momentum transferred at a scattering centre that is well-separated from $t_\text{init}=0$.
\end{enumerate}
In writing \eqref{eq3.6}, we have arranged different terms such that the contributions from the broadening mechanism becomes explicit,
\begin{eqnarray}
	\left(\frac{1}{Q^2}\, P_{g \to c\, \bar{c}} \right)^{\text{med}}_{N=1,\,\text{broad}}
	&=& \frac{1}{2}n_0\, L \int \frac{d{\bf q}}{(2\pi)^2} \vert a_3({\bf q},z) \vert^2\, 
	\left( 1 - \frac{1}{L\Gamma_1} \sin\left[ L\Gamma_1\right]  \right)
	\nonumber \\
	&& \times \left[	 \left(\frac{1}{Q^2}\, P_{g \to c\, \bar{c}} \right)^{\rm vac}_{\bm{\upkappa} \to \bm{\upkappa} + {\bf q}}  -	 
		\left(\frac{1}{Q^2}\, P_{g \to c\, \bar{c}} \right)^{\rm vac}   \right]\, .
		   \label{eq3.17}
\end{eqnarray}
In the incoherent limit, the interference factor \eqref{eq3.8} is unity and the second negative term reduces the probability of the vacuum splitting by the probability that a scattering occurred
$\propto \tfrac{1}{2} n_0 L \sigma_\text{el,3}$.
This  is  compensated by a positive term $\left(\frac{1}{Q^2}\, P_{g \to c\, \bar{c}} \right)^{\rm vac}_{\bm{\upkappa} \to \bm{\upkappa} + {\bf q}} $ that shifts the vacuum-distribution in transverse space. In total, \eqref{eq3.17} implements a probability-conserving rearrangement  of $c\bar{c}$ pairs in transverse momentum space. 

\subsubsection{Gluons propagating from \texorpdfstring{$t_\text{init} = -\infty$}{tinit=-infinity}.}
\label{sec3.4.2}
Given the interpretation of \eqref{eq3.17} in terms of the broadening mechanism, it is natural to conjecture that the difference between \eqref{eq3.6} and \eqref{eq3.17} has an interpretation in terms of medium-induced $c\bar{c}$ radiation in the incoherent limit. To test this conjecture, we design a gedankenexperiment in which broadening is switched off, so that the kinematic dependence of the terms expected for stimulated $c\bar{c}$-radiation can be established in a separate calculation. This is achieved by preparing an \emph{on-shell gluon}
 that cannot split into a $c\bar{c}$-pair without medium modification, and letting it scatter on a spatially extended target. The way to do this in the present formalism is to prepare the gluon in the infinite past at $t_\text{init}=-\infty$. Such a gluon is on-shell in the sense that to zeroth order in opacity, 
 no splitting occurs\footnote{Although the incoming gluon is on-shell  in this respect, one has $Q_1 \geq 2 m_c$ in \eqref{eq3.7}.}:
\begin{eqnarray}
	\left(\frac{1}{Q^2}\, P_{g \to c\, \bar{c}} \right)^{\text{med},\, t_\text{init}=-\infty}_{N=0} = 0  \, .
		  \label{eq3.18}
\end{eqnarray}
One checks \eqref{eq3.18} by  setting $n(\xi) = n_0 = 0$ in \eqref{eq2.2}  and doing all integrals
analytically. To let this on-shell gluon scatter once, we determine the first order opacity correction to \eqref{eq3.16}.  A 
relatively lengthy calculation, based entirely on the technical steps discussed already, yields 
\begin{eqnarray}
	\left(\frac{1}{Q^2}\, P_{g \to c\, \bar{c}} \right)^{\text{med},\, t_\text{init}=-\infty}_{N=1}
	&=&  \frac{1}{2} n_0\, L \int \frac{d{\bf q}}{(2\pi)^2} \vert a_3({\bf q},z) \vert^2\, 
	\nonumber \\
	&& \times \left[	 \left( \frac{1}{Q_1^2} - \frac{1}{Q^2}  \right)^2  \frac{m_c^2}{2z(1-z)}
+ \left( \frac{\left(\bm{\upkappa}    + {\bf q} \right)}{Q_1^2} - \frac{\bm{\upkappa} }{Q^2}  \right)^2  \frac{z^2 + (1-z)^2}{2z(1-z)} \right]
		   \nonumber \\
		   &=&
		   \left[ \left(\frac{1}{Q^2}\, P_{g \to c\, \bar{c}} \right)^{\text{ med}}_{N=1} - \left(\frac{1}{Q^2}\,P_{g \to c\, \bar{c}} \right)^{\text{ med}}_{N=1,\,\text{broad}}\right]_{L \gg \tau_{q\to c\bar{c}}}\,,
		   \label{eq3.19}
\end{eqnarray}
where $L \gg \tau_{g\to c\bar{c}}$ guarantees that the interference factor that is absent when $t_\text{init}=-\infty$ is also $1$ in \eqref{eq3.11} and \eqref{eq3.17}.
In this way, we understand the full $N=1$ opacity correction to the $g\to c\bar{c}$ vacuum splitting in terms of two contributions
\eqref{eq3.17} and \eqref{eq3.19} that have an intuitive probabilistic interpretation in the incoherent limit and that both become negligible 
if the in-medium path length is small compared to the formation time. 
 
\subsection{Numerical results for the \texorpdfstring{$\bm{\upkappa}^2$}{kappasquared}-differential  in-medium splitting function}
  \label{sec3.5}
Numerical evaluation of the $N=1$ opacity correction~\eqref{eq3.6} to the splitting function requires specifying the shape of the elastic scattering cross
section $\vert a({\bf q})\vert^2$ and the value of the opacity $ n_0 L \sigma_\text{el}$. Motivated by the Gyulassy-Wang model~\cite{Gyulassy:1993hr} of partonic  $2\to 2$ scatterings 
with medium constituents, we use Yukawa-type scattering potentials distributed with constant density $n_0$ and regulated in the IR by a Debye screening mass $\mu_D$,
\begin{equation}
	\frac{1}{4} n_0\, L\,  \vert a({\bf q})\vert^2
	\equiv \frac{c_\text{opaque} }{2\mu_D^2} \vert \tilde{a}(\tilde{\bf q})\vert^2 \, , \qquad
	 \vert \tilde{a}(\tilde{\bf q})\vert^2 \equiv \frac{2\pi}{ \left( \tfrac{1}{2} + \tilde{\bf q}^2\right)^2 }
	\, ,
	\label{eq3.20}
\end{equation}
where $c_\text{opaque} $ is the overall normalisation constant. 
Integrating both sides over $\int \tfrac{d{\bf q}}{(2\pi)^2}$ relates $c_\text{opaque} $ to the opacity
\begin{equation}
c_\text{opaque} =  \frac{1}{4} n_0 L \sigma_\text{el}\, . 
\label{eq3.21}
\end{equation}
For the combination of dipole cross sections appearing in
$\sigma_3$, we rescale \eqref{eq3.4} as
\begin{equation}
	\vert \tilde{a}_3(\tilde{\bf q},z)\vert^2 = 
	 -\frac{1}{2N_c} \frac{2\pi}{ \left( \tfrac{1}{2} + \tilde{\bf q}^2\right)^2 } +\frac{N_c}{2} \frac{2\pi\, z^2}{ \left( \tfrac{1}{2} z^2+ \tilde{\bf q}^2\right)^2 }	+\frac{N_c}{2} \frac{2\pi\, (1-z)^2}{ \left( \tfrac{1}{2}(1-z)^2 + \tilde{\bf q}^2\right)^2 }\, .
	\label{eq3.22}
\end{equation}
Here and in the following, we rescale masses and transverse momenta  by $2\mu_D^2$,
 \begin{equation}
 \tilde{m}_\text{c}^2 = \frac{{m}_\text{c}^2}{2\mu_D^2 }\, ,\qquad  \tilde{\bm{\upkappa}}^2 = \frac{{\bm{\upkappa}}^2}{2\mu_D^2},\qquad  
 \tilde{\bf q}^2 = \frac{{\bf q}^2}{2\mu_D^2}\, .
\label{eq3.23}
 \end{equation}
 Expressing the initial gluon energy $E_g$ in units of the characteristic energy scale $\mu_D^2 L$, we define the dimensionless gluon energy $e_g$ via
 \begin{equation}
 	E_g \equiv \frac{1}{2} 2\mu_D^2 L\, e_g\, .
	\label{eq3.24}
 \end{equation}
 \noindent
The $N=1$ opacity correction~\eqref{eq3.6} then depends on $2\mu_D^2$, $e_g$, and the normalisation $c_\text{opaque} $,
\begin{eqnarray}
	\left(\frac{1}{Q^2}\, P_{g \to c\, \bar{c}} \right)^{\rm med}_{N=1}
  &=& c_\text{opaque}  \frac{z(1-z)}{2 \mu_D^2} \int \frac{d\tilde{\bf q}}{(2\pi)^2}  \vert \tilde{a}_3(\tilde{\bf q},z)\vert^2 
	\left( 1 - \frac{1}{L\Gamma_1} \sin\left[ L\Gamma_1\right]  \right)
	\nonumber \\
	&& \times \left[	 \frac{z^2 + (1-z)^2}{\tilde{m}_c^2 + (\tilde{\bm{\upkappa}} + \tilde{\bf q})^2} - \frac{z^2 + (1-z)^2}{\tilde{m}_c^2 + \tilde{\bm{\upkappa}}^2} 
		+  \frac{2 \tilde{m}_c^2 z (1-z)}{\left(\tilde{m}_c^2 + (\tilde{\bm{\upkappa}} + \tilde{\bf q})^2\right)^2} 
	\right.
	\nonumber \\
		&&  \qquad  \left. - \frac{2 \tilde{m}_c^2 z (1-z)}{\left(\tilde{m}_c^2 + \tilde{\bm{\upkappa}}^2\right)^2}   
			+  \tilde{m}_c^2 \left(\frac{1}{\tilde{m}_c^2 + (\tilde{\bm{\upkappa}} + \tilde{\bf q})^2} - \frac{1}{\tilde{m}_c^2 + \tilde{\bm{\upkappa}}^2} \right)^2
		\right.
	\nonumber \\
	&& \qquad \left. + \left(\frac{\tilde{\bm{\upkappa}} + \tilde{\bf q}}{\tilde{m}_c^2 + (\tilde{\bm{\upkappa}} + \tilde{\bf q})^2} - \frac{\tilde{\bm{\upkappa}}}{\tilde{m}_c^2 + \tilde{\bm{\upkappa}}^2} \right)^2
	  \left[z^2 + (1-z)^2 \right] \right]\, .
		   \label{eq3.25}
\end{eqnarray}
The dimensionless gluon energy $e_g$ enters only in the  interference term 
 \begin{eqnarray}
 1 - \frac{1}{L\Gamma_1} \sin\left[ L\Gamma_1  \right]
 &=& 1 - \frac{e_g z (1-z)}{\tilde{m}_\text{c}^2 + \left(\tilde{\bm{\upkappa}} + \tilde{\bf q}\right)^2}  
 \sin\left[ \frac{\tilde{m}_\text{c}^2 + \left(\tilde{\bm{\upkappa}} + \tilde{\bf q}\right)^2}{e_g z (1-z)} \right] \nonumber \\
&&  \longrightarrow \left\{
                \begin{array}{ll}
                  1, \qquad \, e_g \ll  \tfrac{\tilde{m}_\text{c}^2 + \left(\tilde{\bm{\upkappa}} + \tilde{\bf q}\right)^2}{z (1-z)}  \, ,\\
                  0, \qquad \, e_g \gg  \tfrac{\tilde{m}_\text{c}^2 + \left(\tilde{\bm{\upkappa}} + \tilde{\bf q}\right)^2}{z (1-z)}\, .
                \end{array}
              \right.
              \label{eq3.26} 
 \end{eqnarray}
 \begin{figure}[t]
    \centering
           \subfig{a}{\includegraphics[width=.49\textwidth]{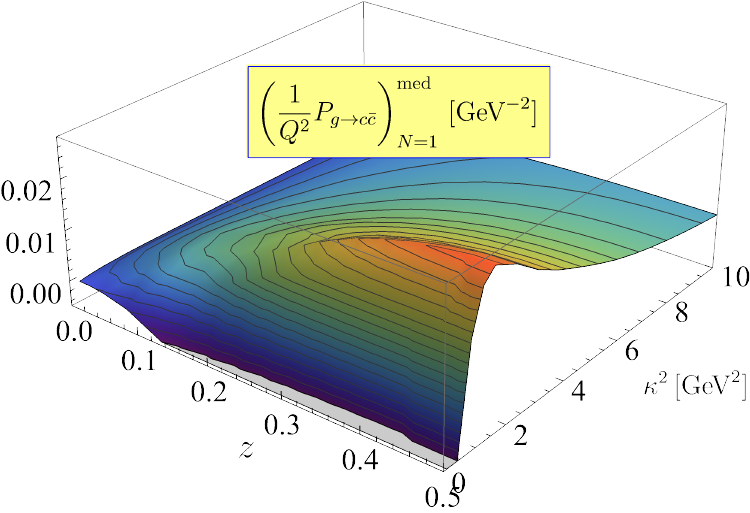}}%
       \subfig{b}{\includegraphics[width=.49\textwidth]{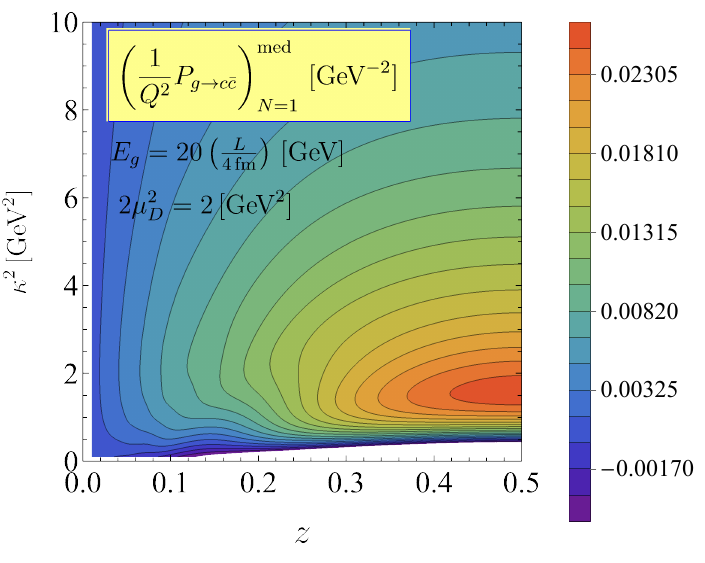}}
     \caption{The medium-modification of the $g\to c\bar{c}$ splitting evaluated in the $N=1$ opacity approximation 
   for `hard' Yukawa-type scattering centres~\eqref{eq3.20}. Each panel shows different viewpoints on the results for typical momentum squared $2 \mu_D^2 = 2\text{ GeV}^2$ transferred from the scattering centre.}
    \label{figN1}
\end{figure}  

\begin{figure}[t]
    \centering
       \subfig{a}{\includegraphics[width=.49\textwidth]{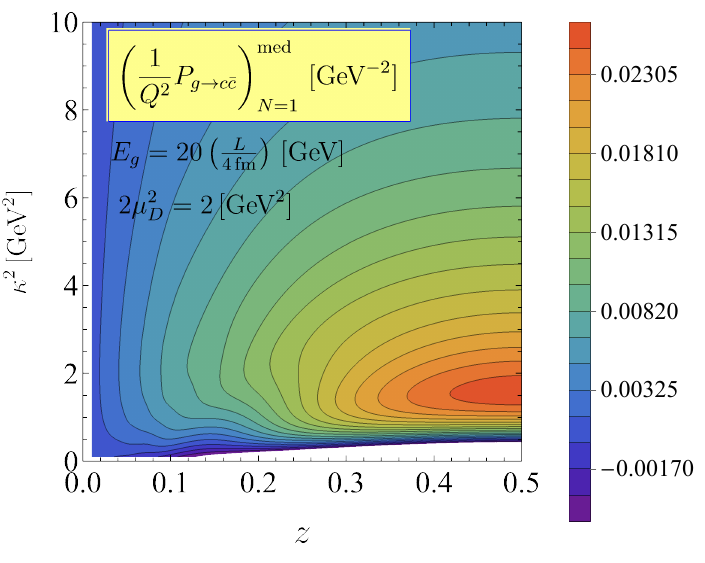}}%
        \subfig{b}{\includegraphics[width=.49\textwidth]{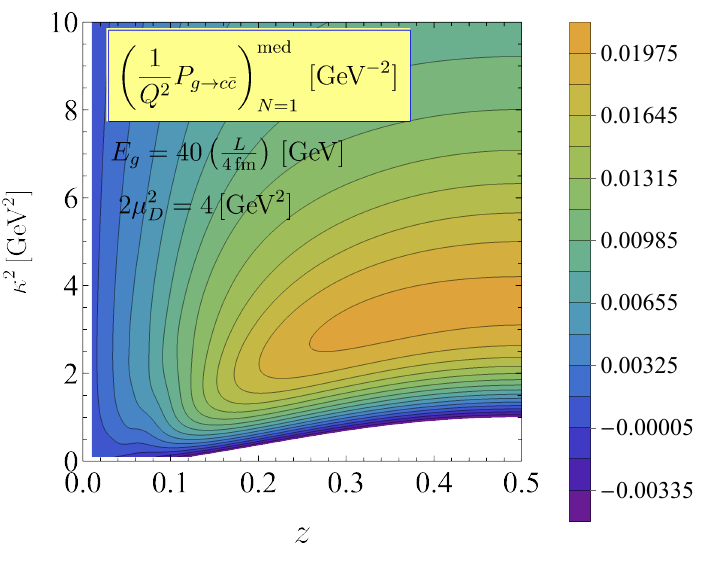}}
        \subfig{c}{\includegraphics[width=.49\textwidth]{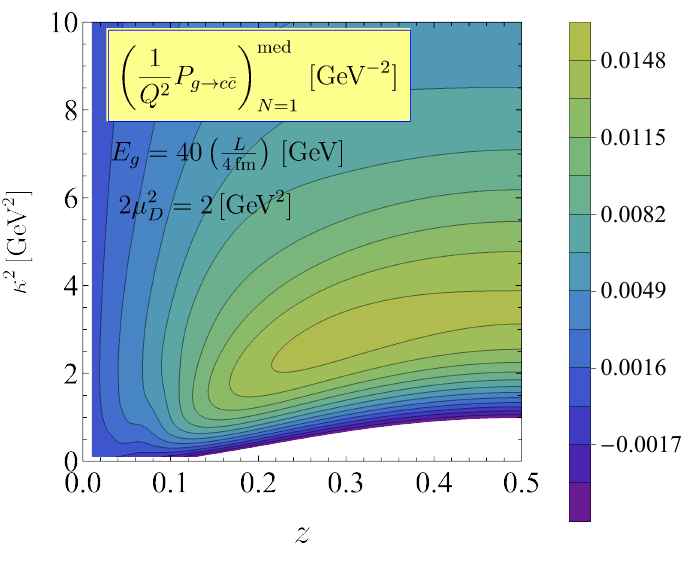}
        }%
        \subfig{d}{\includegraphics[width=.49\textwidth]{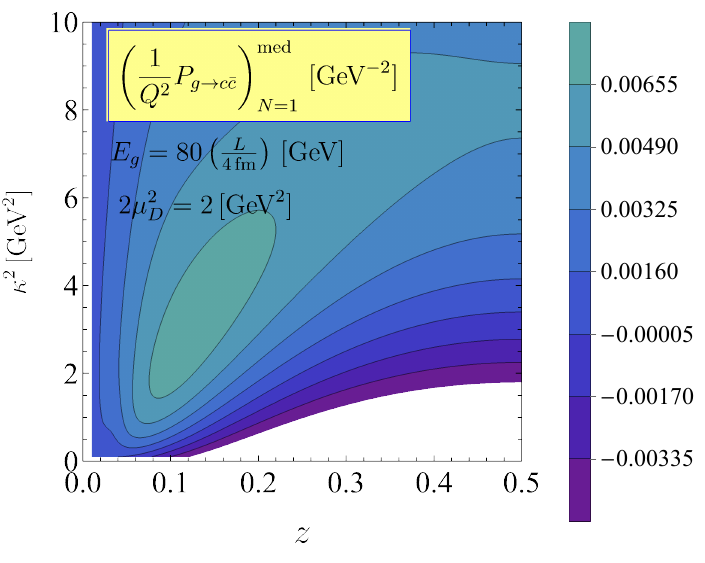}}
     \caption{The medium modification of the $g\to c\bar{c}$ splitting evaluated in the $N=1$ opacity approximation 
   for `hard' Yukawa-type scattering centres~\eqref{eq3.20}. (a) is the same as Fig.~\ref{figN1}(b). Different panels show results for different values of the typical squared momentum $2 \mu_D^2$ transferred from the scattering centre and for different gluon energies $E_g$.}
    \label{figN12}
\end{figure}  

An expansion in powers of opacity is only meaningful for $c_\text{opaque}  \ll 1$.
To determine the opacity, one requires knowledge about the scattering centre density $n_0$, the total elastic cross-section $\sigma_\text{el}$ and the in-medium path-length $L$. In models assuming a weakly coupled thermal plasma, $n_0$, $\sigma_\text{el}$ and $\mu_D$ depend on the plasma temperature only~\cite{Gyulassy:1993hr}, see e.g. Ref.
\cite{Isaksen:2022pkj} for a recent implementation. 
In the following we discuss only the functional shape of \eqref{eq3.25} and fix the normalisation $c_\text{opaque} = 1$ for the numerical results displayed in Figs.~\ref{figN1} and~\ref{figN12}.

Numerical results for the medium-modified splitting function \eqref{eq3.25}  in  Figs.~\ref{figN1} and~\ref{figN12} display the tell-tale signs of momentum broadening. The medium moves $c\bar{c}$-pairs from very small to larger relative transverse momentum, thus depleting the vacuum distribution at very small $\upkappa$ . 
The typical shift of transverse momentum is found to  peak at 
${\upkappa}^2 \simeq 2\, \mu_D^2$, which is one motivation for rescaling masses and transverse momenta by the scale $2\, \mu_D^2$ in \eqref{eq3.23} (see also Appendix~\ref{appc}).  Fig.~\ref{figN12}  also indicates that the medium modification of the splitting function is positive over a much larger region of phase space than the region in which it is negative. This is a first indication that the total yield of $c\bar{c}$-pairs can be enhanced due to medium-effects. We shall quantify this enhancement 
 in section~\ref{sec6} after having fixed the normalisation of the medium-modified splitting function.

According to \eqref{eq3.26}, destructive interference increases with increasing $e_g$ and is more pronounced for symmetric splittings and for small ${\upkappa}^2$. The results in Fig.~\ref{figN12} clearly display this effect: as $e_g$ is increased (equivalently, as $E_g$ is increased with fixed $2\mu_D^2$), the size of the medium modification decreases.
In general, a medium of fixed $L$ and $2 \mu_D^2$ has a maximal gluon energy $E_g$ below which the medium modification of the splitting function is approximately independent of $E_g$ and above which the medium modification is significantly reduced due to interference effects.

\begin{figure}[t]
    \centering
       \includegraphics[width=.59\textwidth]{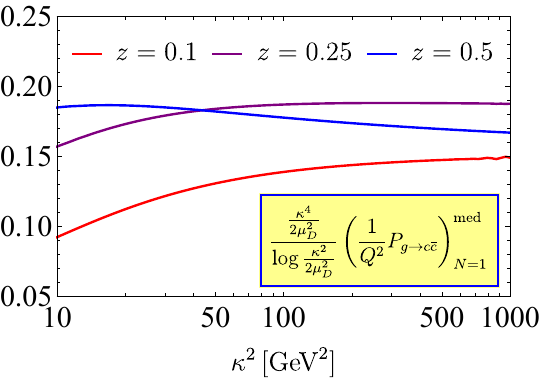}
    \caption{Large-${\upkappa}^2$-dependence of the medium-modified splitting function for $N=1$ hard scattering approximation, illustrating the asymptotic behaviour \eqref{eq3.28}, for particular values $E_g=20\,\left(\frac{L}{4\,\text{fm}}\right)\,\text{GeV}$ and $2\mu_D^2 = 2\, \text{GeV}^2$.}
    \label{fig4b}
\end{figure}

\subsection{Large \texorpdfstring{$\bm{\upkappa}^2/(2\mu_D^2)$}{kappasquared/(2 muDsquared)} limit}
\label{sec3.6}
As discussed in the introduction, one expects on general grounds that the medium modification $\left(\frac{1}{Q^2}\, P_{g \to c\, \bar{c}} \right)^{\rm med}_{N=1}$ can be a numerically large power correction, but that it cannot contribute with large logarithms to the medium-modified parton shower. This can be verified by analysing \eqref{eq3.25}.  One first notes that  \eqref{eq3.25} is free of soft singularities $\propto 1/z$, so there cannot be a logarithmic enhancement in the longitudinal phase space $dz$. This is also seen clearly from  Figs.~\ref{figN1} and~\ref{figN12}. The remaining question is whether the transverse phase space $d{\upkappa}^2$ can yield a logarithmic enhancement. For the vacuum splitting  $\left(\frac{1}{Q^2}\, P_{g \to c\, \bar{c}} \right)^{\rm vac} \propto \frac{1}{Q^2}\propto \tfrac{1}{{\upkappa}^2}$, this is the case. But for the medium-modification, we find in the limit of large ${\upkappa}^2$ the leading  term in \eqref{eq3.6} 
is $\propto\frac{1}{Q^4}\propto\frac{1}{{\upkappa}^4}$,
\begin{equation}
\left(\frac{1}{Q^2}\, P_{g \to c\, \bar{c}} \right)^{\rm med}_{N=1}  \approx \frac{1}{Q^4}\frac{z^2+(1-z)^2}{z(1-z)} \left[\frac{1}{4}n_0 L\int \frac{d{\bf q}}{(2 \pi)^2}|a_3({\bf q},z)|^2 \frac{4 ({\bf q}\cdot\bm{\upkappa})^2}{m_c^2+\bm{\upkappa}^2}\right]\,.
\label{eq3.27}\
\end{equation}
For the Yukawa-type scattering potential \eqref{eq3.20},
the integral in the square brackets is logarithmically divergent and therefore the asymptotic behaviour of the $N=1$ hard scattering splitting function is
\begin{align}
\left(\frac{1}{Q^2}\, P_{g \to c\, \bar{c}} \right)^{\rm med}_{N=1} 
& \propto  \frac{2\mu_D^2}{{\bm \upkappa}^4}  \log\left[\frac{{\bm \upkappa}^2}{2\mu_D^2}\right]
\label{eq3.28}
\end{align}
Fig.~\ref{fig4b} shows numerical evidence for this asymptotic behaviour. The prefactor of \eqref{eq3.28} depends on $z$ according to \eqref{eq3.27}.
The logarithmic factor $\log\left[\frac{{\bm \upkappa}^2}{2\mu_D^2}\right]$ 
arises from the power-law scattering potential \eqref{eq3.20}. It does not appear  for the Gaussian potential model studied in Appendix~\ref{appc}. 

It follows from \eqref{eq3.28} that, up to possible logarithmic corrections, the phase space integrals appearing in \eqref{eq1.3} yield 
\begin{equation}
 \int_{Q^2_l}^{Q^2_h} dQ^2 \int_0^1 dz  \left(\frac{1}{Q^2}\, P_{g \to c\, \bar{c}} \right)^{\rm med}_{N=1} \propto \mathcal{O}(\mu_D^2) \left( \frac{1}{Q_l^2} -  \frac{1}{Q_h^2} \right)\, .
 \label{eq3.29}
\end{equation}
Since $\mu_D^2$ is the only momentum scale in the problem, one concludes on general grounds that it must set the numerical size of this integral. 
The power correction~\eqref{eq3.29} will be important for our discussion of no-branching probabilities in section~\ref{sec6}.

\section{Multiple soft scattering approximation}
  \label{sec4}
In the $N=1$ opacity expansion, we have associated the opacity $c_\text{opaque}$ with the probability that a 
scattering occurs. As opacity increases,  higher orders  in opacity become more important.  In contrast to \eqref{eq3.25}, 
 $P^\text{med}_{g\to c\bar{c}}$ will not continue to increase linearly with $c_\text{opaque}$. 
 We note that \eqref{eq2.6} resums all orders in opacity and is not limited to small $c_\text{opaque}$. Contributions from all orders in opacity are kept in 
 a saddle-point approximation of the path integral  which we discuss now.
   
 \subsection{Saddle-point approximation}
 \label{sec4.1}
For small ${\bf r}$ we have $	n(\xi)\sigma({\bf r}) = \frac{1}{2}\hat{\bar{q}} \,{\bf r}^2$, where 
\begin{equation}
    \hat{\bar{q}} =n(\xi)  \int \frac{d{\bf q}}{(2\pi)^2}\, \vert a({\bf q}) \vert^2 {\bf q}^2\, ,
    \label{eq4.1}
\end{equation}
The dipole cross section \eqref{eq2.4} vanishes for vanishing dipole size ${\bf r}$ and it increases monotonously for small ${\bf r}$. The path-integral ${\cal K}$ 
in \eqref{eq2.6} is therefore dominated by paths close to the saddle point ${\bf r} = 0$ and we may seek an approximation by expanding it around small distances ${\bf r}$. 
The simplest such approximation\footnote{For a simple illustration of the physics captured in this  approximation, one may back-propagate
 the final phase $\exp\left[{-i\, \bm{\upkappa} \cdot {\bf r}_\text{out}}\right]$ with the absorption factor written  in \eqref{eq2.2}.  Assuming a static medium of finite extent, 
 $\hat{\bar{q}}(\xi) = \hat{\bar{q}}$, and $z\ll 1$ one finds
\begin{eqnarray}
\int d{\bf r}_\text{out} e^{ - \tfrac{1}{4} \int_0^{L} d\xi\, \hat{q}(\xi)\, {\bf r}_\text{out}^2 }\, 
		  e^{-i\, \bm{\upkappa} \cdot{\bf r}_\text{out} } = \frac{4 \pi}{\hat{q}\, L }
		  \exp\left[ - \frac{\bm{\upkappa}^2}{\hat{q}\, L } \right]\, .
		  \label{eq4.2}
\end{eqnarray}
Here, the width of the $\bm{\upkappa}^2$-distribution broadens linearly with increasing in-medium path-length $L$, as expected for a transverse Brownian motion induced by multiple soft random momentum transfers. This is consistent with the interpretation of $\hat{q}$ as the average squared momentum transfer per unit path length. }
 is to make a quadratic ansatz for the small-distance behaviour of the dipole cross section~\cite{Zakharov:1996fv} (for a discussion of subtleties related to this ansatz, see Appendix~\ref{appc}) 
\begin{equation}
  n(\xi)\, \sigma_3({\bf r},z) \simeq \frac{1}{2}\, \hat{\bar{q}}(\xi)\,\left( C_F- N_c z(1-z) \right)    {\bf r}^2 \equiv 
  	 \frac{1}{2}\, \hat{q}(\xi,z)\, {\bf r}^2\, .
	\label{eq4.3}
\end{equation}

With  \eqref{eq4.3}, ${\cal K}$ in \eqref{eq2.6} becomes the path integral of a two-dimensional harmonic oscillator
${\cal K}_\text{osc}$. For a static medium, $\hat{q}(\xi,z) =\hat{q}(z)$, it takes the explicit form
\begin{equation}
	{\cal K}_\text{osc}\left({\bf r}_\text{in},t;{\bf r}_\text{out},\bar{t} \vert \mu \right) 
		= \frac{A}{\pi\, i} \exp\left[ i\, A\, B\, \left( {\bf r}_\text{in}^2 + {\bf r}_\text{out}^2\right) - 2\, i\, A\, 
		{\bf r}_\text{in}\cdot{\bf r}_\text{out}\right]\, ,
		\label{eq4.4}
\end{equation}
where 
\begin{equation}
	A = \frac{\mu\Omega}{2 \sin\left[ \Omega \left(\bar{t}-t \right)\right]}\, ,\qquad
	B = \cos\left[ \Omega \left(\bar{t}-t \right)\right]
	\label{eq4.5}
\end{equation}
are written in terms of the complex-valued oscillator frequency
\begin{equation}
  \Omega = \sqrt{\frac{c(z) C_F \hat{\bar{q}} }{2i\mu}}\,,\,\,\,\,\,\, c(z) \equiv 1 - \frac{N_c}{C_F} z(1-z) .
	\label{eq4.6}
\end{equation}
With the ansatz \eqref{eq4.3}, all transverse integrals of \eqref{eq2.2} can be done analytically.

\subsection{Analytical expression for the splitting function}
\label{sec4.2}
Following~\cite{Wiedemann:2000tf}, we split the two longitudinal  integrals in \eqref{eq2.2}  into six parts 
\begin{equation}
	\int_{t_\text{init}}^{t_\infty} \int_t^{t_\infty}  = \int_{t_\text{init}}^{0} \int_t^{0}  +  \int_{t_\text{init}}^{0} \int_0^L
+  \int_{t_\text{init}}^{0} \int_L^{t_\infty}  + \int_0^L \int_t^L +  \int_0^L \int_L^{t_\infty} +  \int_L^{t_\infty}  \int_t^{t_\infty} \, .
 	\label{eq4.7}
\end{equation}
We label the resulting contributions to \eqref{eq2.2} by the six terms 
\begin{equation}
	\left(\frac{1}{Q^2}\, P_{g \to c\, \bar{c}} \right)^{\text{tot}} = I_1 + I_2 +  I_3 + I_4 +  I_5 + I_6\, .
 	\label{eq4.8}
\end{equation}
The first three terms would enter in a calculation with $t_\text{init} = - \infty$, as discussed in section~\ref{sec3.4.2}.  Here,  however, we focus on the case that the parent gluon is produced within the medium ($t_\text{init} = 0$). In this case, $I_1 = I_2 = I_3 = 0$. 

Since there is no medium at longitudinal positions larger than $L$, the contribution $I_6$ involves only free propagation. 
From a calculation closely related to that of section~\ref{sec2.2}, one finds
\begin{equation}
	I_6 = \left(\frac{1}{Q^2}\, P_{g \to c\, \bar{c}} \right)^{\text{vac}} =   \frac{1}{2 Q^4} \left[ \left( m_c^2 + \bm{\upkappa}^2 \right) \frac{z^2 + (1-z)^2}{z(1-z)}  + 2 m_c^2 \right]\, .
 	\label{eq4.9}
\end{equation}
The medium-modification of the splitting function for multiple soft scattering is therefore given by the two contributions
\begin{equation}
	\left(\frac{1}{Q^2}\, P_{g \to c\, \bar{c}} \right)^{\text{med}}_{\text{m.s.}} =I_4 +  I_5\, .
 	\label{eq4.10}
\end{equation}
Inserting \eqref{eq4.4} into \eqref{eq2.2}, simplifying the analytical expressions and changing to dimensionless variables, one finds 
for $\hat{{q}}(\xi) = \hat{{q}}\, \theta\left( L-\xi \right)$ after a lengthy calculation (see Appendix~\ref{app:Isoftderiv})
\begin{align}
I_4=& \frac{1}{e_g \hat{q} L c(z)}\, \mathfrak{Re} \int_0^1 d\mathfrak{y} \frac{-i\tilde{\Omega}}{ \sin\tilde{\Omega} \mathfrak{y}}
e^{-i \frac{\tilde{m}_\text{c}^2}{2 \tilde{\mu}} \mathfrak{y} }	 		 \Bigg\{  \tilde{m}_c^2\left( \text{Ei}\left[ \frac{\tilde{\bm{\upkappa}}^2\tan\tilde{\Omega}\mathfrak{y}}{i 2\tilde{\mu}\tilde{\Omega}} \right] - \text{Ei}\left[\frac{\tilde{\bm{\upkappa}}^2 \tan\tilde{\Omega} \mathfrak{y}}{  i 2\tilde{\mu}\tilde{\Omega} - c(z)(1-\mathfrak{y})  \tan\tilde{\Omega} \mathfrak{y}} \right] \right)\nonumber\\
&-\frac{2i\tilde{\Omega}\tilde{\mu} }{\sin \tilde{\Omega} \mathfrak{y}}	(z^2 + (1-z)^2)\Bigg( \text{Ei}\left[ \frac{\tilde{\bm{\upkappa}}^2\tan\tilde{\Omega}\mathfrak{y}}{i 2\tilde{\mu}\tilde{\Omega}} \right] - \text{Ei}\left[\frac{\tilde{\bm{\upkappa}}^2 \tan\tilde{\Omega} \mathfrak{y}}{  i 2\tilde{\mu}\tilde{\Omega} -c(z)(1-\mathfrak{y})  \tan\tilde{\Omega} \mathfrak{y}} \right] \nonumber\\
&\hspace{4cm}-\exp\left[\frac{\tilde{\bm{\upkappa}}^2\tan\tilde{\Omega} \mathfrak{y}}{ i 2\tilde{\mu}\tilde{\Omega}   } \right]  + \frac{\exp\left[ \frac{\tilde{\bm{\upkappa}}^2 \tan\tilde{\Omega} \mathfrak{y}}{ i 2\tilde{\mu}\tilde{\Omega} - c(z)(1-\mathfrak{y})  \tan\tilde{\Omega} \mathfrak{y}} \right]}{1-\frac{1-\mathfrak{y}}{i2 \tilde{\mu} \tilde{\Omega} }c(z) \tan \tilde{\Omega} \mathfrak{y}} \Bigg)\Bigg\}\, ,\label{eq4.11}\\
I_5=&\frac{1}{e_g \hat{q} L}\, \mathfrak{Re}\,  \int_{0}^{1} d\bar{\mathfrak{y}}\, \frac{-i}{\cos \tilde{\Omega} \bar{\mathfrak{y}}}
e^{- i \frac{\tilde{m}_\text{c}^2}{2 \tilde{\mu}}\bar{\mathfrak{y}}  + \frac{\tilde{\bm{\upkappa}}^2\tan \tilde{\Omega} \bar{\mathfrak{y}} }{i2\tilde{\mu} \tilde{\Omega}}}
     \frac{1}{(\tilde{m}_c^2+\tilde{ \bm{\upkappa}}_c)}
     \left[ \frac{\tilde{\bm{\upkappa}}^2}{\cos\tilde{\Omega}\bar{ \mathfrak{y}}} ( z^2 + (1-z)^2 )+\tilde{m}_\text{c}^2\right]\, .\label{eq4.12}
\end{align}
where Ei denotes the exponential integral and the compact expressions  \eqref{eq4.11} and \eqref{eq4.12} were obtained by changing to dimensionless integration variables
$\mathfrak{y} = \frac{\bar{t}-t}{L}$, $\overline{\mathfrak{y}} =  \frac{L-t}{L}$ and introducing the 
dimensionless variables ($\hat{q} = C_F \hat{\bar{q}}$)
\begin{equation}
e_g = \frac{2E_g}{\hat{q} L^2}\,,\quad \tilde{m}_\text{c}^2 = \frac{m_\text{c}^2}{\hat{q} L}\, ,\quad
	\tilde{\bm{\upkappa}}^2 = \frac{\bm{\upkappa}^2}{\hat{q}L}\,
,\quad 	\tilde{\Omega} = \Omega L\,
,\quad 	\tilde{\mu} = \frac{\mu}{\hat{q} L^2}\,
 .	\label{eq4.13}
\end{equation}

\begin{figure}[t]
    \centering
       \subfig{a}{\includegraphics[width=.49\textwidth]{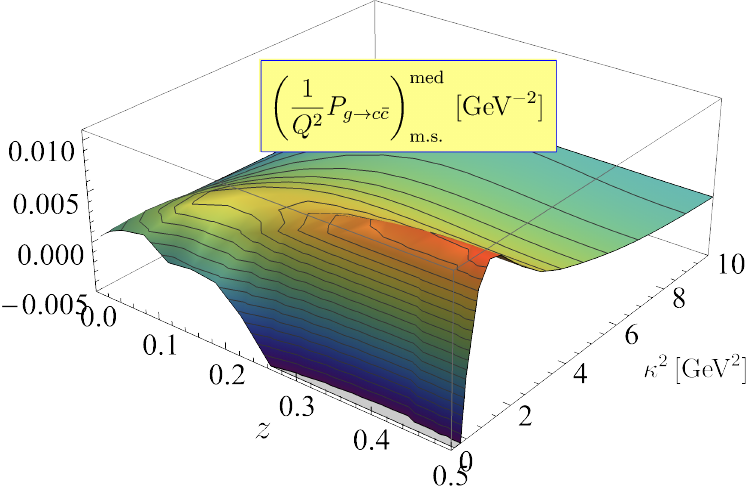}}%
        \subfig{b}{\includegraphics[width=.49\textwidth]{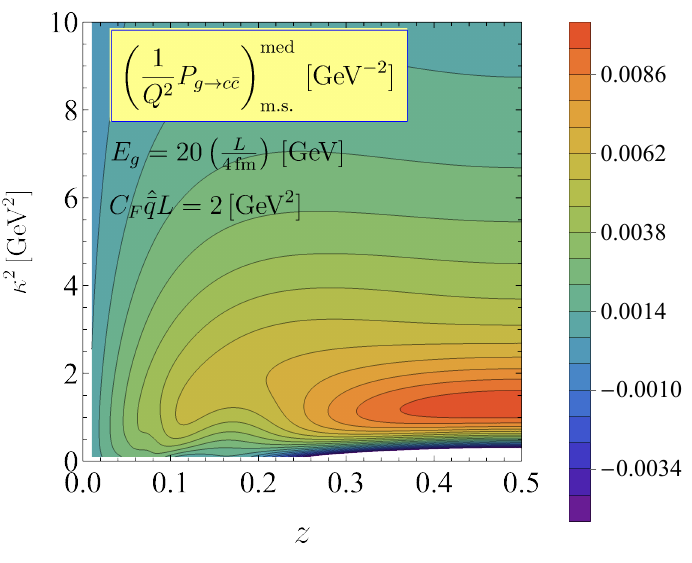}}
    \caption{The medium-modified $g\to c\bar{c}$ splitting function evaluated in the saddle point approximation~\eqref{eq4.10}-\eqref{eq4.13}. The two panels show different representations of the same calculation. }
    \label{figMS1}
\end{figure}
\begin{figure}[t]
    \centering
       \subfig{a}{\includegraphics[width=.49\textwidth]{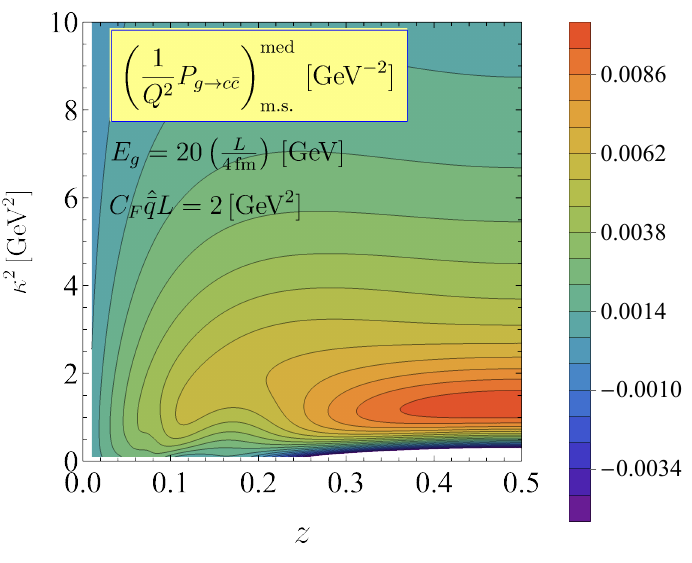}}%
        \subfig{b}{\includegraphics[width=.49\textwidth]{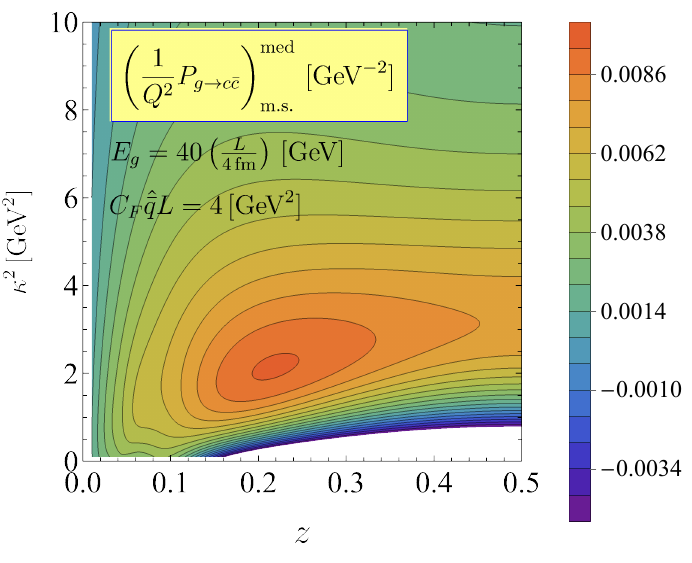}}
        \subfig{c}{\includegraphics[width=.49\textwidth]{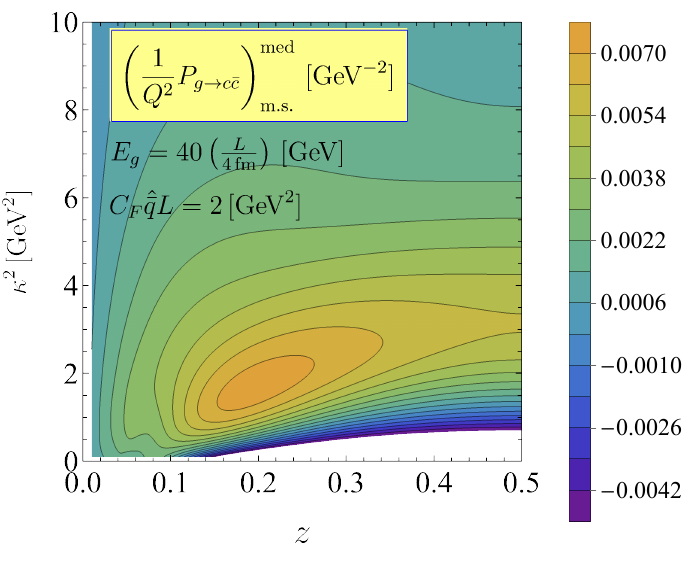}}%
        \subfig{d}{\includegraphics[width=.49\textwidth]{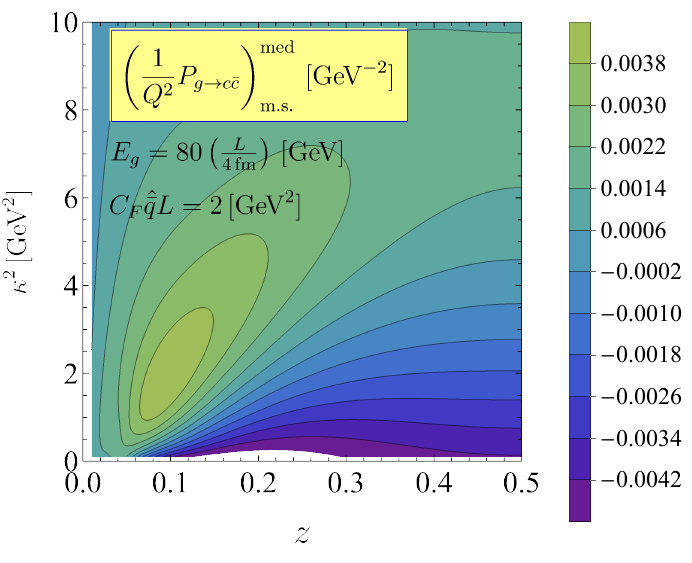}}
    \caption{The medium-modified $g\to c\bar{c}$ splitting function evaluated in the saddle point approximation~\eqref{eq4.10}-\eqref{eq4.13}. (a) is the same as Fig.~\ref{figMS1}(b). Different panels show results for different values of the typical momentum squared $\hat{q}L$ transferred from the medium and for different gluon energies $E_g$. }
    \label{figMS2}
\end{figure}

\subsection{Numerical results for a static brick}
\label{sec4.3}
Equations \eqref{eq4.11} and \eqref{eq4.12} can be evaluated numerically and they determine the medium modification \eqref{eq4.10} of the $g\to c\bar{c}$ splitting for a static medium in the multiple soft scattering approximation. In general, \eqref{eq4.10} is a function of $\bm{\upkappa}^2$, $z$, $m_c^2$, the gluon energy $E_g$ (which enters though $e_g$ and $\mu = E_g z (1-z)$), the quenching parameter $\hat{q}$, and the in-medium path length $L$. It can be expressed  in terms of a single dimensionful quantity  $\hat{q}L$ and the dimensionless parameters $e_g,\tilde{\bm{\upkappa}}^2,\tilde{m}_\text{c}^2$ and $z$.

Numerical results for the medium-modified splitting function in the multiple soft scattering approximation are shown in Figs.~\ref{figMS1} and~\ref{figMS2}. 
To ease the physical interpretation, these results are not displayed as a function of the  dimensionless parameter $e_g$, but instead for physical gluon energies in units of $\tfrac{L}{\text{4\,fm}} \text{GeV}$. 
They reveal the same features observed in the $N=1$ opacity calculation in Figs.~\ref{figN1} and~\ref{figN12}. In particular, one observes momentum broadening
on a scale $\langle \bm{\upkappa}^2\rangle \sim \hat{q}L$, and the splitting function
is enhanced over a wide range of $\bm{\upkappa}$ and $z$.
There is also numerical evidence for a formation time-dependent suppression: 
the medium modification decreases for increasing $E_g$ since the formation time becomes larger than the in-medium path length $L$.

\begin{figure}[t]
    \centering
       \subfig{a}{\includegraphics[width=.49\textwidth]{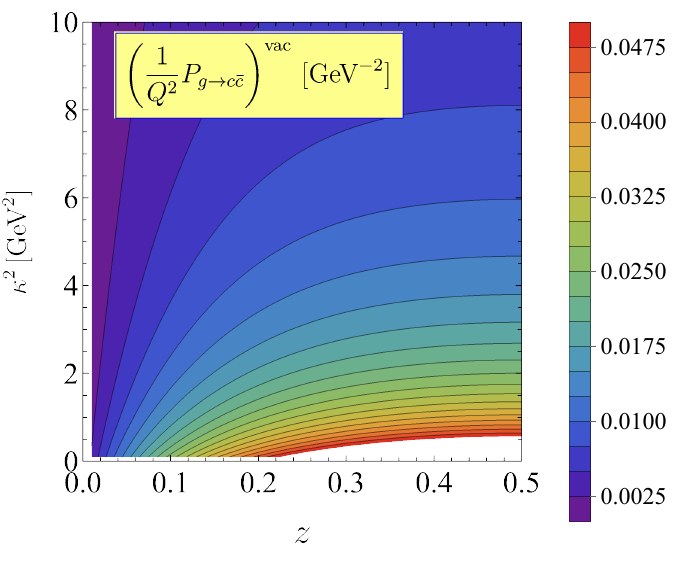}}%
 \subfig{b}{\includegraphics[width=.49\textwidth]{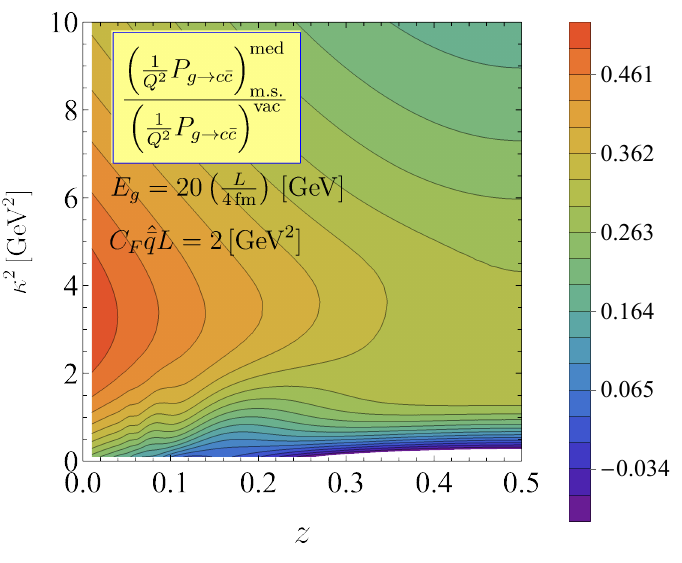}}
    \caption{(a)  The vacuum splitting function \eqref{eq2.14} in units of  $\text{GeV}^{-2}$.  (b) For an averaged squared momentum transfer 
    $\hat{q}L = 2\, \text{GeV}^{2}$ from the medium, the $g\to c\bar{c}$ splitting is enhanced within a wide range of intermediate transverse momentum $\bm{\upkappa}$.
    }
    \label{figMSratio}
\end{figure}

In displaying results of the $N=1$ opacity expansion in section~\ref{sec3.4}, we had fixed the opacity $c_\text{opaque}=1$ in \eqref{eq3.25} by hand. 
This is different for the saddle point approximation studied here, where both the shape and norm of $\left(\frac{1}{Q^2}\, P_{g \to c\, \bar{c}} \right)^{\text{med}}_{\text{m.s.}} $ are calculated and depend on $\hat{q}L$ and $e_g$. The absolute scale displayed in Figs.~\ref{figMS1} and \ref{figMS2} is physically meaningful. To understand whether this scale is sizeable or negligible, we compare $\left(\frac{1}{Q^2}\, P_{g \to c\, \bar{c}} \right)^{\text{med}}_{\text{m.s.}} $ in Fig.~\ref{figMSratio} to the vacuum splitting function. The vacuum  
splitting function is depleted for very small $\bm{\upkappa}^2$ due to the medium modification, and enhanced for a broad range of intermediate 
transverse momenta $\bm{\upkappa}^2 \sim \hat{q}L$. Qualitatively this can be understood as the combination of broadening that moves $c\bar{c}$ pairs from small to larger $\bm{\upkappa}^2$, and enhanced $c\bar{c}$ radiation. At very large $\bm{\upkappa}^2$, the medium modification shows the $\propto 1/ \bm{\upkappa}^4$-dependence characteristic of a power correction while the vacuum contribution shows the $\propto 1/ \bm{\upkappa}^2$-dependence characteristic of a leading logarithmic contribution. Therefore, the ratio shown in Fig.~\ref{figMSratio}(b) asymptotes to zero in the limit of large $\bm{\upkappa}^2$.

\subsection{The \texorpdfstring{$g\to c\bar{c}$}{g->ccbar} splitting in an expanding medium}

In this section we provide the results for the medium-modified splitting function for an arbitrary profile $\hat{q}(\xi)$ with finite support over $0<\xi<L$.
We first introduce the dimensionful quantities characterising the typical momentum, energy and length scales associated with~$\hat{q}(\xi)$, 
\begin{align}
  \left<q^2\right>_\text{med} = C_F\int_{0}^L d\xi \hat{\bar{q}}(\xi), \quad
  E_\textsc{bdmps}= C_F\int_{0}^L d\xi \hat{\bar{q}}(\xi)\xi, \quad
\bar{L} \equiv \frac{2E_\textsc{bdmps}}{\left<q^2\right>_\text{med}}\, .
\label{eq4.14}
\end{align}
Note that in this section $L$ is an arbitrary cutoff, while $\bar L<L$ is the characteristic length scale of the $\hat{q}$ distribution.
The use of a finite cutoff $L$ is necessary for splitting the integrals as in \eqref{eq4.7}.

Once again we will write integrals $I_4$ and $I_5$ in dimensionless units. In particular, we define $\mathfrak{y} = t/\bar L$ and $\bar{\mathfrak{y}} = \bar{t}/\bar L$ and use the rescaled variables:
\begin{align}
&e_g = \frac{E_g}{E_\textsc{bdmps}}\,,\quad \tilde{m}_\text{c}^2 = \frac{m_\text{c}^2}{\left<q^2\right>_\text{med}}\, ,\quad
	\tilde{\bm{\upkappa}}^2 = \frac{\bm{\upkappa}^2}{\left<q^2\right>_\text{med}}\,,\nonumber\\
 &\tilde{\Omega}(\xi)= \Omega(\xi)\bar{L}\,
,\quad 	2\tilde{\mu} = \frac{2\mu}{2E_\textsc{bdmps}},\quad \alpha = \frac{L}{\bar{L}}\,.\label{eq4.15}
\end{align}

\begin{figure}
    \centering
    \subfig{a}{\includegraphics[width=.49\textwidth]{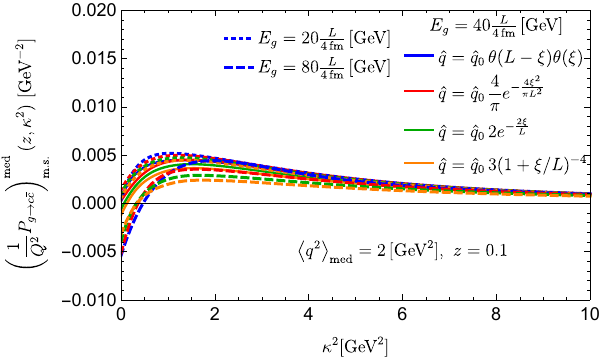}}%
    \subfig{b}{\includegraphics[width=.49\textwidth]{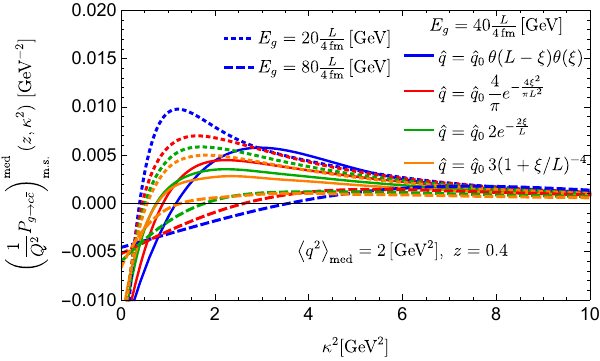}}
    \subfig{c}{\includegraphics[width=.49\textwidth]{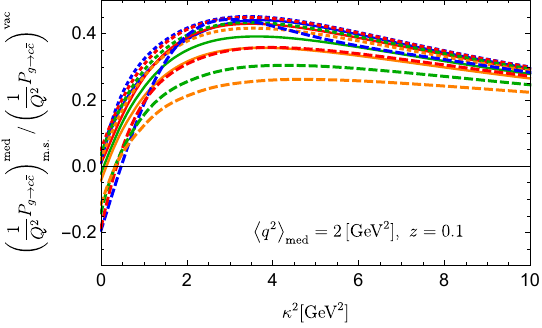}}%
    \subfig{d}{\includegraphics[width=.49\textwidth]{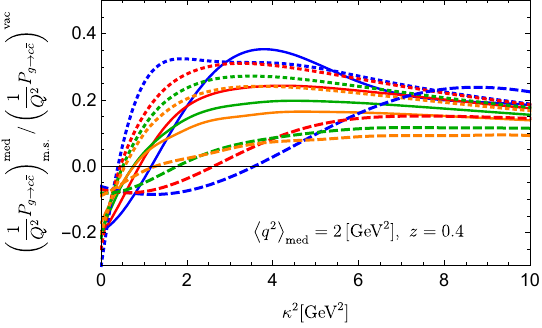}}
 \caption{ Medium-modified splitting functions for three different expanding media (red, green and orange curves) compared to an equivalent static medium (blue curves). Results for three different gluon energies are compared for (a) $z=0.1$ and (b) $z=0.4$ as a function of ${\bm \upkappa}^2$. Panels (c) and (d) show the results of (a) and (b) divided by the vacuum splitting function.
    }\label{fig:generalP}
\end{figure}
Leaving the details of the derivation to the Appendix~\ref{app:Igensoftderiv} we present the final result: 
\begin{align}
		I_4&=\mathfrak{Re}\, \frac{1}{e_g^2\left<q^2\right>_\text{med} }\, \int_{0}^{\alpha} d\mathfrak{y} \int_\mathfrak{y}^{\alpha} d\bar{\mathfrak{y}}\, 
    \exp\left[{- i \frac{\tilde{m}_c^2}{2 \tilde{\mu}} (\bar{\mathfrak{y}}-\mathfrak{y}) - i\frac{\tilde{\bm{\upkappa}}^2}{2\tilde{\mu}}\{\partial_{\bar{\mathfrak{y}}}\log \tilde{D}(\bar{\mathfrak{y}},\mathfrak{y})   -\int_{\bar{\mathfrak{y}}}^{\alpha}  d\tilde{\xi} \tilde{\Omega}^2(\tilde{\xi}) \}^{-1}}\right]\,
		\nonumber \\
		&\qquad \times \Bigg[ 2 i \tilde{\mu}\frac{(1-z)^2 + z^2}{z(1-z)} \left( \frac{-i \tilde{\bm{\upkappa}}^2}{2 \tilde{\mu}}
		\frac{\partial_{\bar{\mathfrak{y}}} \log \tilde{D}(\bar{\mathfrak{y}},\mathfrak{y}) }{\{\partial_{\bar{\mathfrak{y}}}\log \tilde{D}(\bar{\mathfrak{y}},\mathfrak{y})   -\int_{\bar{\mathfrak{y}}}^{\alpha} d\tilde{\xi} \tilde{\Omega}^2(\tilde{\xi})\}^2} + 
		\frac{\int_{\bar{\mathfrak{y}}}^{\alpha} d\tilde{\xi} \tilde{\Omega}^2(\tilde{\xi})}
		{\partial_{\bar{\mathfrak{y}}}\log \tilde{D}(\bar{\mathfrak{y}},\mathfrak{y})   -\int_{\bar{\mathfrak{y}}}^{\alpha} d\tilde{\xi} \tilde{\Omega}^2(\tilde{\xi})}\right) \nonumber\\
		&\qquad \qquad \qquad +  \frac{\tilde{m}_c^2}{z(1-z)}\tilde{D}(\bar{\mathfrak{y}},\mathfrak{y})
		\Bigg]
		\frac{1}{ \tilde{D}(\bar{\mathfrak{y}},\mathfrak{y})^2 \left(\partial_{\bar{\mathfrak{y}}}\log \tilde{D}(\bar{\mathfrak{y}},\mathfrak{y})   -\int_{\bar{\mathfrak{y}}}^{\alpha}  d\tilde{\xi} \tilde{\Omega}^2(\tilde{\xi}) \right)}\, ,\label{eq4.16}\\
		I_5&=\mathfrak{Re}\, \frac{1}{e_g\left<q^2\right>_\text{med}}\, \int_{0}^{\alpha} d\mathfrak{y} \frac{-i}{\tilde{m}_c^2+\tilde{\bm{\upkappa}}^2}
    \exp\left[{- i \frac{\tilde{m}_c^2}{2 \tilde{\mu}} (\alpha-\mathfrak{y})- i\frac{\tilde{\bm{\upkappa}}^2}{2\tilde{\mu}}\{ \left.\partial_{\bar{\mathfrak{y}}}\log \tilde{D}(\bar{\mathfrak{y}},\mathfrak{y})\right|_{\bar{\mathfrak{y}}=\alpha}\}^{-1}} \right]
		  \nonumber \\		  		
       &\qquad\qquad \times \left[ \frac{ \tilde{\bm{\upkappa}}^2  \left(z^2 + (1-z)^2 \right)}{\left.\partial_{\bar{\mathfrak{y}}}\tilde{D}(\bar{\mathfrak{y}},\mathfrak{y})\right|_{\bar{\mathfrak{y}}=\alpha}}  +  \tilde{m}_c^2  \right] \frac{1}{\left.\partial_{\bar{\mathfrak{y}}} \tilde{D}(\bar{\mathfrak{y}},\mathfrak{y})\right|_{\bar{\mathfrak{y}}=\alpha}}
       \, ,\label{eq4.17}
\end{align}
where $\tilde D(\bar{\mathfrak{y}},\mathfrak{y})$ solves the second order differential equation
\begin{equation}
  \frac{\partial^2}{\partial^2 \tilde{\xi}}\tilde{D}(\tilde{\xi},\mathfrak{y}) = - \tilde{\Omega}^2(\tilde{\xi}) \tilde{ D}(\tilde{\xi},\mathfrak{y}),\quad \tilde D(\mathfrak{y},\mathfrak{y})=0,\quad \left.\frac{\partial \tilde D(\tilde\xi,\mathfrak{y})}{\partial\tilde \xi}\right|_{\tilde\xi=\mathfrak{y}}=1,
  \label{eq4.18}
\end{equation}
with $\tilde{\Omega}^2(\tilde{\xi}) = \frac{1}{2i\tilde{\mu}} c(z)\frac{ \hat{q}(\tilde{\xi})\bar L}{\left<q^2\right>_\text{med}}$.  If $\hat q$ is constant, then $L=\bar L$, $\tilde D(\bar{\mathfrak{y}},\mathfrak{y}) = \tilde \Omega^{-1}\sin \tilde \Omega( \bar{\mathfrak{y}}-\mathfrak{y})$ and we recover the results of the previous section.

Fig.~\ref{fig:generalP} shows results for $\hat{q}(\xi)$ profiles of different expansion scenarios (see Fig.~\ref{fig:interference_factor}(a)) compared to results for an equivalent static brick with the same values of $\left<q^2\right>_\text{med}$ and $E_\textsc{bdmps}$. Generically a faster expansion (orange) leads to a reduction of the medium modification, since in that case the hottest phase will be at a time before the splitting occurs. At smaller energies or small $z$, however, the splitting time is so short that all scenarios lead to almost identical results. Even for larger energies and more equal splitting fractions, the expanding scenarios and a static medium show 
the same qualitative dependencies on ${\bm \upkappa}^2$ and the same order of magnitude effect. For some applications,
it may therefore be sufficient to explore expanding scenarios by calculating the medium modification for an equivalent static brick. If higher accuracy of the result is desired, the general formulas \eqref{eq4.16}, \eqref{eq4.17} presented here make it possible to avoid resorting to the simpler model of a static brick.

\section{Medium-enhanced \texorpdfstring{$g\to c\bar{c}$}{g->ccbar}-production }
\label{sec6}
The number of heavy quark--anti-quark pairs within gluon jets has been calculated first in the early days of QCD in Refs.~\cite{Mueller:1985zz,Mueller:1985zp}. This calculation has two parts. First, one determines analytically the number of gluons of offshellness $Q^2$ inside the gluon jet within the so-called modified leading logarithmic approximation. Then, one multiplies this number with the $g\to c\bar{c}$ splitting function and one integrates over the available phase space. 

The modern way of arriving at the number of heavy quark--anti-quark pairs is arguably via Monte Carlo simulations of jets in multi-purpose event generators. Since $c\bar{c}$ pairs can only be observed via their hadronic or leptonic decay products, determining the yield of $c\bar{c}$ pairs in event generators has the advantage of studying hadronic final states whose detectability can be assessed. A Monte Carlo study lies outside the scope of the present work. However, to prepare for the future use of our calculations in Monte Carlo event generators, we shall discuss in section~\ref{sec6.1} 
how $P_{g\to c\bar{c}}^\text{med}$ affects the 
Sudakov factor~\eqref{eq1.3} which is a key ingredient for describing the parton shower evolution. Following the line of logic presented in this section, in a companion paper~\cite{Attems:2022otp} we have used the medium modification $P_{g\to c\bar{c}}^\text{med}$ calculated here along with a standard Monte Carlo event generator to estimate the impact of this modified $g\to c\bar{c}$ splitting in a modified parton shower.
This has allowed us to discuss experimental strategies for how to detect the 
medium-enhanced $g\to c\bar{c}$ production in realistic jet event samples in heavy-ion collisions.

\subsection{A reweighting prescription for the medium-modified \texorpdfstring{$g\to c\bar{c}$}{g->ccbar} branching probability }
\label{sec6.1}

We have found in section~\ref{sec3} that $P^{\text{med}}_{g\to c\bar{c}} \sim \mathcal{O}\left(\tfrac{\langle {\bf q}^2\rangle_\text{med}}{Q^2}\right) \sim  \mathcal{O}\left(\tfrac{ m_c^2} {Q^2}\right) $. This motivates us to include
$P^{\text{med}}_{g\to c\bar{c}} $ in the Sudakov factor \eqref{eq1.3} on equal par with the mass term $\propto \tfrac{m_c^2}{Q^2}$ of the vacuum splitting function. 
We write the total no-branching probability, including medium-modifications, as
\begin{eqnarray}
	S^{\text{tot}}_{g\to X\cup \lbrace c\bar{c}\rbrace} &=& \exp \left[ - \frac{\alpha_s}{2\pi}  \int \frac{dQ^2}{Q^2}\, dz\, \left( P^{\text{vac}}_{g\to c\bar{c}}  + P^{\text{med}}_{g\to c\bar{c}} 
	+ P^{\text{vac}}_{g\to X}  +  P^{\text{med}}_{g\to X}  \right) \right]
	\nonumber \\
	&=& S^{\text{tot}}_{g\to c\bar{c}} \, S^{\text{tot}}_{g\to X}\, .
		\label{eq7.2}
\end{eqnarray}
Here $g \to X$ includes all channels the gluon can split into, except the $c\bar{c}$ one, and $g\to X\cup \lbrace c\bar{c}\rbrace$ denotes the splitting into any channel including the $c\bar{c}$ one.

We ask how the inclusion of $P^{\text{med}}_{g\to c\bar{c}} $ can be expected to change the yield of $c\bar{c}$-pairs per parton shower. To this end, we consider the probability $\left( 1- S^{\text{tot}}_{g\to c\bar{c}} \right)\, S^{\text{tot}}_{g\to X}$
that the next splitting of the gluon is into a $c\bar{c}$ pair. Compared to the same process in the vacuum, the medium changes this probability by a factor
\begin{eqnarray}
	\frac{\left( 1- S^{\text{tot}}_{g\to c\bar{c}} \right)\, S^{\text{tot}}_{g\to X}}{
	\left( 1- S^{\text{vac}}_{g\to c\bar{c}} \right)\, S^{\text{vac}}_{g\to X} } 
		=
	\frac{\int  \frac{dQ^2}{Q^2} dz \left( P_{g\to c\bar{c}}^{\rm vac} + P_{g\to c\bar{c}}^{\rm med}\right) }{\int  \frac{dQ^2}{Q^2} dz P_{g\to c\bar{c}}^{\rm vac} }
	 \exp\left[- { \frac{\alpha_s}{2\pi}\!\int\! \frac{dQ^2}{Q^2} dz  P_{g\to X}^{\rm med} } \right].
	\label{eq7.3}
\end{eqnarray}
Here, we have assumed\footnote{This is justified if the exponent 
\begin{equation}
	\frac{\alpha_s}{2\pi} \int_{4 m_c^2}^{Q_h^2}  \frac{dQ^2}{Q^2} \int_0^1 dz P_{g\to c\bar{c}}^\text{vac} = \frac{\alpha_s}{2\pi} \left( \frac{1}{3} \log \frac{Q^2_h}{4 m_c^2} + \frac{Q_h^2 - 4 m_c^2}{4 Q_h^2}  \right)
	\label{eq7.4}
\end{equation}
is much smaller than one. The exponent \eqref{eq7.4} reaches a value $\approx 0.13$ for $\alpha_s = 0.3$ and $Q_h^2 = 10^4\, \text{GeV}^2$, and it is smaller for smaller $Q_h^2$ and/or smaller $\alpha_s$. So, for all resolution scales $Q_h^2 < 10^4\, \text{GeV}^2$ reachable in the parton showers accessible in heavy ion collisions, this assumption is justified. 
} that the branching into a $c\bar{c}$ pair is sufficiently rare, so that $\left( 1- S^{\text{tot}}_{g\to c\bar{c}} \right)$ and $\left( 1- S^{\text{vac}}_{g\to c\bar{c}} \right)$ can be expanded to first order in the exponent.
On general grounds, one expects that the medium modification of all splitting functions is  suppressed by additional powers of $1/Q^2$ \cite{Kastella:1989vd}. We have seen this explicitly for the $g\to c\bar{c}$ splitting function in \eqref{eq3.29}.
Since
the same is true for the medium modification of other gluon splittings $g\to X$ this allows us to estimate
\begin{equation}
\frac{S^{\text{tot}}_{g\to X}}{ S^{\text{vac}}_{g\to X}}
=  e^{ - \frac{\alpha_s}{2\pi}\!\int_{Q_l^2}^{Q_h^2}\! \frac{dQ^2}{Q^2} dz  P_{g\to X}^{\rm med} } 
\sim e^{ - \tfrac{\alpha_s}{2\pi} \# \left(\tfrac{\hat{q}L}{Q_l^2} - \tfrac{\hat{q}L}{Q_h^2}  \right) } = 1 - \mathcal{O}(\alpha_s)\, .
	\label{eq7.5}
\end{equation}
As this term depends only on $P_{g\to X}^{\rm med} $ and not on $P_{g\to X}^{\rm vac}$, it is a power correction
$\mathcal{O}\left(\alpha_s\tfrac{\hat{q}L}{Q^2}\right)$ one can expand in,  irrespective of the range over which the $Q^2$-integral is performed.  From \eqref{eq7.3} and \eqref{eq7.5}, to leading order in $\alpha_s$ the medium modifies the probability that the next splitting is 
$g\to c\bar{c}$ by a factor
\begin{equation}
1+ \frac{\int dQ^2   \int dz  \left( \frac{1}{Q^2} P_{g\to c\bar{c}} \right)^{\rm med} }{\int dQ^2   \int dz  \left( \frac{1}{Q^2} P_{g\to c\bar{c}} \right)^{\rm vac} }\, .
\label{eq7.6a}
\end{equation}
We wrote equations \eqref{eq7.2}, \eqref{eq7.3} and \eqref{eq7.6a}
with indefinite integrals since branching and no-branching probabilities can be given for any phase space region. In particular, keeping the $Q^2$- and $z$-dependence fully differential, we define the weight
\footnote{Similar to other implementations of medium-modified Monte Carlo parton showers, this reweighting proposal is based on a simple replacement $P^{\rm vac} \to P^{\rm vac} + P^{\rm med}$ in a vacuum parton shower. In particular, the upper and lower bounds ($Q_h^2$ and $Q_l^2$) in the Sudakov factor \eqref{eq1.3} are specified as in the vacuum, i.e., splitting functions are sampled only for $\bm\upkappa^2$-values in the range $Q_l^2 z (1-z) - m_c^2 < \bm\upkappa^2 < Q_h^2 z (1-z) - m_c^2$. 
However, physically, scattering on the medium can place $c\bar{c}$ pairs outside the range allowed by vacuum kinematics.
In principle, this can lead to an artificial depletion or enhancement of the yield of $c\bar{c}$ pairs produced in such MC simulations. 
As explained in section~\ref{sec2.3}, the medium modification of the splitting function \eqref{eq2.2} always increases the total yield of $c\bar{c}$ pairs, if integrated over all phase space. }
\begin{equation}
    w^\text{med}_{g\to c\bar{c}} (E_g,\bm{\upkappa}^2,z) =  1 + 
    \frac{ \left( \frac{1}{Q^2} P_{g\to c\bar{c}} \right)^{\rm med}(E_g,\bm{\upkappa}^2,z) }{\left( \frac{1}{Q^2} P_{g\to c\bar{c}} \right)^{\rm vac}(\bm{\upkappa}^2,z) }\, .
    \label{eq7.6b}
\end{equation}
Reweighting Monte Carlo-generated events with $w^\text{med}_{g\to c\bar{c}}$ is therefore one way to implement the enhanced $c\bar{c}$-production and the transverse momentum broadening described by $P^\text{med}_{g\to c\bar{c}}$.
The weight $w^\text{med}_{g\to c\bar{c}}$ deviates from unity by the ratio plotted in Fig.~\ref{figMSratio}(b).
This weight does not account for effects of parton energy loss on the outgoing $c\bar{c}$-pair 
or on other components of the parton shower. For this, a more complete parton shower formulation is required that includes the effects of medium-modified $g\to gg$, $q\to qg$ and $g\to q\bar{q}$ splittings. Such a medium-modified parton shower formulation would also allow one to quantify the accuracy of truncating \eqref{eq7.5} at leading
$\mathcal{O}(\alpha_s)$ which is at the basis of the proposed reweighting with \eqref{eq7.6b}.

\begin{figure}
    \centering
    \includegraphics[width=0.8\textwidth]{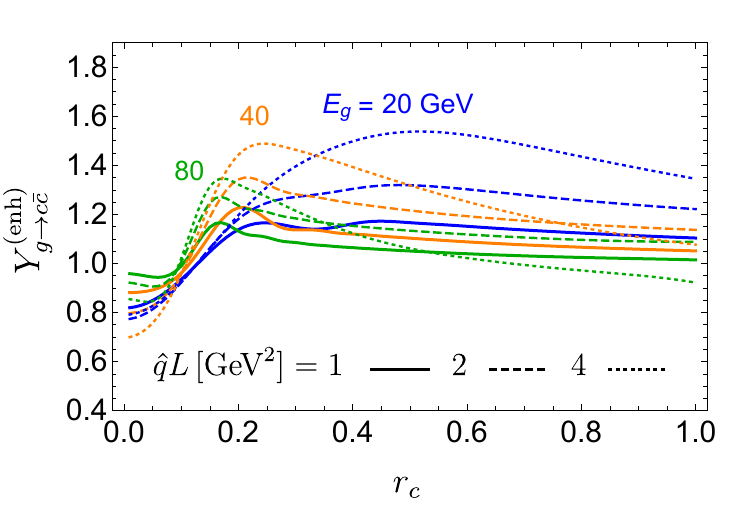}
    \caption{The factor $Y_{g\to c\bar{c}}(r_c) $ of  \eqref{eq7.6c} that characterises the relative enhancement of $g\to c\bar{c}$ splittings into pairs with angular separation $r_c$ in the QCD plasma, in the multiple soft scattering approximation. Different colours show different gluon energies and different line styles show different values of $\hat{q} L$.}
    \label{fig8}
\end{figure}

\subsection{The medium-modified \texorpdfstring{$g\to c\bar{c}$}{g->ccbar} in a typical jet opening cone}
\label{sec6.2}

The medium-modified $g\to c\bar{c}$ splitting functions plotted in sections~\ref{sec3} and \ref{sec4} 
display the signatures of both medium-induced momentum broadening and enhanced $c\bar{c}$-radiation. 
Here, we illustrate how these two effects manifest themselves in the phase space region available in a typical jet opening cone.

To this end, we consider splittings $g\to c\bar{c}$ for fixed gluon energy $E_g$, and we relate the transverse momentum $\bm{\upkappa}$ to the angular separation $r_c = \sqrt{\Delta \eta^2 + \Delta\phi^2}$ between the $c$ and $\bar{c}$ via~\footnote{This is valid when the transverse momentum $\bm{\upkappa}$ is smaller than the $c$ and $\bar{c}$ momenta $p_c$ and $p_{\bar{c}}$ so that
\begin{equation}
    r_c = \sin^{-1} \left(\frac{\vert \bm{\upkappa}\vert}{|p_c|}\right)+ \sin^{-1} \left(\frac{\vert \bm{\upkappa}\vert}{|p_{\bar c} |}\right) \sim \vert \bm{\upkappa}\vert \left(\frac{1}{|p_c|} + \frac{1}{|p_{\bar c} |} \right)\,.
\end{equation}}
\begin{equation}
    \vert \bm{\upkappa}\vert = r_c E_g z (1-z)\, .
\end{equation}
After changing integration variables in \eqref{eq7.6a},
\begin{equation}
    \int dQ^2 \int dz \left[ \dots \right] =  \int dz \int \frac{d\bm{\upkappa}^2}{z(1-z)} \left[ \dots \right] = E_g^2 \int dz \int dr_c^2 \, z (1-z) \left[ \dots \right]\, ,
\end{equation}
we consider the modification factor of the $c\bar{c}$ yield
\begin{equation}
	Y_{g\to c\bar{c}}(r_c,E_g) = 1+ \frac{ \int dz \, z(1-z) \left( \frac{1}{Q^2} P_{g\to c\bar{c}} \right)^{\rm med} \vert_{\vert \bm{\upkappa}\vert = r_c E_g z(1-z)} }{ \int dz \, z(1-z) \left( \frac{1}{Q^2} P_{g\to c\bar{c}} \right)^{\rm vac} \vert_{\vert \bm{\upkappa}\vert = r_c E_g z(1-z)} }\, .
		\label{eq7.6c}
\end{equation}
The modification factor $Y_{g\to c\bar{c}}(r_c,E_g)$ depends on the gluon energy and is differential in the angular separation $r_c$ between charm and anti-charm quark, but is integrated over the entire longitudinal phase space $dz$. 
As seen in Fig.~\ref{fig8}, the plasma can deplete the number of nearly-collinear $c\bar{c}$ pairs at small $r_c$. This indicates that momentum broadening is more efficient in moving $c\bar{c}$ pairs out of this small phase space region than 
medium-induced $c\bar{c}$ radiation is in topping up the yield. As $r_c$ is increased, enhanced radiation soon dominates over the effects 
of momentum broadening and $Y_{g\to c\bar{c}}(r_c)$ rises rapidly above unity, see Fig.~\ref{fig8}. 
With increasing momentum transfer from the medium, i.e., with increasing $\hat{q}L$, 
the overall enhancement of the $c\bar{c}$ yield at large $r_c$ increases. This is more pronounced for smaller gluon energies. 

We emphasise however that further work is required to go from Fig.~\ref{fig8} to an experimentally-accessible distribution of $c\bar{c}$ pairs within a jet cone.
In particular, one needs to understand how gluons undergoing $g\to c\bar{c}$ splitting are distributed in 
$E_g$, $\bm{\upkappa}^2$ and $z$ within a reconstructed jet, and to relate these $c\bar{c}$ pairs to measurable hadronic or leptonic decay products. A first proof-of-principle that this is indeed possible will be given in a 
companion paper~\cite{Attems:2022otp}.

\section{Splittings into beauty and light-flavoured quarks}
\label{sec8b}
The derivation of the medium-modified splitting function \eqref{eq2.2} is based on the close-to-eikonal approximation which assumes that $z\, E_g,\, (1-z) E_g \gg \bm{\upkappa}\, , m_q$. Within this range of applicability, \eqref{eq2.2} holds irrespective of the quark mass. In particular, it applies to $g \to b\bar{b}$ splitting if the charm mass is replaced by the $b$-quark mass $m_b=4.18\,\text{GeV}$~\cite{ParticleDataGroup:2020ssz}.

\begin{figure}
    \centering
\subfig{a}{\includegraphics[height=.39\textwidth]{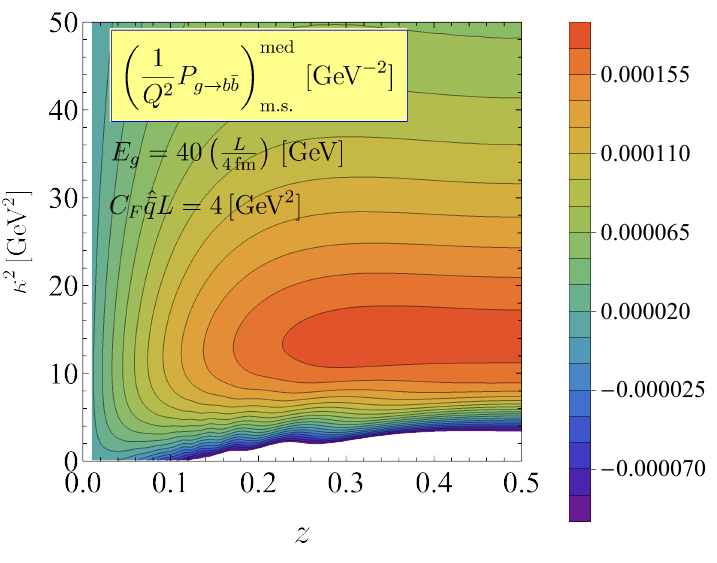}}%
\subfig{b}{\includegraphics[height=.39\textwidth]{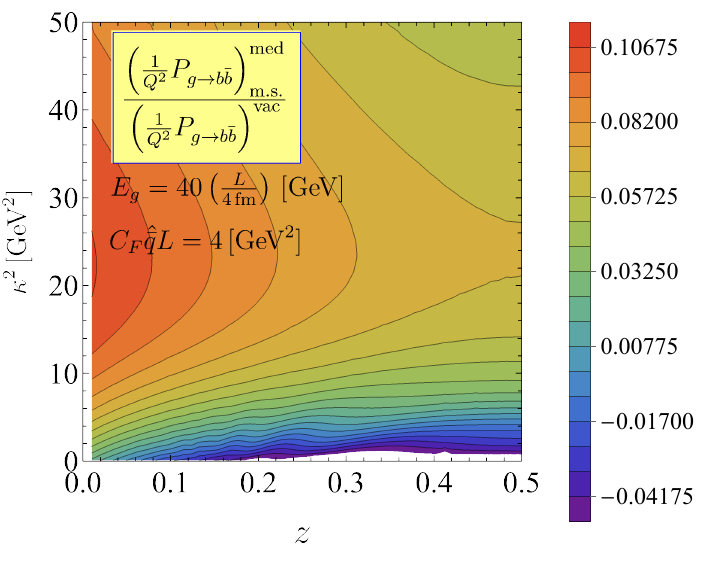}}
    \caption{(a) The medium-modified splitting function for $g\to b\bar{b}$ with $m_b = 4.18$ GeV, and (b) divided by the vacuum splitting function. 
  }
    \label{figbb}
\end{figure}

As seen in sections~\ref{sec3.4} and~\ref{sec4.2}, the quark mass enters the splitting function \eqref{eq2.2} in units of  the squared momentum transfer from the medium,
i.e., $\tfrac{m_q^2 }{2\mu_D^2}$ or $\tfrac{m_q^2}{\hat{q}L}$. 
For the medium enhancement of $g\to b\bar{b}$ to be numerically comparable to that of $g\to c\bar{c}$, one therefore expects that the $\hat{q}L$ transferred from the medium must be a factor of $m_b^2/m_c^2 \approx 11$ larger. 
Fig.~\ref{figbb} shows numerical results
for the $g \rightarrow b\bar{b}$ splitting function. Comparing to Fig.~\ref{figMSratio} confirms that the medium modification of $g\rightarrow b\bar{b}$ is significantly smaller than for $g\rightarrow c\bar{c}$.
For the parameter range favoured in heavy ion phenomenology, i.e., $\hat{q}L < 4\, \text{GeV}^2$, medium modifications of $g\to b\bar{b}$ are thus expected to be 
mild,
with modifications 
in the few-percent range. We finally note that the wiggles in Fig.~\ref{figbb} at small $\upkappa$ become more pronounced for larger $e_g$, a feature which we did not observe for charm quarks in a similar range of $E_g$ and $\hat{q}L$, though we did not explore this parameter range in detail.

Let us comment on the difference between light- and heavy-flavoured quark--anti-quark pairs in more detail. For heavy quark production, the $1\to 2$ contribution from gluon splitting is phase space suppressed compared to the $2\to 2$ contribution, as discussed in the introduction. Also, within vacuum parton showers, the dominant contribution to  jet multiplicity and many other jet characteristics comes from the splittings $q\to q\, g$ and $g\to g\, g$ that are enhanced by both a collinear and a soft logarithm. The splitting $g\to q\bar{q}$ does not have a soft singularity and it makes a subleading, single logarithmic contributions. As discussed in the introduction, what makes heavy quarks special is that $g\to c\bar{c}$ or $g \to b\bar{b}$ can be isolated experimentally.
In contrast to heavier quarks,  strange quarks need not originate from gluon splittings on perturbative scales but it can be also produced during the 
hadronisation process. In MC event generator simulations of hadronic collisions with Lund hadronisation model, for instance, it is a consequence of the non-perturbatively small 
strange quark mass that most strange hadrons arise from string fragmentation.  As a consequence, the techniques of Ref.~\cite{Ilten:2017rbd} do not apply to identifying gluons splitting into light-flavoured $q\bar{q}$-pairs. While equation \eqref{eq2.2} applies also to light-flavoured quark-- anti-quark pairs, 
we do not have arguments at present that such an addition is numerically relevant or that it could lead to phenomenologically distinct signatures.

\section{Derivation of the medium-modified \texorpdfstring{$g\to c\bar{c}$}{g->ccbar} splitting function}
\label{sec7}
In this section, we derive the medium-modified $g\to c\bar{c}$ splitting in time-ordered close-to-eikonal perturbation theory. As we shall explain, this amounts to
determining for the $g\to c\bar{c}$ multiple scattering amplitude $\mathcal{M}(t)$ the squared amplitude in the form 
\begin{align}
  \int_0^{t_\infty}\!\!\!\!\! dt \int_0^{t_\infty}\!\!\!\!\!  d \bar{t}\,  e^{i \tfrac{m_c^2}{2 E_g z (1-z)} \left(t - \bar{t}\right)}\,  \mathcal{M}(t) \mathcal{M}^\dagger(\bar{t}) 
 = 2 \Re \int_0^{t_\infty}\!\!\!\!\!  dt \int_{t}^{t_\infty}\!\!\!\!\!  d \bar{t}\,  e^{i \tfrac{m_c^2}{2 E_g z (1-z)} \left(t - \bar{t}\right)}\, \mathcal{M}(t) \mathcal{M}^\dagger (\bar{t})\, .
 \label{eq77.1}
 \end{align}
 Diagrammatically, we denote $\mathcal{M}(t) \mathcal{M}^\dagger (\bar{t})$ as
\begin{equation}
  \vcenter{\hbox{\includegraphics[width=0.4\linewidth]{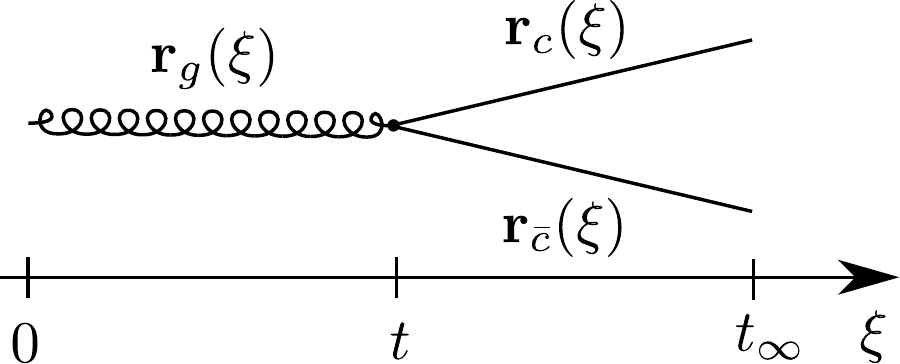}}}
   \vcenter{\hbox{\includegraphics[width=0.4\linewidth]{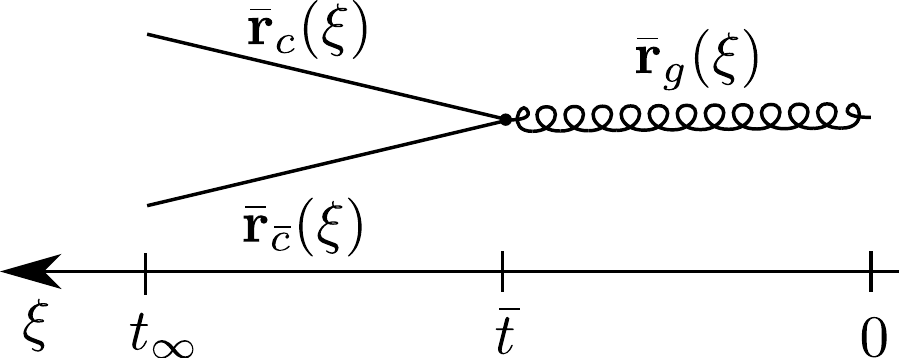}}}\, .
   \label{eq77.2}
\end{equation}
Here, the $\xi$-arrows from $0$ to $t_\infty$ illustrate the time-ordering in amplitude $\mathcal{M}(t)$ and complex conjugate amplitude $\mathcal{M}^\dagger (\bar{t})$.

The gluon has incoming transverse momentum $\bk_g$ and incoming energy $E_g$, the outgoing $c$- and $\bar{c}$-quarks carry final transverse momenta $\bk_c$ and 
$\bk_{\bar{c}}$ and energy fractions $E_c = z E_g$ and $E_{\bar{c}} = (1-z) E_g$, respectively. These momenta define the prepared in-state at time $\xi = 0$ and the observed 
out-state at time $\xi = t_\infty$ of the multiple scattering diagram. They are therefore the same for amplitude and complex conjugate amplitude.

In contrast, transverse spatial coordinates arise as Fourier conjugates of transverse momenta and are integrated over on amplitude level. This explains why the radii
denoted in \eqref{eq77.2} are generally different in $\mathcal{M} (t)$ and $\mathcal{M}^\dagger (\bar{t})$. Also, in time-ordered perturbation theory, the $g\to c\bar{c }$-splitting
occurs generally at different times $t$ and $\bar{t}$ in amplitude and complex conjugate amplitude. Due to the reordering \eqref{eq77.1}, it is sufficient to consider the case
$t \leq \bar{t}$.

\subsection{Diagrammatic rules for close-to-eikonal multiple-scattering}
\label{sec7.1}
There are several derivations of the BDMPS-Z formalism that start from describing multiple parton scattering on a spatially extended target 
in terms of  path-ordered Wilson lines ${\cal P} \exp\left[ -i \int_{\xi_1}^{\xi_2} A(\br(\xi),\xi)\, d\xi \right]$, associated to each trajectory 
$(\br(\xi),\xi)$ of a parton traversing the medium from time $\xi_1$ to time $\xi_2$, see e.g.~\cite{Kovner:2003zj,Apolinario:2014csa,Dominguez:2019ges}. 
In these formulations, $A$ denotes the coloured vector potential 
with which the parton interacts along its trajectory and which is given in the representation of the parton projectile. 
Medium properties are then defined in terms of correlation functions (\emph{target averages}) of $A$. In particular, two-point correlation functions of $A$
are assumed to be translationally invariant in the transverse direction and local (w.r.t. colour and momentum transfer) in the longitudinal direction. 

Here, we bypass the explicit formulation in terms of Wilson lines and target averages by 
specifying the diagrammatic rules that arise once one has expanded the path-ordered Wilson lines to arbitrary powers in $A$ and all target averages
have been performed. The following diagrammatic rules for evaluating $\mathcal{M}(t)\, \mathcal{M}^\dagger(\bar{t})$ are consistent with the 
derivations in~\cite{Wiedemann:2000za,Kovner:2003zj,Apolinario:2014csa,Dominguez:2019ges}:

\begin{enumerate}
\item \underline{Transverse phases at end-points of diagrams}. The in- and out-going quark and gluon states in \eqref{eq77.2} are associated with free plane transverse wave functions
 \begin{equation}
  \left( e^{-i \br_g(0)\cdot \bk_g} e^{-i \br_c(t_\infty)\cdot \bk_c} e^{-i \br_{\bar{c}}(t_\infty)\cdot \bk_{\bar{c}}} \right)\times
  \left( e^{i \bar{\br}_g(0)\cdot\bk_g} e^{i \bar{\br}_c(t_\infty)\cdot\bk_c} e^{i \bar{\br}_{\bar{c}}(t_\infty)\cdot\bk_{\bar{c}}}  \right)
   \label{eq77.3}\, .
   \end{equation}
\item \underline{Free propagation of phases}. The free evolution of transverse phases from transverse position $\br(\xi_1)$ at time $\xi_1$ to position 
$\br(\xi_2)$ at time $\xi_2$ is given by the free propagator
  \begin{align}
    G_0\big[\br;\xi_1, \xi_2|E\big] 
   & = \frac{E}{2\pi i (\xi_2-\xi_1)} \exp \left[- \frac{E \left( \br(\xi_2)-\br(\xi_1) \right)^2 }{2i(\xi_2-\xi_1)} 
   \right]\, .
   \label{eq77.4} 
  \end{align}
 For instance, forward evolution of the gluon wave-function from $0$ to $\xi$ yields
 \begin{align}
  \vcenter{\hbox{\includegraphics[width=0.2\linewidth]{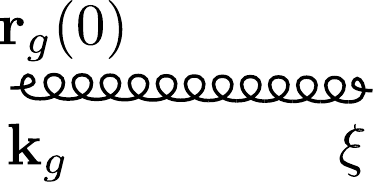}}}
 & = \int d\br_g(0) G_0\big[\br_g;0, \xi |E_g\big] e^{-i\br_g(0)\cdot\bk_g}  
 = e^{-i\br_g(\xi)\cdot \bk_g} e^{-i \frac{\bk_g^2}{2E_g}(\xi -0)}\, .
  \label{eq77.5}
\end{align}
Free propagation is seen to shift the transverse coordinate in the transverse phase and it leads to a multiplicative longitudinal phase. The longitudinal phase 
takes the form of a \emph{transverse energy}  $\tfrac{\bk_g^2}{2E_g}$ times a longitudinal distance. 
\item \underline{Treatment of mass-terms in longitudinal phases.}
For massive quarks, the free propagator \eqref{eq77.4} includes a mass-term which leads in
the free evolution of the phase $e^{-i\br_c(0)\cdot \bk_c}$ to a mass-dependent longitudinal phases  $e^{i \frac{\bk_c^2 + m_c^2}{2E_c}\xi} $\cite{Armesto:2003jh}. 
However, these mass-dependent phases cancel between $\mathcal{M}(t)$  and $\mathcal{M}^\dagger(\bar{t})$ for all times $\xi > \bar{t}$ and $\xi < t$. 
The remaining phase difference is an overall factor 
\begin{equation}
	e^{i \tfrac{m_c^2}{2 E_c}t}   e^{i \tfrac{m_c^2}{2 E_{\bar{c}}}t} e^{-i \tfrac{m_c^2}{2 E_c}\bar{t}}   e^{-i \tfrac{m_c^2}{2 E_{\bar{c}}}\bar{t}}  
	 = e^{i \tfrac{m_c^2}{2 E_g z (1-z)} (t-\bar{t})}\, .
	 \label{eq77.7}
\end{equation}
Since this factor is unaffected by medium-interactions, we have included it in the definition of \eqref{eq77.1}. With this choice, 
the quark propagators take the apparently mass-less form \eqref{eq77.4}.
\item \underline{Scattering centres.}  
We denote a single interaction between medium and partonic projectile by diagrams such as 
 \begin{align}
   \label{eq77.8}   \vcenter{\hbox{\includegraphics[width=0.4\linewidth]{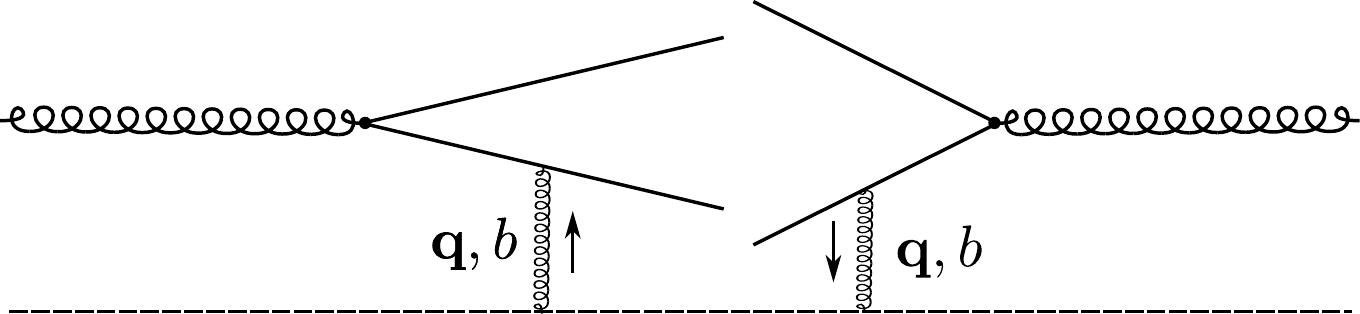}}} 
   \Longrightarrow \vcenter{\hbox{\includegraphics[width=0.4\linewidth]{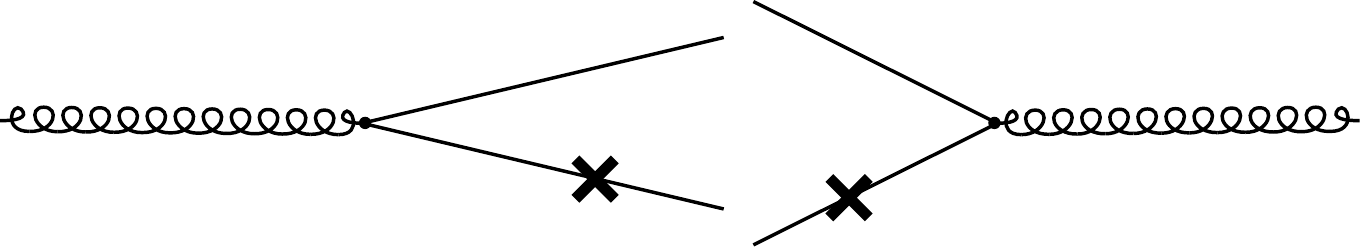}}}\, ,\\
   \label{eq77.9}    \vcenter{\hbox{\includegraphics[width=0.4\linewidth]{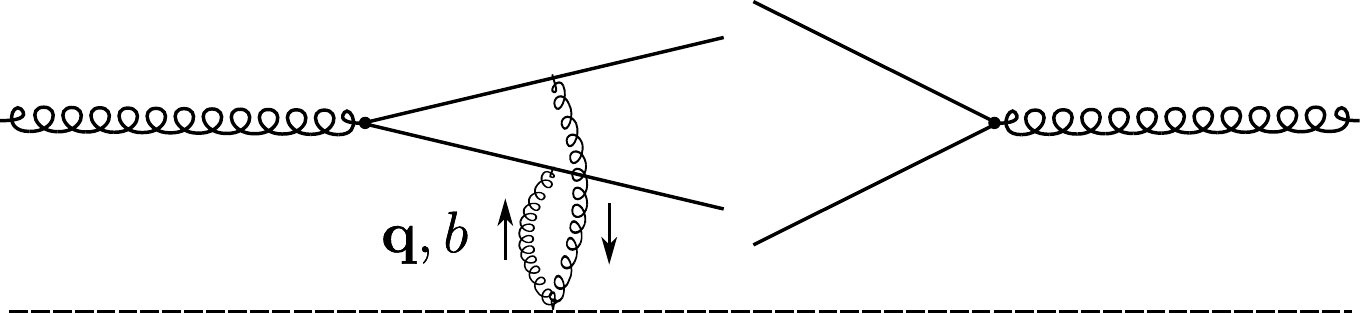}}}
   \Longrightarrow \vcenter{\hbox{\includegraphics[width=0.4\linewidth]{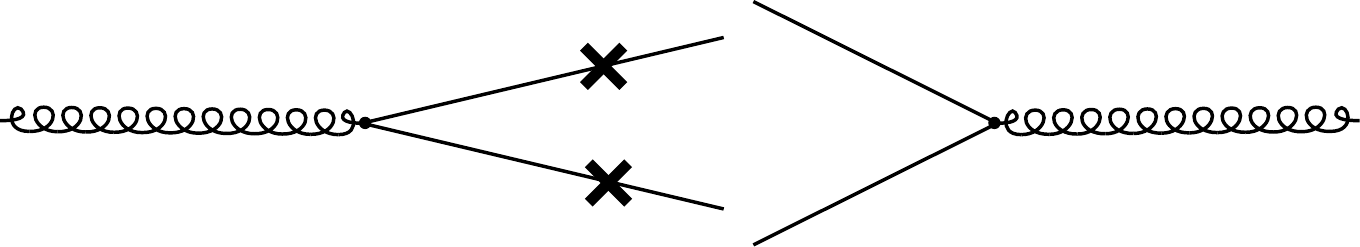}}}\, ,\\
   \label{eq77.10} \vcenter{\hbox{\includegraphics[width=0.4\linewidth]{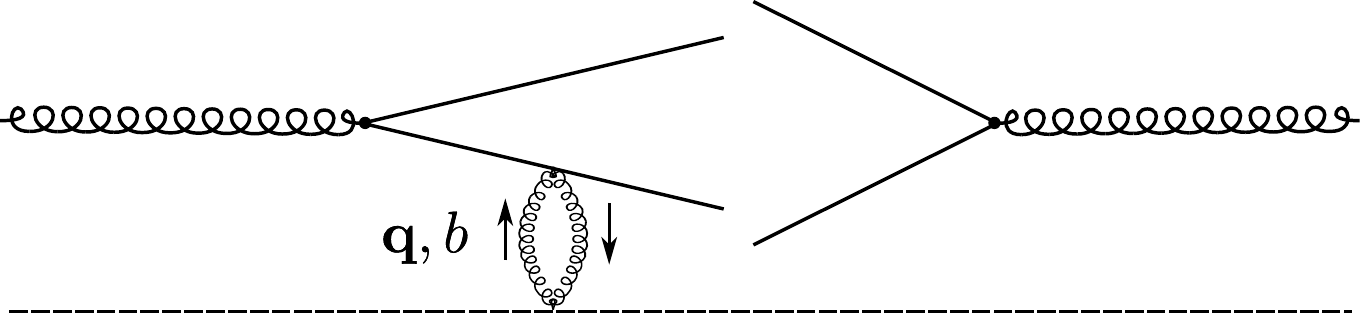}}}
   \Longrightarrow \vcenter{\hbox{\includegraphics[width=0.4\linewidth]{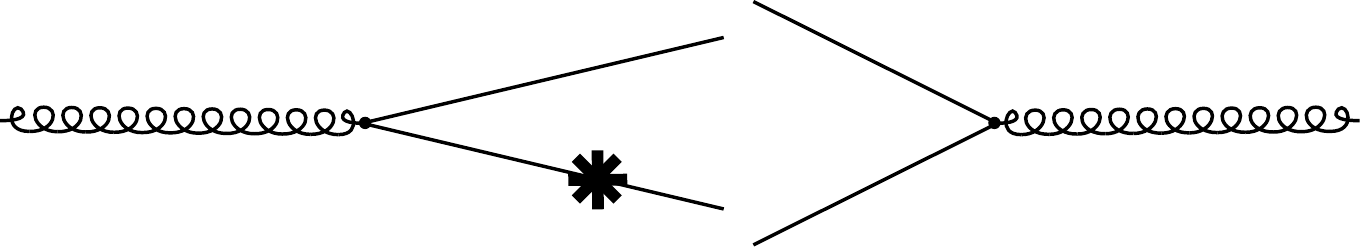}}}\, .
 \end{align}
Here, the extra line on the left-hand side represents a constituent of the medium that exchanges a gluon of colour $b$ and transverse momentum $\bq$ with the partonic
projectile. We assume that this interaction is local in the longitudinal direction 
and that there is no preferred transverse position and direction for momentum transfer. On the cross section level, gluon exchange with a single scattering centre as depicted in \eqref{eq77.8} - \eqref{eq77.10} can then be written as a multiplicative factor
  \begin{equation}
    \int d\xi n(\xi) \int \frac{d\bq}{(2\pi)^2} |a(\bq)|^2 e^{-i\bq\cdot [\br_1(\xi)-\br_2(\xi)]}\times F_\text{comb}\times F_\text{colour}\, ,\label{eq77.11}
  \end{equation}
where $| a(\bq)|^2$ can be interpreted as the elastic cross section associated to the medium constituent and the density $n(\xi)$ determines the probability with which the interaction  occurs at $\xi$. The transverse coordinates $\br_1(\xi)$, $\br_2(\xi)$ denote 
the positions of the projectile partons with which the two exchanged gluons interact. 

The combinatorial factors $F_\text{comb} $ is obtained by 
\begin{itemize}
\item associating a factor $i$ ($-i$) for each gluon exchange with the quark 
or gluon (anti-quark) line in the amplitude $\mathcal{M}$. 
The complex conjugate factor arises in the conjugate amplitude $\mathcal{M}^\dagger$.
\item associating an additional factor $\tfrac{1}{2}$ 
whenever two gluon exchanges occur between the same scattering centre and the same parton line. 
\end{itemize}
For instance, 
\begin{equation}
F_\text{comb} = \left\{ \begin{array}{cc}  i\times(-i)=  1 &\quad  \hbox{for type \eqref{eq77.8}} \\
 i\times(-i)=  1  & \quad  \hbox{for type \eqref{eq77.9}} \\
 \tfrac{1}{2} i\times i = - \tfrac{1}{2} &\quad  \hbox{for type \eqref{eq77.10}} 
\end{array} \right. \, .
\end{equation}

\item \underline{Colour and colour averages.}
The colour factor $F_\text{colour}$ in \eqref{eq77.11} amounts to inserting an $SU(N_c)$ generator in the fundamental ($T^b_{ij}$) or 
adjoint ($i\, f^{abc}$) representation, depending on whether the gluon exchange is with a quark or a gluon in the projectile. The 
fundamental generators will appear in the colour trace in the order in which they arise along the fermion line, 
see e.g. \eqref{eq77.26} below.

To calculate \eqref{eq77.1}, we sum over the colour of all final state quarks and we average over the colour $a$ of the incoming gluon, using 
$\frac{1}{N_c^2-1} \sum^{N_c^2-1}_{a=1}$. We also sum over the colours of all gluons exchanged between medium and partonic projectile.

\item \underline{Structure of splitting vertex.} 
The elementary $g\to c\bar{c}$ amplitude to leading $\mathcal{O}(1/E_g)$ 
is proportional to \cite{Kovchegov:1999kx,Lappi:2016oup}
\begin{align}
	\bar{u}^{(s)}(p_c) \epsilon_\mu^\perp(\lambda) \gamma^\mu v^{(r)}(p_{\bar{c}}) & \propto \left[ \delta_{s,-r} m_c \left( 1 + s\lambda \right) 
		\right. \nonumber \\
		& \quad \left.  - 2 \delta_{s,r}\, s\,  \{ z \delta_{s,\lambda} - (1-z) \delta_{s,-\lambda}  \} \,
		     \bm{\upepsilon}(\lambda)\cdot\left({\bf p}_r + (1-2z) {\bf P}\right) \right]\, ,
		\label{eq77.12}
\end{align}
where
\begin{equation}
	{\bf p}_r \equiv \frac{1}{2} \left({\bf p}_c - {\bf p}_{\bar{c}}\right)\, ,\qquad  {\bf P} \equiv \frac{1}{2} \left({\bf p}_c + {\bf p}_{\bar{c}}\right)\, .
	\label{eq77.13}
\end{equation}
Here, the outgoing charm and anti-charm quarks are described by spinors $\bar{u}^{(s)}(p_c)$, $v^{(r)}(p_{\bar{c}})$ with spin $s,r = \pm 1$. The 
two-dimensional vector ${\bm{\upepsilon}}(\lambda) = (1,i\lambda)$ with $\lambda = \pm 1$,  defines the transverse polarization of the gluon,
\begin{equation}
\epsilon^\perp_\mu(\lambda) = \frac{1}{\sqrt{2} }\left(0,{\bm{\upepsilon}}(\lambda),-\frac{{\bm{\upepsilon}}(\lambda)\cdot{\bf P}}{E_g}\right)\, .
\end{equation} 
Medium-induced scattering on the quark lines does not change the spinor structure to leading $\mathcal{O}(1/E_g)$. 
This is so, since spin-flip is mediated by 
the transverse components of the Dirac $\gamma$-functions while to leading order in $E_g$, the numerator of the fermion propagator is 
$\slashed{p}_c = E_c \left(\gamma^0 - \gamma^3\right)$. Therefore, in the presence of scattering centres, the $r-$, $s-$ and $\lambda$-dependence of
\eqref{eq77.12} remains unchanged. However, the momenta $ {\bf p}_c$, ${\bf p}_{\bar{c}}$ that arrive at the $g\to c\bar{c}$ vertex at
time $t$ differ in general from the momenta $ {\bf k}_c$, ${\bf k}_{\bar{c}}$ in the final state. Since the 
transverse momenta at time $t$ in the amplitude can be read off from the phases $e^{-i {\bf p}_c\cdot {\bf r}_c(t)} e^{-i {\bf p}_{\bar{c}} \cdot{\bf r}_{\bar{c}}(t) }$, 
it is technically advantageous to replace 
\begin{equation}
	{\bf p}_r \longrightarrow i \frac{\partial}{\partial {\bf r}_r(t)}\, ,\qquad {\bf P} \longrightarrow i \frac{\partial}{\partial {\bf R}(t)} \quad
	\text{in \eqref{eq77.12}}\, ,
	\label{eq77.14}
\end{equation}
where ${\bf R} = {\bf r}_c + {\bf r}_{\bar{c}}$ and ${\bf r}_r =  {\bf r}_c - {\bf r}_{\bar{c}}$. 
In the complex conjugate amplitude, the transverse quark momenta arrive at the $g\to c\bar{c}$ vertex at later time $\bar{t}$ with generally different 
transverse momenta. They are read out by the derivatives 
\begin{equation}
	\bar{\bf p}_r \longrightarrow -i \frac{\partial}{\partial \bar{\bf r}_r(\bar{t})}\, ,\qquad \bar{\bf P} \longrightarrow -i \frac{\partial}{\partial \bar{\bf R}(\bar{t})} \, ,
	\label{eq77.15}
\end{equation}
where $\bar{\bf R} = \bar{\bf r}_c + \bar{\bf r}_{\bar{c}}$ and $\bar{\bf r}_r =  \bar{\bf r}_c - \bar{\bf r}_{\bar{c}}$.
The amplitude \eqref{eq77.12} at time $t$ times the complex conjugate amplitude at later time $\bar{t}$ can then be written as
\begin{align}
	& \sum_{r,s,\lambda = \pm 1} 
	\bar{u}^{(s)}(p_c) \epsilon_\mu^\perp(\lambda) \gamma^\mu v^{(r)}(p_{\bar{c}})\, \bar{v}^{(r)}(\bar{p}_{\bar{c}})\, {\epsilon^*}_\nu^\perp(\lambda) \gamma^\nu\,  
	u^{(s)}(\bar{p}_c)\nonumber \\
	& \propto \left\{  m_c^2 + \left[z^2 + (1-z)^2\right] 
	\left( \frac{\partial}{\partial {\bf r}_r(t)} + (1-2z)   \frac{\partial}{\partial {\bf R}(t)} \right) 
	\left( \frac{\partial}{\partial \bar{\bf r}_r(\bar{t})} + (1-2z)   \frac{\partial}{\partial \bar{\bf R}(\bar{t})} \right)     \right\}
	\label{eq77.16}
\end{align}
This follows from inserting \eqref{eq77.14} [\eqref{eq77.15}]  into \eqref{eq77.12} [the conjugate of \eqref{eq77.12}] and carrying out the sums over 
spins and polarisations. 

For notational simplicity, we do not specify the normalisation in \eqref{eq77.12}, \eqref{eq77.16}. The overall normalisation of \eqref{eq77.1} will be fixed 
finally by a physical argument, see section~\ref{sec2.2}.
\end{enumerate}

\subsection{Target average for times \texorpdfstring{$\xi<t < \bar{t}$}{xi<t<tbar}  }
\label{sec7.2}
We consider the general case that an arbitrary number $m$ of scattering centres interacts with the gluon at times $0<\xi_1<\xi_2<\ldots<\xi_m < t$ 
prior to splitting of the gluon in the amplitude. We show first how the target average for the first scattering at $\xi_1$ follows from the above-mentioned diagrammatic rules.
We then turn to reiterating this result for $m$-fold scattering. 

We make the colour structure of the first scattering centre explicit and parameterise the colour structure induced by all subsequent scatterings
in a general colour matrix $M^{de}$
\begin{equation}
  \includegraphics[width=0.8\linewidth]{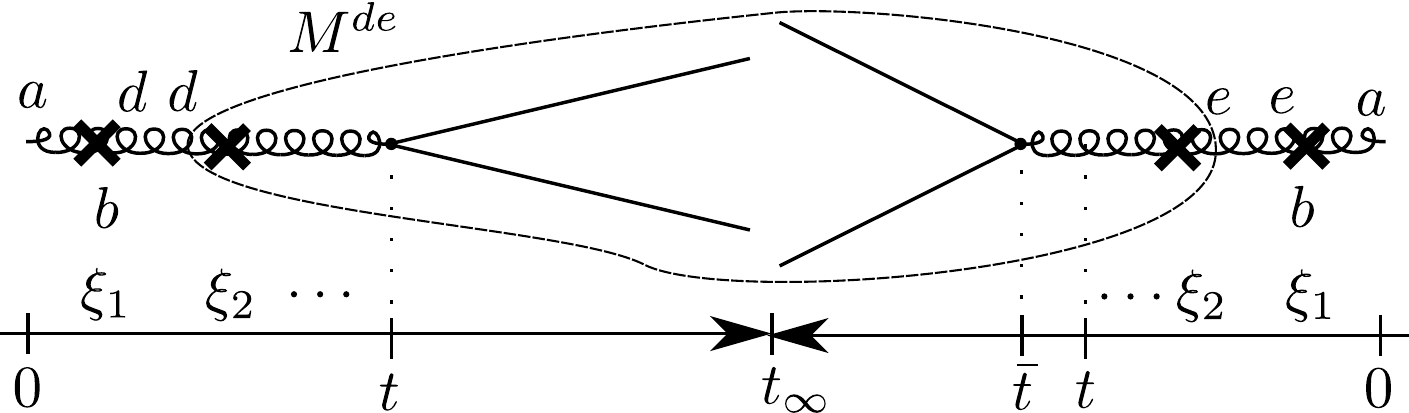}
  \label{eq77.17}
\end{equation}
The sums over the incoming colour $a$ and the exchanged colour $b$ indices  can then be carried out~\cite{Haber:2019sgz}, 
\begin{equation}
 i (-i) \left( i f^{ab d} \right) M^{de} \left( i f^{eba} \right)  = C_A\, \delta_{ed} M^{ed} = C_A\,  M^{dd} \, .
\end{equation}
In addition, the double gluon exchange at $\xi_1$ in the amplitude yields a contribution 
\begin{equation}
  \vcenter{\hbox{\includegraphics[width=0.5\linewidth]{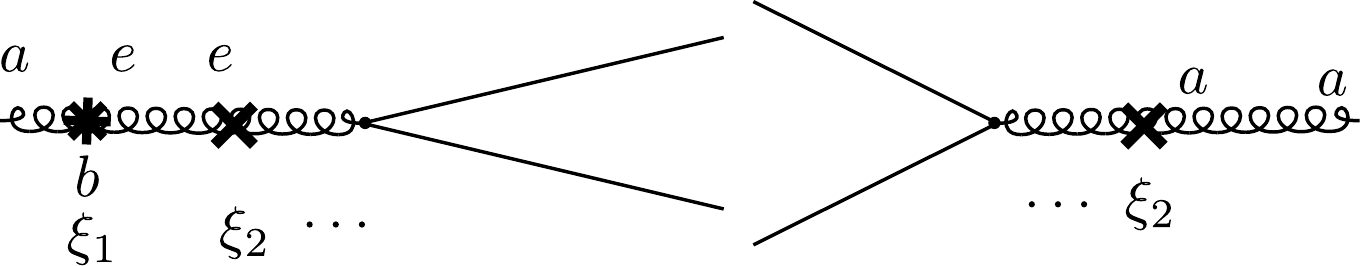}}} 
  \label{eq77.19} 
\end{equation}
with colour factor $- \frac{1}{2}  \left( if^{abd}  \right)\left( i f^{dbe} \right) M^{ea} =- \frac{1}{2} C_A\, M^{ee}$, and an identical contribution from a double gluon exchange
with the complex conjugate amplitude. In both classes of diagrams \eqref{eq77.17} and \eqref{eq77.19}, 
the colour-average over two gluon lines yields an adjoint Casimir $C_A = N_c$, times a remaining colour structure 
$M^{dd}$ that is again diagonal in colour. This colour average can then be reiterated for the second scattering centre at $\xi_2$, etc. 

Evolution of the initial transverse gluon phase  $e^{-i \br_g(0)\cdot \bk_g}  \,  e^{i \bar{\br}_g(0)\cdot\bk_g}$ in \eqref{eq77.7} from initial time $\xi=0$ up to (but not including) the second scattering centre at  $\xi_2$ yields 
\begin{align}
&\vcenter{\hbox{\includegraphics[width=0.25\linewidth]{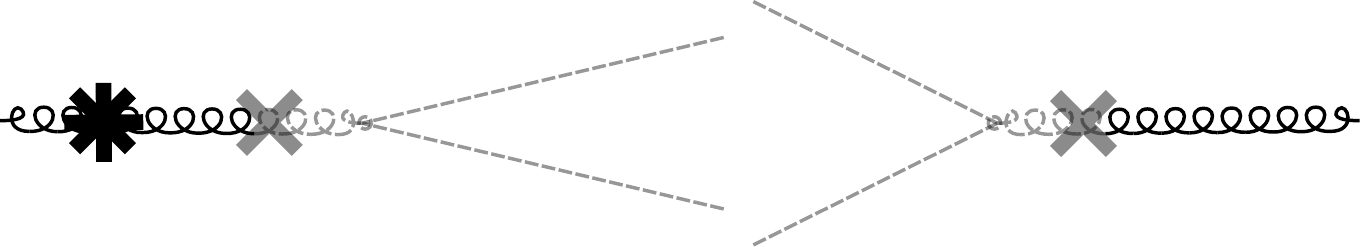}}} +
\vcenter{\hbox{\includegraphics[width=0.25\linewidth]{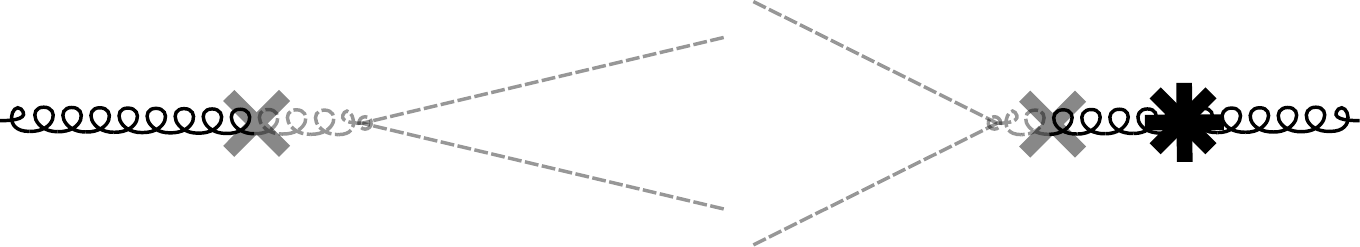}}} +
\vcenter{\hbox{\includegraphics[width=0.25\linewidth]{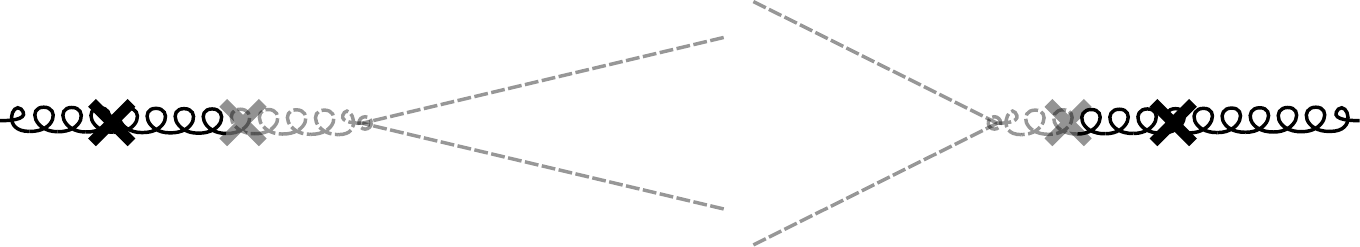}}} \nonumber\\
  &= \left[ C_A \delta_{de} \int_0^{\xi_2} d\xi_1  n(\xi_1)\int \frac{d\bq}{(2\pi)^2} |a(\bq)|^2\int d\br_g(0) d\br_g(\xi_1) d\bar{\br}_g(0) 
  d\bar{\br}_g(\xi_1) \right. \nonumber \\
  &\qquad \left. \times e^{-i\bk_g \cdot [\br_g(0)-\bar{\br}_g(0)]} G_0\big[\br_g; 0, \xi_1|E_g\big]\,G_0\big[\bar{\br}_g; \xi_1,0|E_g\big] 
  \left( e^{-i\bq\cdot[\br_g(\xi_1)-\bar{\br}_g(\xi_1)]} -1\right)  \right] \nonumber \\
 &\qquad  \times G_0\big[\br_g; \xi_1,\xi_2|E_g\big]\,G_0\big[\bar{\br}_g; \xi_2,\xi_1|E_g\big] \, M^{de} \nonumber \\
   &\equiv  \int d\br_g(\xi_1)d\bar{\br}_g(\xi_1)e^{-i \bk_g\cdot[\br_g(\xi_1)-\bar{\br}_g(\xi_1)]} \left[ \int_0^{\xi_2}d\xi_1 n(\xi_1)\, (-C_A \sigma\big[\br_g(\xi_1) - \bar{\br}_g(\xi_1)\big]) \right]
   \nonumber \\
    &\qquad  \times G_0\big[\br_g; \xi_1,\xi_2|E_g\big]\,G_0\big[\bar{\br}_g; \xi_2,\xi_1|E_g\big]  \, M^{dd} \, .
    \label{eq77.20}
\end{align}
Here, the first equation follows from the diagrammatic rules  \eqref{eq77.5} and \eqref{eq77.11} with  colour prefactors determined from \eqref{eq77.17}-\eqref{eq77.19}.
Performing the Gaussian integrals over $d\br_g(0)  d\bar{\br}_g(0)$, we wrote the second equation with the help of the dipole cross section 
$\sigma(\br ) \equiv \int \tfrac{d\bq}{(2\pi)^2} |a(\bq)|^2 (1-e^{-i\bq \cdot\br})$. 
The matrix $M^{dd}$ is a notational shorthand for all diagrammatic contributions occurring for $\xi \geq \xi_2$. In particular, $M^{dd}$ is a function of $\br_g(\xi_2)$ and 
$\bar{\br}_g(\xi_2)$.

We now introduce the transverse centre-of-mass coordinates for a two-body system consisting of the gluons in amplitude
and complex conjugate amplitude, 
\begin{align}
 \hat{\bm{\uprho}}(\xi) \equiv \br_g(\xi) - \bar{\br}_g(\xi) \, , \qquad 
 \bar{\bm{\uprho}}(\xi) \equiv \br_g(\xi) + \bar{\br}_g(\xi)\, ,
 \label{eq77.21}
\end{align}
with $d\bar{\bm{\uprho}}(\xi) d\hat{\bm{\uprho}}(\xi) = 2^2\, d\br_g(\xi)d\bar{\br}_g(\xi)$.
The propagators from $\xi_1$ to $\xi_2$ in \eqref{eq77.20} can then be written as free path-integrals with boundary conditions at $\xi_1$ and $\xi_2$, 
  \begin{align}
    G_0\big[\br_g;\xi_1, \xi_2|E_g\big] \, G_0\big[\bar{\br}_g;\xi_2, \xi_1|E_g\big]  & = \int \mathcal{D}\br_g  \mathcal{D}\bar{\br}_g
    \exp\left[ \frac{iE_g}{2} \int_{\xi_1}^{\xi_2} d\xi \, \left( \dot{\br}_g^2(\xi) -  \dot{\bar{\br}}_g^2(\xi) \right) \right] \nonumber \\
    &  = \int \mathcal{D}\hat{\bm{\uprho}}  \mathcal{D} \bar{\bm{\uprho}}
    \exp\left[ \frac{iE_g}{2} \int_{\xi_1}^{\xi_2} d\xi \,  \dot{ \bar{\bm{\uprho}}}(\xi). \dot{ \hat{\bm{\uprho}}}(\xi) \right] \, .
      \label{eq77.22} 
  \end{align}
As a consequence of translational invariance of the medium-average in the transverse plane, the matrix $M^{dd}= M^{dd}( \hat{\bm{\uprho}}(\xi) )$ 
in \eqref{eq77.20} can not depend on $ \bar{\bm{\uprho}}(\xi_1)$. Therefore, integration over $d\bar{\bm{\uprho}}(\xi_1)$ is trivial  and it is given by
  \begin{align}
   \frac{1}{2^2} \int d\bar{\bm{\uprho}}(\xi_1)  G_0\big[\br;\xi_1, \xi_2|E\big] \, G_0\big[\bar{\br};\xi_2, \xi_1|E\big]  
   = \delta^{(2)} \left(  \hat{\bm{\uprho}}(\xi_1)  -  \hat{\bm{\uprho}}(\xi_2) \right)\, .      \label{eq77.23} 
  \end{align}
In this way, the transverse distance $ \hat{\bm{\uprho}}(t)=  \hat{\bm{\uprho}}(\xi)$ remains unchanged for  $\xi<t$.

In the absence of a scattering centre at $\xi_1$, the first scattering centre would be at $\xi_2$. This case would be described by dropping the term in
brackets $\left[ \dots \right]$ from  \eqref{eq77.20}. Therefore, by recursively reiterating $m$ times the term $\left[ \dots \right]$ in  \eqref{eq77.20},
$m$-fold scattering is taken into account.  Since $M^{dd}$ in \eqref{eq77.20} remains diagonal in colour after scattering, and since $\hat{\bm{\uprho}}$ is frozen in time $\xi$,
summing over arbitrary $m$-fold scatterings between  $0< \xi < t$ yields
\begin{align}
& \sum_{m=0}^\infty  \int_0^{t}d\xi_m \int_{0}^{\xi_{m} }d\xi_{m-1} \ldots \int_{0}^{\xi_2} d\xi_1  n(\xi_{m})\ldots n(\xi_2) n(\xi_1) \nonumber \\
& \qquad \times
 \frac{1}{2^2} \int d\bar{\bm{\uprho}}(t) d\hat{\bm{\uprho}}(t) e^{-i \bk_g \cdot \hat{\bm{\uprho}}(t)} \left[ -C_A \sigma\left(\hat{\bm{\uprho}}(t)\right) \right]^{m} 
 M^{dd}\left(\hat{\bm{\uprho}}(t)\right)
 \nonumber \\
 & = \frac{1}{2^2} \int d\bar{\bm{\uprho}}(t) d\hat{\bm{\uprho}}(t) e^{-i\bk_g\cdot \hat{\bm{\uprho}}(t) } e^{-\int^{t}_0 d\xi n(\xi) C_A \sigma\left[ \hat{\bm{\uprho}}(t)\right]}
  \bigotimes
  \vcenter{\hbox{\includegraphics[width=0.3\linewidth]{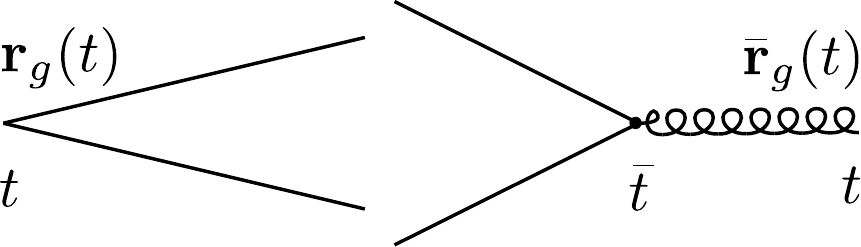}}}\, .
  \label{eq77.24}
\end{align}
Here, the effects of arbitrary many scatterings on the initial gluon phase $e^{-i \bk_g \cdot \hat{\bm{\uprho}}}$ prior to time $t$ are given by the 
exponentiated dipole cross section with colour prefactor $C_A=N_c$. The matrix $M^{dd}\left(\hat{\bm{\uprho}}(t)\right)$ parameterises all diagrammatic contributions occurring for $\xi > t$. It is represented diagrammatically as indicated in the last line. 

The factor $e^{-\int^{t}_0 d\xi n(\xi) C_A \sigma\left[ \hat{\bm{\uprho}}(t)\right]}$ in \eqref{eq77.24} is the well-known target average for two adjoint
Wilson lines separated by a transverse distance $\hat{\bm{\uprho}}(t)$, see e.g.~\cite{Kovner:2003zj}. The present subsection explains how this known result is obtained
from the diagrammatic rules in section~\ref{sec7.1}.

\subsection{Target average for times \texorpdfstring{$t < \xi< \bar{t}$}{t<xi<tbar} to leading \texorpdfstring{$\mathcal{O}\left(\tfrac{1}{N_c^2}\right)$}{O(1/Ncsquared)}.  }
\label{sec7.3}
In close analogy to our analysis in the previous subsection, we consider now diagrams including an arbitrary number of $m$ scatterings located at times
 $t<\xi_1<\xi_2<\ldots<\xi_m < \bar{t}$. Denoting the scattering at $\xi_1$  explicitly, and leaving all other scattering centres unspecified, 
 we find to leading $\mathcal{O}\left(\tfrac{1}{N_c^2}\right)$ 
 the following diagrammatic contributions:
 \begin{align}
   &  \vcenter{\hbox{\includegraphics[width=0.25\linewidth]{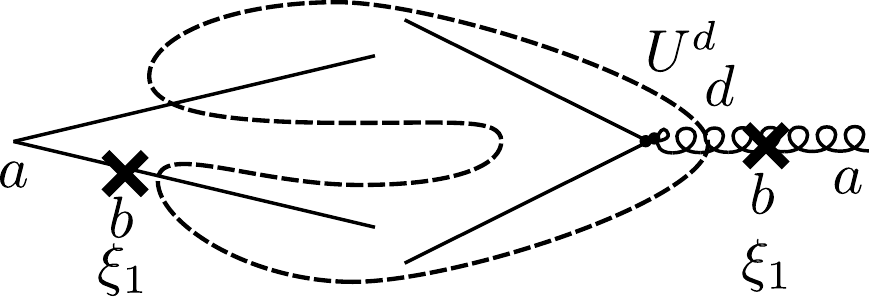}}}\quad 
   + \quad   \vcenter{\hbox{\includegraphics[width=0.25\linewidth]{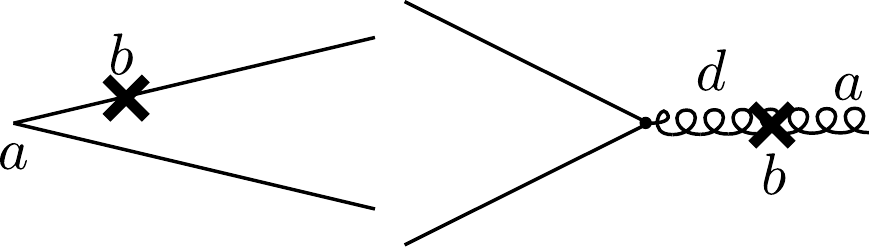}}}
  \nonumber \\
   &+ \quad \vcenter{\hbox{\includegraphics[width=0.25\linewidth]{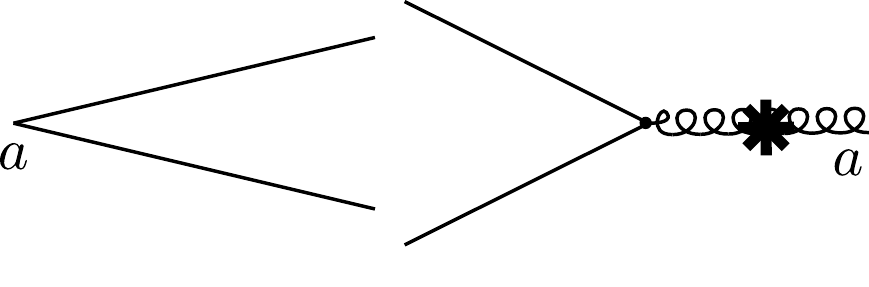}}} \quad 
   + \quad  \vcenter{\hbox{\includegraphics[width=0.25\linewidth]{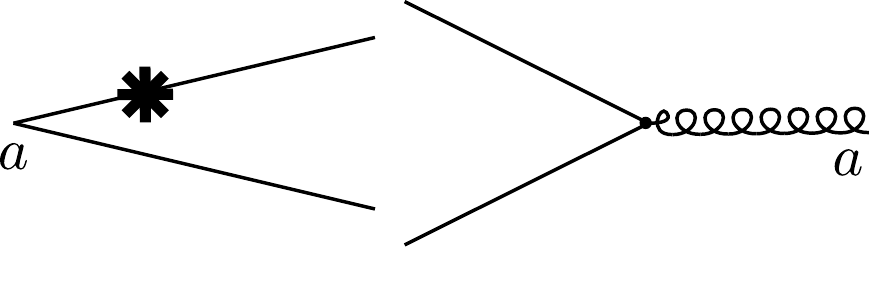}}} 
   \quad + \quad   \vcenter{\hbox{\includegraphics[width=0.25\linewidth]{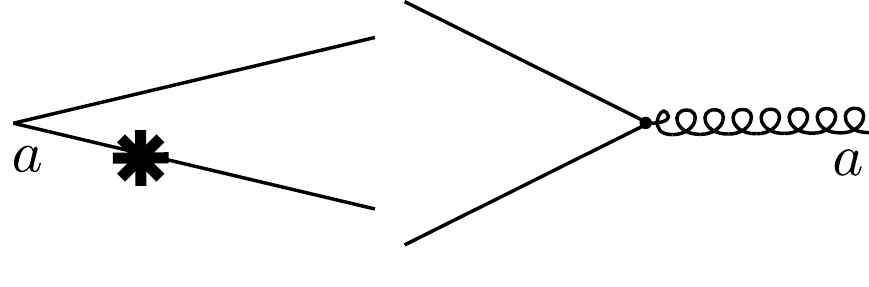}}} 
   \label{eq77.25}
 \end{align}
We first determine the colour factors associated to these diagrams. For the diagrams in the first line, we introduce a shorthand $U^d$ 
that parameterise the colour structure of the subsequent $(m-1)$ scatterings.  For each value of the adjoint index $d$, 
$U^d$ is an element in the fundamental representation of $SU(N_c)$. The colour structure of the diagram with scattering on anti-charm line in the amplitude and gluon in the complex conjugate amplitude (the first diagram in \eqref{eq77.25}) reads then~\cite{Haber:2019sgz}
 \begin{align}
 (-i)(-i)  \text{Tr}\, \left[ T^b T^a U^{d} \right]\left( i f^{d b a } \right)  
                       = \frac{1}{2}N_c \text{Tr}\, \left[ T^{d} U^{d} \right] \, .\label{eq77.26}
\end{align}
The same factor is obtained for the second diagram in \eqref{eq77.25}. 
In the absence of a first scattering centre at $\xi_1$, the colour factor of the first  two diagrams  in \eqref{eq77.25} would be $\text{Tr}\, \left[ T^{a} U^{a} \right]$. Therefore, 
averaging over the first exchanged colour $b$ leads to a prefactor $\frac{1}{2}N_c$ times the same colour structure for $(m-1)$-fold scattering. This averaging can be 
recursively reiterated for each subsequent scattering centre. 

According to \eqref{eq77.19}, the first diagram in the second line of \eqref{eq77.25}
has a colour factor $-\frac{1}{2}N_c$ times the colour structure for an $(m-1)$-fold scattering. 
The sum of the second and third term in the second line yields a colour factor $2 \times (-\tfrac{1}{2})C_F = -\frac{1}{2}N_c 
\times \left( 1 + \mathcal{O}\left(\tfrac{1}{N_c^2}\right) \right)$.
Therefore, to leading order $\mathcal{O}\left(\tfrac{1}{N_c^2}\right)$, 
we find
\begin{align}
	 \frac{1}{2^2} \int d\bar{\bm{\uprho}}(t) \text{\eqref{eq77.25}} 
	 &= 
	\frac{1}{2^2} \int d\bar{\bm{\uprho}}(t) \int_{t}^{\xi_2}d\xi_1 n(\xi_1) \int \frac{d\bq}{(2\pi)^2}|a(\bq)|^2\, \int d\br_c(\xi_1) d\br_{\bar{c}}(\xi_1) d\bar{\br}_g(\xi_1) 
	\nonumber\\ 
  &\qquad\times G_0\big[\br_c; t, \xi_1|E_c \big] G_0\big[{\br}_{\bar{c}};t,\xi_1|E_{\bar{c}}\big]  G_0\big[\bar{\br}_g;\xi_1,t|E_g\big] \nonumber \\
  & \qquad\times \frac{N_c}{2} \, \left( e^{-i\bq\cdot[\br_{\bar{c}}(\xi_1) - \bar{\br}_g(\xi_1)]} + e^{-i\bq\cdot[\br_c(\xi_1)-\bar{\br}_g(\xi_1)]}  - 2\right)\, \nonumber \\
   &\qquad\times G_0\big[\br_c; \xi_1,\xi_2|E_c \big] G_0\big[{\br}_{\bar{c}};\xi_1,\xi_2|E_{\bar{c}}\big]  G_0\big[\bar{\br}_g;\xi_2,\xi_1|E_g\big] \text{Tr}\, \left[ T^{d} U^{d} \right]\,.
   \label{eq77.27}
\end{align}
Here,  the integration over $ \tfrac{1}{2^2} \int d\bar{\bm{\uprho}}(t)$ is taken from \eqref{eq77.24}. 
The three terms in the next-to-last line are the phases associated to the first, second and the remaining three diagrams of \eqref{eq77.25}. The factor $\left[ T^{d} U^{d} \right]$
is a notational shorthand for all diagrammatic contributions occurring for $\xi \geq \xi_2$. 

To simplify \eqref{eq77.27}, we change the integration variables to 
\begin{align}
  \label{eq77.28} \bar{\bm{\uprho}}(\xi) &\equiv z \br_{c}(\xi) +(1-z) \br_{\bar{c}}(\xi)+ \bar{\br}_g(\xi)\, , \\
  \label{eq77.29} \hat{\bm{\uprho}}(\xi) &\equiv z \br_{c}(\xi) +(1-z) \br_{\bar{c}}(\xi)- \bar{\br}_g(\xi)\, , \\
  \label{eq77.30} \br_r(\xi) &\equiv  \br_{c}(\xi) -\br_{\bar{c}}(\xi)\, .
\end{align}
With the boundary condition  $\br_c(t) = \br_{\bar{c}}(t)=\br_g(t)$, this is consistent with the previous definitions \eqref{eq77.21}. The spatial integration at each
time comes with a Jacobian factor 4, $d\br_c d\br_{\bar{c}} d\bar{\br}_g = \frac{1}{2^2}d\bar{\bm{\uprho}}d\hat{\bm{\uprho}}d\br_r$. The combination of the
three free propagators in \eqref{eq77.27} can then be written as
\begin{align}
& G_0\big[\br_c; t, \xi_1|E_c \big] G_0\big[{\br}_{\bar{c}};t,\xi_1|E_{\bar{c}}\big]  G_0\big[\bar{\br}_g;\xi_1,t|E_g\big]  \nonumber \\
& = \int \mathcal{D}\br_c \mathcal{D}\br_{\bar{c}} \mathcal{D}\bar{\br}_g 
    \exp\left[ \frac{iE_g}{2} \int_{\xi_1}^{\xi_2} d\xi \, \left( z\, \dot{\br}_c^2(\xi) + (1-z) \dot{\br}_{\bar{c}}^2(\xi) -  \dot{\bar{\br}}_g^2(\xi) \right) \right] \nonumber \\
    &  = \int \mathcal{D}\hat{\bm{\uprho}}  \mathcal{D} \bar{\bm{\uprho}} \mathcal{D}\br_r
    \exp\left[ \frac{iE_g}{2} \int_{\xi_1}^{\xi_2} d\xi \,  
    \left( \dot{ \bar{\bm{\uprho}}}(\xi). \dot{ \hat{\bm{\uprho}}}(\xi) + z(1-z) \dot{\br}_r^2(\xi) \right) \right] \, .
\end{align}
Here, the path-integral in $ \mathcal{D}\hat{\bm{\uprho}}  \mathcal{D} \bar{\bm{\uprho}} $ is the same as \eqref{eq77.22}. Since the medium-average does not
introduce any dependence on the centre-of-mass coordinate $\bar{\bm{\uprho}}$, all $d\hat{\bm{\uprho}}(\xi)$-integrations are trivial and they imply that the
relative distance $\hat{\bm{\uprho}}$ remains frozen in, $\hat{\bm{\uprho}}(\xi) = \hat{\bm{\uprho}}(\bar{t})$ for all $\xi \leq \bar{t}$, see \eqref{eq77.23}.  
This allows us to write 
\begin{align}
	\text{\eqref{eq77.27}} 
	 &= 
	\frac{1}{2^2} \int d\bar{\bm{\uprho}}(\bar{t}) \frac{N_c}{2} \int_{t}^{\xi_2}d\xi_1 n(\xi_1) \int \frac{d\bq}{(2\pi)^2}|a(\bq)|^2\, \int d\br_r(\xi_1) \, \mathcal{K}_0\big[\br_r; t, \xi_1|\mu \big] 
	\nonumber \\
  & \quad\times \left( e^{-i\bq\cdot[(1-z)\br_r(\xi_1) + \hat{\bm{\uprho}}(\bar{t}) ]} + e^{-i\bq\cdot[-z \br_r(\xi_1) + \hat{\bm{\uprho}}(\bar{t}) ]}  - 2\right)\, 
  \mathcal{K}_0\big[\br_r; \xi_1, \xi_2|\mu \big] \, {\rm Tr}\left[ T^{d} U^{d} \right]
  \nonumber \\
   & = \frac{1}{2^2} \int d\bar{\bm{\uprho}}(\bar{t})\, \left[ - \int d\br_r(\xi_1) \, \mathcal{K}_0\big[\br_r; t, \xi_1|\mu \big]  \right. \nonumber \\
   & \quad \times \int_{t}^{\xi_2}d\xi_1 n(\xi_1) 
   \underbrace{ \left.  \left( \frac{N_c}{2} \sigma((1-z)\br_r(\xi_1) + \hat{\bm{\uprho}}(\bar{t})) 
   + \frac{N_c}{2} \sigma( -z\br_r(\xi_1) + \hat{\bm{\uprho}}(\bar{t}))  \right) \right]  }_{\equiv \sigma_3\left(\br_r(\xi_1),\hat{\bm{\uprho}}(\bar{t}),z \right)} \nonumber \\
    & \quad \times  \mathcal{K}_0\big[\br_r; \xi_1, \xi_2|\mu \big] \, {\rm Tr}\left[ T^{d} U^{d} \right]  \label{eq77.32}
\end{align}
where
  \begin{align}
     \mathcal{K}_0\big[\br_r; \xi_1, \xi_2| \mu \big]  & = \int \mathcal{D}\br_r
    \exp\left[ i \int_{\xi_1}^{\xi_2} d\xi \, \left( \frac{\mu}{2} \dot{\br}_r^2(\xi) \right) \right] \, ,
       \label{eq77.33free} 
  \end{align}
  is the free path-integral with effective mass $\mu=z(1-z)E_g$.
 One notes
 that the term in $\left[ \dots \right]$-brackets in \eqref{eq77.32} is a building block for the subsequent $(m-1)$-scatterings parameterised by ${\rm Tr}\left[ T^{d} U^{d} \right]$.
 Following a set of steps similar to those used to get from \eqref{eq77.20} to \eqref{eq77.24}, this allows one to exponentiate the longitudinal integral over $n(\xi) \sigma_3$
    \begin{align}
  \frac{1}{2^2} \int d\bar{\bm{\uprho}}(t) \sum_{m=0}^\infty \vcenter{\hbox{\includegraphics[width=0.25\linewidth]{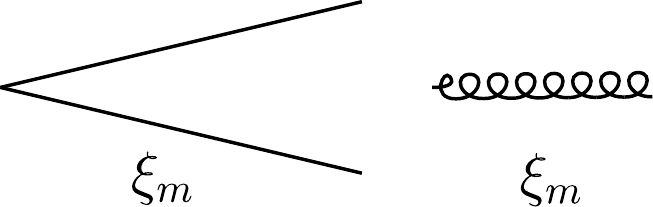}}} 
  & = \frac{1}{2^2} \int d\bar{\bm{\uprho}}(\bar{t})\,  \mathcal{K}\big[\br_r; t, \bar{t}| \hat{\bm{\uprho}}(\bar{t}),\mu \big] \label{eq77.32full}
   \end{align}
  where
   \begin{align}
 & \mathcal{K}\big[\br_r; \xi_1, \xi_2| \hat{\bm{\uprho}}(\bar{t}) ,\mu \big]  & = \int \mathcal{D}\br_r
    \exp\left[ i \int_{\xi_1}^{\xi_2} d\xi \, \left( \frac{\mu}{2} \dot{\br}_r^2(\xi) - \frac{n(\xi)\sigma_3\big(\br_r(\xi),\hat{\bm{\uprho}}(\bar{t}),z \big)}{i} \right) \right] \, .
       \label{eq77.33} 
  \end{align}
 Combining \eqref{eq77.24} and \eqref{eq77.32full}, the target average \eqref{eq77.2} over all times can therefore be written as
\begin{align}
 &  \frac{1}{2^2} \int d\bar{\bm{\uprho}}(\bar{t}) d\hat{\bm{\uprho}}(\bar{t}) e^{-i\bk_g\cdot \hat{\bm{\uprho}}(\bar{t}) } e^{-\int^{t}_0 d\xi n(\xi) C_A \sigma\left[ \hat{\bm{\uprho}}(\bar{t})\right]}
 \mathcal{K}\big[\br_r; t, \bar{t}| \hat{\bm{\uprho}}(\bar{t}) ,\mu \big] \nonumber \\
 & \qquad \bigotimes
 \vcenter{\hbox{\includegraphics[width=0.25\linewidth]{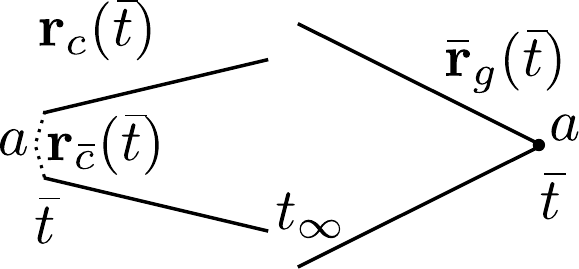}}} \, .
  \label{eq77.34}
\end{align}

\subsection{Target average for times \texorpdfstring{$\xi>  \bar{t}$}{xi>tbar} to leading \texorpdfstring{$\mathcal{O}\left(\tfrac{1}{N_c^2}\right)$}{O(1/Ncsquared)}.  }
\label{sec7.4}
To include multiple scatterings in the late-time diagram in \eqref{eq77.34}, we start by combining the
four transverse coordinates $\br_c$, $\br_{\bar{c}}$, $\bar{\br}_c$, $\bar{\br}_{\bar{c}}$ introduced in \eqref{eq77.2} for $\xi \geq \bar{t}$ into
\begin{align}
  \label{eq77.35}  \bar{\bm{\uprho}}(\xi) &\equiv z \br_{c}(\xi) +(1-z) \br_{\bar{c}}(\xi) + z \bar{\br}_{c}(\xi) +(1-z)\bar{ \br}_{\bar{c}}(\xi)\, , \\
   \label{eq77.36} \hat{\bm{\uprho}}(\xi) &\equiv z \br_{c}(\xi) +(1-z) \br_{\bar{c}}(\xi) -z \bar{\br}_{c}(\xi) -(1-z)\bar{ \br}_{\bar{c}}(\xi)\, ,\\
  \label{eq77.37} \br_r(\xi) &\equiv  \br_{c}(\xi) -\br_{\bar{c}}(\xi)\, ,\\
  \label{eq77.38} \bar{\br}_r(\xi) &\equiv  \bar{\br}_{c}(\xi) -\bar{\br}_{\bar{c}}(\xi)\, .
\end{align}
For $\xi = \bar{t}$, these definitions are consistent with \eqref{eq77.28}-\eqref{eq77.30}. It will be useful to further combine the last two radii to
\begin{align}
  \label{eq77.39} \delta\br_r(\xi) &\equiv  \br_r(\xi) -  \bar{\br}_r(\xi) \, ,\\
  \label{eq77.40} {\bf R}(\xi) &\equiv  \br_r(\xi) +  \bar{\br}_r(\xi) \, ,
\end{align}
so that the outgoing phase \eqref{eq77.3} at $t_\infty$ takes the form
 \begin{equation}
   e^{-i \br_c(t_\infty)\cdot \bk_c} e^{-i \br_{\bar{c}}(t_\infty)\cdot \bk_{\bar{c}}} e^{i \bar{\br}_c(t_\infty)\cdot\bk_c} e^{i \bar{\br}_{\bar{c}}(t_\infty)\cdot\bk_{\bar{c}} } 
   =  e^{-i 2 \mathbf{K} \cdot \rhoh(t_\infty)}  e^{-i {\bm{\upkappa}} \cdot \delta\br_r (t_\infty)} \, ,
   \label{eq77.41}
   \end{equation}
   where 
   \begin{align}
	{\bf K} & \equiv \frac{1}{2} \left(\bk_c + \bk_{\bar{c}} \right) \, ,\label{eq77.42} \\
	\bm{\upkappa} & \equiv \left(1-z\right) \bk_c - z \bk_{\bar{c}}  \label{eq77.43}\, .
\end{align}
\subsubsection{Case \texorpdfstring{$m=1$}{m=1}}
\label{sec7.4.1}
We consider first the case $m=1$ that only one scattering centre occurs at late time $\bar{t} < \xi_1 < t_\infty$.  
The diagrams 
\begin{align}
 & \vcenter{\hbox{\includegraphics[width=0.2\linewidth]{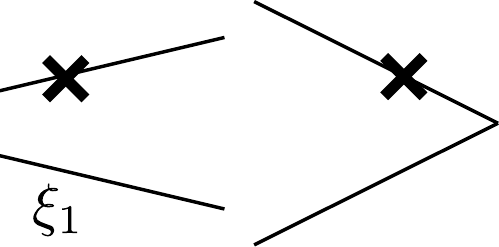}}} 
  +\vcenter{\hbox{\includegraphics[width=0.2\linewidth]{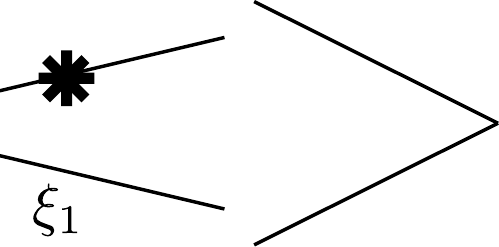}}} 
  +\vcenter{\hbox{\includegraphics[width=0.2\linewidth]{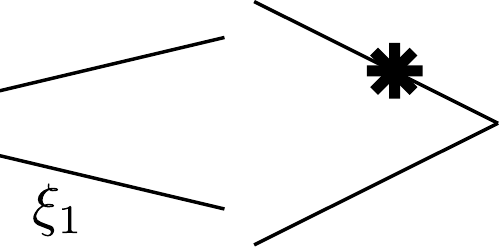}}}
  \label{eq77.44}
\end{align}
come with colour factor $C_F = \tfrac{N_c}{2}\left[1 + \mathcal{O}\left(\tfrac{1}{N_c^2}\right) \right]$, $- \tfrac{C_F}{2}$ and  $- \tfrac{C_F}{2}$, respectively, times a colour trace associated to
$m-1=0$-fold scattering. 
This follows from ${\rm Tr}\left[ T^a T^b T^b T^a\right] = C_F  \, {\rm Tr}\left[ T^a T^a\right]$~\cite{Haber:2019sgz}.
The scattering centre inserted at $\xi_1$ thus leads to a contribution
\begin{align}
 \eqref{eq77.44}   &\Longrightarrow  - \int d\xi_1 n(\xi_1) \int \frac{d\bq}{(2\pi)^2}|a(\bq)|^2 C_F \left(1- e^{-i\bq\cdot[\br_c(\xi_1) - \bar{\br}_c(\xi_1)]}\right) \nonumber \\
 &= - \int d\xi_1 n(\xi_1) C_F \sigma \left( (1-z)\delta \br_r(\xi_1) + \hat{\bm{\uprho}}(\xi_1) \right)\, .
 \label{eq77.45}
\end{align}
In close analogy, we find 
\begin{align}
  &\vcenter{\hbox{\includegraphics[width=0.2\linewidth]{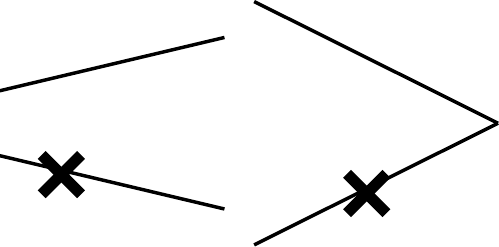}}} 
  +\vcenter{\hbox{\includegraphics[width=0.2\linewidth]{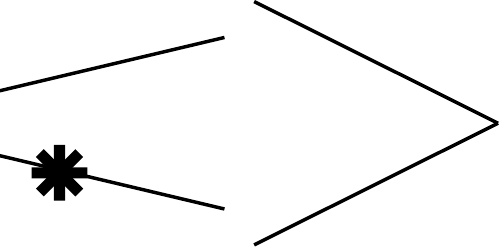}}} 
  +\vcenter{\hbox{\includegraphics[width=0.2\linewidth]{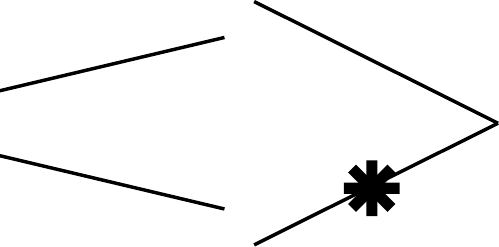}}} \nonumber \\
  & \Longrightarrow  - \int d\xi_1 n(\xi_1) C_F \sigma[-z \delta \br_r( \xi_1) + \rhoh(\xi_1)]\, ,
  \label{eq77.46}
\end{align}
where the different argument of the dipole cross section arises from replacing the phase $e^{-i\bq\cdot[\br_c(\xi_1) - \bar{\br}_c(\xi_1)]}$ in \eqref{eq77.45} by
$e^{-i\bq\cdot[\br_{\bar c}(\xi_1) - \bar{\br}_{\bar c}(\xi_1)]}$. To leading order in $N_c$, we find
\begin{equation}
\eqref{eq77.45} + \eqref{eq77.46} 
= - \int d\xi_1 n(\xi_1) \sigma_3\left(\delta\br_r(\xi_1),\hat{\bm{\uprho}}(\xi_1),z \right)\, ,
\label{eq77.47}
\end{equation}
with $\sigma_3$ defined in \eqref{eq77.32}. 

\subsubsection{Neglecting \texorpdfstring{$\mathcal{O}\left(\tfrac{1}{N_c^2}\right)$}{O(1/Ncsquared)}-suppressed terms}
In addition to the diagrams considered in section~\ref{sec7.4.1}, the case of a single scattering centre at late times includes 
four off-diagonal diagrammatic contributions 
that come with an $\mathcal{O}\left(\tfrac{1}{N_c^2}\right)$-suppressed colour factor
$ {\rm Tr}\left[ T^a T^b T^a T^b \right] = -\tfrac{1}{2N_c} {\rm Tr}\left[ T^a T^a \right] $ 
\begin{align}
 & \vcenter{\hbox{\includegraphics[width=0.2\linewidth]{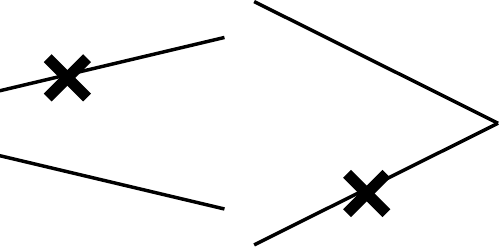}}} 
+  \vcenter{\hbox{\includegraphics[width=0.2\linewidth]{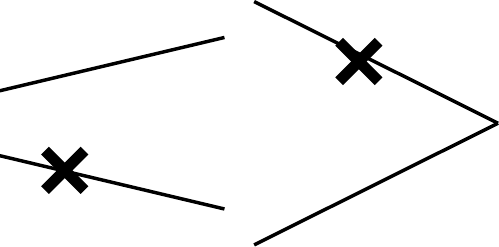}}} 
+  \vcenter{\hbox{\includegraphics[width=0.2\linewidth]{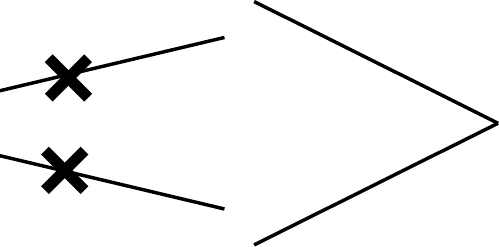}}} 
+  \vcenter{\hbox{\includegraphics[width=0.2\linewidth]{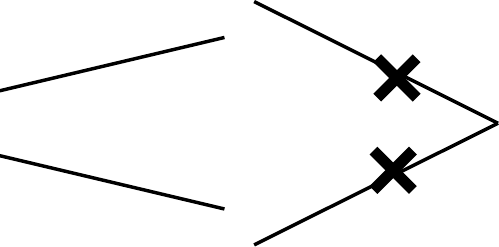}}}  \nonumber \\
&  \Longrightarrow \int d\xi_1 n(\xi_1)\int \frac{d\bq}{(2 \pi)^2}|a(\bq)|^2 \left( - \frac{1}{2N_c}  \right) \nonumber \\
  &\times\left\{ - e^{-i\bq\cdot[\br_c(\xi_1)-\bar{\br}_{\bar{c}}(\xi_1)]} - e^{-i\bq\cdot[\br_{\bar{c}}(\xi_1)-\bar{\br}_{{c}}(\xi_1)]} + e^{-i\bq\cdot[\br_c(\xi_1)-{\br}_{\bar{c}}(\xi_1)]} +e^{-i\bq\cdot[\bar{\br}_c(\xi_1)-\bar{\br}_{\bar{c}}(\xi_1)]}   \right\} \, .
  \label{eq77.48}
\end{align}
To leading order in $N_c$, these contributions can be neglected.\footnote{
The case $g\to q\bar{q}$ considered here has simpler 
target averages in the large $N_c$-limit than the case of photon splitting considered in ~\cite{Dominguez:2019ges}.  The reason is that the primary  $\gamma \to q\bar{q}$-exchange comes without a colour generator $T^a$. As a consequence, 
for $\gamma \to q\bar{q}$, the corresponding diagrams in \eqref{eq77.44}, 
\eqref{eq77.46} and \eqref{eq77.48} 
come all with the same colour factor ${\rm Tr}\left[ T^b T^b \right] $ 
and \eqref{eq77.48} 
cannot be neglected. In contrast, in the present case, the colour factors
${\rm Tr}\left[ T^a T^b T^b T^a\right]$ in  \eqref{eq77.44}, \eqref{eq77.46} differ from the colour factor ${\rm Tr}\left[ T^a T^b T^a T^b \right]$ in  \eqref{eq77.48}. 
This explains why in the analysis of  $\gamma \to q\bar{q}$ a non-reducible average over four Wilson lines arises even to leading order in $N_c$ while our final expression will not contain such a more complicated target average to leading  $\mathcal{O}\left(\tfrac{1}{N_c^2}\right)$ .}

\subsubsection{General case of \texorpdfstring{$m\geq 1$}{m>=1} scatterings to leading \texorpdfstring{$\mathcal{O}\left(\tfrac{1}{N_c^2}\right)$}{O(1/Ncsquared)}}
We now consider the colour algebra for $m$-fold scattering at late times, $\bar{t} < \xi_1 < \xi_2 < \dots < \xi_m < t_\infty$. 
One checks that if at least one of the $m$ interactions is off-diagonal in the sense of \eqref{eq77.48}, then the corresponding diagram is 
$\mathcal{O}\left(\tfrac{1}{N_c^2}\right)$-suppressed.  As a consequence, to leading $\mathcal{O}\left(\tfrac{1}{N_c^2}\right)$, we need to consider only diagrams that are of the type \eqref{eq77.44}
or \eqref{eq77.46} for \emph{each} of the $m$ scattering centres. By performing the colour average first at $\xi_m$, then at $\xi_{m-1}$ etc, one checks easily that each of the $m$
scattering centres is associated with a factor $\propto C_F$ of the form \eqref{eq77.47}.

We next write the combination of the four free quark propagators that evolve the system from $\bar{t}$ to $\xi_1 > \bar{t}$ where the interaction \eqref{eq77.47} occurs
\begin{align}
& G_0\big[\br_c; t, \xi_1|E_c \big] G_0\big[{\br}_{\bar{c}};t,\xi_1|E_{\bar{c}}\big]  G_0\big[\bar{\br}_c;\xi_1,t|E_c\big] G_0\big[\bar{\br}_{\bar{c}};\xi_1,t|E_{\bar{c}}\big]  \nonumber \\
& = \int \mathcal{D}\br_c \mathcal{D}\br_{\bar{c}} \mathcal{D}\bar{\br}_c \mathcal{D}\bar{\br}_{\bar c} 
    \exp\left[ \frac{iE_g}{2} \int_{\xi_1}^{\xi_2} d\xi \, \left( z\, \dot{\br}_c^2(\xi) + (1-z) \dot{\br}_{\bar{c}}^2(\xi) -  
    	z \dot{\bar{\br}}_c^2(\xi)  - (1-z)\dot{\bar{\br}}_{\bar c}^2(\xi)  \right) \right] \nonumber \\
    &  = \int \mathcal{D}\hat{\bm{\uprho}}  \mathcal{D} \bar{\bm{\uprho}} \mathcal{D}\delta\br_r \mathcal{D}{\bf R}
    \exp\left[ \frac{iE_g}{2} \int_{\xi_1}^{\xi_2} d\xi \,  
    \left( \dot{ \bar{\bm{\uprho}}}(\xi). \dot{ \hat{\bm{\uprho}}}(\xi) + z(1-z) \dot{\bf R}(\xi). \dot{\delta\br}_r(\xi) \right) \right] \, .
    \label{eq77.49}
\end{align}
As noted already in  \eqref{eq77.22}, 
it follows from  transverse translational invariance that scattering  cannot introduce a 
$ \bar{\bm{\uprho}}(\xi)$-dependence. Therefore,  all $d \bar{\bm{\uprho}}(\xi)$-integrations are trivial and the transverse 
distance $\hat{\bm{\uprho}}(\xi) $ remains frozen to a  
fixed value for all times $\xi$. This logic was explained in \eqref{eq77.23} and the $\bar{\bm{\uprho}}(\xi)$- and $\hat{\bm{\uprho}}(\xi)$-
dependence of \eqref{eq77.49} ensures that the same logic applies also at late times $\xi > \bar{t}$.

An analogous logic applies here to the treatment of the pair of coordinates ${\bf R}(\xi)$ and ${\delta\br}_r(\xi) $ in the free late time evolution \eqref{eq77.49}. 
These appear in the same combination in \eqref{eq77.49} as the pair 
$ \bar{\bm{\uprho}}(\xi)$ and $\hat{\bm{\uprho}}(\xi) $. Therefore, since the target average \eqref{eq77.47} and the final transverse phase 
\eqref{eq77.41} do not introduce any ${\bf R}$-dependence,  all $d{\bf R}(\xi)$-integrations can be done trivially and they imply that the distance 
${\delta\br}_r({\bar t}) = \delta\br_r({\xi}) $ is frozen in for all times $\xi \geq \bar{t}$. This allows one to exponentiate the sum over arbitrary many late time scatterings
\begin{align}
 \sum_{m=0}^\infty  \vcenter{\hbox{\includegraphics[width=0.2\linewidth]{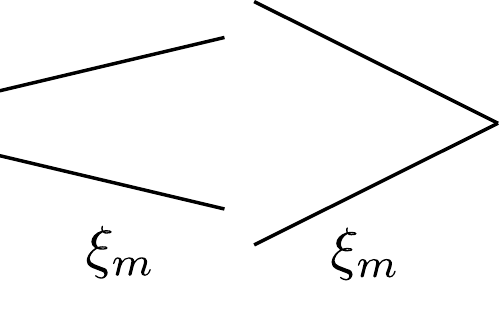}}} 
 & =e^{- \int_{\bar{t}}^{t_\infty} d\xi n(\xi) \sigma_3 (\br_r(\bar{t}), \rhoh(\bar{t}), z) }
 e^{-i 2 \mathbf{K} \cdot \rhoh(\bar{t})}  e^{-i \bm{\upkappa} \cdot \br_r (\bar{t})} \, ,
   \label{eq77.50} 
\end{align}
where we have used the boundary condition ${\delta\br}_r({\bar t}) = {\br}_r({\bar t})$.

\subsection{Main result to leading \texorpdfstring{$\mathcal{O}\left(\tfrac{1}{N_c^2}\right)$}{O(1/Ncsquared)}}
\label{sec7.5}

To write our main result for the production cross section  \eqref{eq77.1}, \eqref{eq77.2}, we now combine the derivation of the differential operator 
\eqref{eq77.16} that accounts for the elementary $g\to c\bar{c}$ vertices in amplitude and complex conjugate amplitude, and  
the target averages derived in sections~\ref{sec7.2}, \ref{sec7.3} and \ref{sec7.4} for times
$\xi < t$, $t < \xi < \bar{t}$  and $\bar{t}< \xi$  respectively. To this end, we insert 
the target average \eqref{eq77.50} for arbitrary many late time 
scatterings into \eqref{eq77.34}, and we supplement this target average over all times with the differential operator  \eqref{eq77.16} for the vertex functions.
For the $g\to c\bar{c}$ multiple scattering cross section of a gluon with initial transverse momentum $\bk_g$, this yields
\begin{align}
\frac{d\sigma_{g\to c\bar{c} }}{d{\bf K}\, d\bm{\upkappa}}
 = &  \mathcal{N} 
 2 \Re \int_0^{t_\infty}\!\!\!\!\!  dt \int_{t}^{t_\infty}\!\!\!\!\!  d \bar{t}\,  e^{i \tfrac{m_c^2}{2 E_g z (1-z)} \left(t - \bar{t}\right)}
 \int d\hat{\bm{\uprho}}(\bar{t}) d\br_r(\bar{t})\, 
 e^{-i\bk_g\cdot \hat{\bm{\uprho}}(\bar{t}) } e^{-\int^{t}_0 d\xi n(\xi) C_A \sigma\left[ \hat{\bm{\uprho}}(\bar{t})\right]} \nonumber \\
 & \qquad  \times e^{- \int_{\bar{t}}^{t_\infty} d\xi n(\xi) \sigma_3 (\br_r(\bar{t}), \rhoh(\bar{t}), z) }
 e^{-i 2\mathbf{K} \cdot \rhoh(\bar{t})}  e^{-i \bm{\upkappa} \cdot \br_r (\bar{t})} \nonumber \\
  & \qquad \times  \left(  m_c^2 + \left[z^2 + (1-z)^2\right] \frac{\partial}{\partial {\bf r}_r(t)}.\frac{\partial}{\partial {\bf r}_r(\bar{t})} \right)
 \mathcal{K}\big[\br_r; t, \bar{t}| \hat{\bm{\uprho}}(\bar{t}) ,\mu \big] \, .
  \label{eq77.51}
\end{align}
The simplification of the vertex structure \eqref{eq77.16} requires some comment. First, to leading order in $N_c$, target averages and phases are independent of ${\bf R}(\bar{t})$ and $\bar{\bf R}(\bar{t})$, and the corresponding derivatives in \eqref{eq77.16} therefore vanish. Second, the dependence on $\br_r(\bar{t})$ in \eqref{eq77.50} is a dependence on $\delta\br_r(\bar{t}) = \br_r(\bar{t})$, and therefore, we can replace in \eqref{eq77.16} 
$\frac{\partial}{\partial \bar{\bf r}_r(\bar{t})} \to - \frac{\partial}{\partial {\bf r}_r(\bar{t})}$. This derivative acts on the phase \eqref{eq77.50} but -- after partial
integration --  acts on $\mathcal{K}$ in \eqref{eq77.51}. 

The  squared amplitude \eqref{eq77.1} defines the cross section \eqref{eq77.51} up to a flux factor. This factor, as well as  a total transverse area $\tfrac{1}{2^2} \int d\bar{\bm{\uprho}}(\bar{t})$ and the multiplicative normalisation of \eqref{eq77.16} is absorbed in the overall norm $\mathcal{N}$.  For the purpose of this paper, it is convenient to fix $\mathcal{N}$ a posteriori by a physics argument given in section~\ref{sec2.2}.

Equation~\eqref{eq77.51} accounts for several physically distinct mechanisms in which medium-induced scattering affects the $g \to c\bar{c}$ process:
\begin{enumerate}
\item Momentum broadening changes the outgoing average $c\bar{c}$-pair momentum ${\bf K}$.
\item Momentum broadening changes the relative $c\bar{c}$-pair momentum $\bk_c - \bk_{\bar{c}}$.  
\item The production yield of $c\bar{c}$-pairs per parent gluon changes in the presence of a medium.
\end{enumerate}

\subsubsection{The medium-modified  \texorpdfstring{$g\to c\bar{c}$}{g->ccbar} splitting function}
In general, parton splitting functions depend on the longitudinal momentum fraction $z$ carried by the charm quark, and they depend on
the invariant mass $Q^2$ of the $c\bar{c}$ pair which can be expressed in terms of $z$, $m_c$ and the relative $c\bar{c}$-pair 
momentum $\bk_c - \bk_{\bar{c}}$. However, parton splitting functions do not depend on ${\bf K}$ since they do not depend on the 
absolute orientation of the emitted partons. This motivates us to eliminate the first of the three above-mentioned mechanisms by 
integrating over ${\bf K}$,
\begin{align}
\left(\frac{1}{Q^2}\, P_{g \to c\, \bar{c}} \right)^{\text{tot}}  & = 
\int d{\bf K} \frac{d\sigma_{g\to c\bar{c} }}{d{\bf K}\, d\bm{\upkappa}} \nonumber \\
&  =   \mathcal{N}  
 2 \Re \int_0^{t_\infty}\!\!\!\!\!  dt \int_{t}^{t_\infty}  d\bar{t}\,  e^{i \tfrac{m_c^2}{2 E_g z (1-z)} \left(t - \bar{t}\right)}
 \int  d\br_{\text{out}}\, 
 e^{- \int_{\bar{t}}^{t_\infty} d\xi n(\xi) \sigma_3 (\br_{\text{out}}, z) }
e^{-i \bm{\upkappa} \cdot \br_{\text{out}} } \nonumber \\
  & \qquad \times 
    \left(  m_c^2 + \left[z^2 + (1-z)^2\right] \frac{\partial}{\partial {\bf r}_{\text{in}}}.\frac{\partial}{\partial {\bf r}_{\text{out}} } \right)
 \mathcal{K}\big[\br_{\text{in}}=0, t; \br_{\text{out}}, \bar{t}| \mu\big] \, .
  \label{eq77.52}
\end{align}
We refer to this expression  as the \emph{medium-modified $g\to c\bar{c}$ splitting function}, and we substantiate this interpretation of \eqref{eq77.52} 
for an appropriately chosen normalisation \eqref{eq2.2}  
in section~\ref{sec2}.
Since the ${\bf K}$-integration sets $\hat{\bm{\uprho}}(\bar{t}) = 0$ in the integrand of \eqref{eq77.51}, and since $\mu = E_g\, z\, (1-z)$ is the only remaining
energy scale in the path integrand, we have introduced here the notational simplifications
\begin{align}	
	\sigma_3 (\br_r, z) & \equiv  \sigma_3 (\br_r, 0, z) = \frac{N_c}{2} \sigma\left((1-z)\br_r \right)  + \frac{N_c}{2} \sigma\left( z\br_r  \right)  \, ,
	  \label{eq77.53} \\
	  \mathcal{K}\big[\br_{\text{in}}, t; \br_{\text{out}}, \bar{t}| \mu \big] &\equiv  \mathcal{K}\big[\br_r; t, \bar{t}| 0 ,\mu \big]\, .
	  	  \label{eq77.54} 
\end{align}
The boundary conditions of the path integral $\mathcal{K}$ are $\br_{\text{in}} \equiv \br_r(t) = 0$ and $\br_{\text{out}}\equiv \br_r(\bar{t})$. An $\mathcal{O}(1/N_c^2)$-correction to \eqref{eq77.53} will be motivated in the following subsection \ref{sec7.6}.

Integration over ${\bf K}$ removes the dependence of \eqref{eq77.51} on an absolute transverse direction. As a consequence, the result \eqref{eq77.52} does 
not depend on the transverse momentum $\bk_g$ of the incoming gluon. Moreover, 
\begin{equation}
	\bm{\upkappa} = \bk_c = \frac{1}{2}\left(\bk_c  - \bk_{\bar{c}} \right) \quad \hbox{in the centre-of-mass frame ${\bf K}=0$.}
	  \label{eq77.55}
\end{equation}
Therefore, \eqref{eq77.52} describes the medium-modified yield of $c\bar{c}$-pairs from gluon splittings, measured differentially with
respect to their relative pair momentum in their centre-of-mass frame.

\subsubsection{The medium-modified \texorpdfstring{$g\to c\bar{c}$}{g->ccbar} production yield}
Integrating \eqref{eq77.52} over $d\bm{\upkappa}$, one obtains a phase-space integrated quantity that counts the number of $c\bar{c}$-pairs 
per incoming parent gluon irrespective of outgoing transverse momentum,
\begin{align}
\int d\bm{\upkappa}\, 
\int d{\bf K} \frac{d\sigma_{g\to c\bar{c} }}{d{\bf K}\, d\bm{\upkappa}} 
&  =   \mathcal{N}  
 2 \Re \int_0^{t_\infty}\!\!\!\!\!  dt \int_{t}^{t_\infty}  d\bar{t}\,  e^{i \tfrac{m_c^2}{2 E_g z (1-z)} \left(t - \bar{t}\right)}
\nonumber \\
  & \qquad \times 
    \left(  m_c^2 + \left[z^2 + (1-z)^2\right] \frac{\partial}{\partial {\bf r}_{\text{in}}}.\frac{\partial}{\partial {\bf r}_{\text{out}} } \right)
 \mathcal{K}\big[\br_{\text{in}}=0, t; \br_{\text{out}}=0, \bar{t}| \mu\big] \, .
  \label{eq77.56}
\end{align}
This expression is consistent with the collision kernel that describes the rate of $c\bar{c}$-pair production in QCD effective kinetic theory
~\cite{Caron-Huot:2010qjx,Arnold:2008iy}. 
For the following, it is noteworthy that the yield \eqref{eq77.56} depends only on medium-induced scattering at intermediate time 
$t < \xi <\bar{t}$ while the target averages at early ($\xi < t$) and late ($\xi > \bar{t}$) times do not enter. This finding is consistent with the 
probabilistic physics interpretation that scattering at $\xi < t$ changes the transverse momentum of the incoming gluon without 
affecting the probability of $g\to c\bar{c}$ splitting, and scattering at late time $\xi > \bar{t}$ modifies the transverse momenta of the 
outgoing $c$- and $\bar{c}$-quarks without affecting their yield. In this sense, $c\bar{c}$-pairs are formed only at intermediate times $t < \xi <\bar{t}$. 
Expression \eqref{eq77.52} is the transverse-momentum differential version of the collision kernel \eqref{eq77.56}. As explained in 
sections~\ref{sec3}, it allows for a more differential discussion of gluon formation time.

\subsection{Comments on \texorpdfstring{$\mathcal{O}\left(\tfrac{1}{N_c^2}\right)$}{O(1/Ncsquared)}-corrections}
\label{sec7.6}
The results summarised in section~\ref{sec7.5} have been derived to leading order in the $1/N_c^2$ expansion. Here we 
discuss shortly the nature of subleading $\mathcal{O}\left(\tfrac{1}{N_c^2}\right)$-corrections. These are characteristically different 
for target averages at different times:
\begin{enumerate}
\item \emph{Target averages at times $\xi < t$, section~\ref{sec7.2} }:\\
The factor $e^{-\int^{t}_0 d\xi n(\xi) C_A \sigma\left[ \hat{\bm{\uprho}}(t)\right]}$ in \eqref{eq77.24} for a pair of
eikonal gluon lines is exact in $N_c$.  
\item \emph{Target averages at times $t < \xi < \bar{t}$, section~\ref{sec7.3} }:\\
The complete  subleading $\mathcal{O}\left(\tfrac{1}{N_c^2}\right)$ correction to \eqref{eq77.27} takes the form
\begin{align}
&  \vcenter{\hbox{\includegraphics[width=0.4\linewidth]{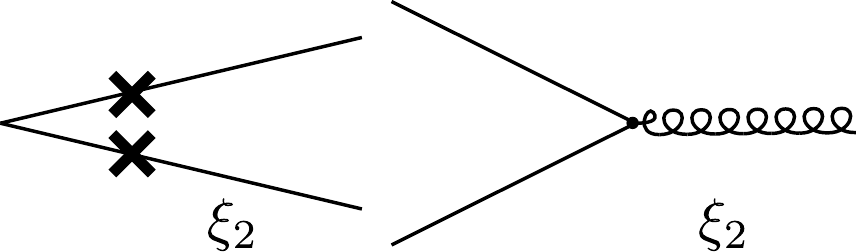}}} 
+ \frac{1}{2N_c} \vcenter{\hbox{\includegraphics[width=0.3\linewidth]{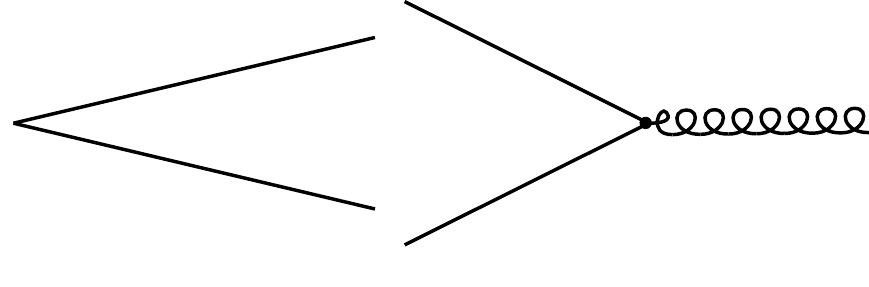}}} \nonumber \\
&= 
	\frac{1}{2^2} \int d\bar{\bm{\uprho}}(t) \int_{t}^{\xi_2}d\xi_1 n(\xi_1) \int \frac{d\bq}{(2\pi)^2}|a(\bq)|^2\, \int d\br_c(\xi_1) d\br_{\bar{c}}(\xi_1) d\bar{\br}_g(\xi_1) 
	\nonumber\\ 
  &\qquad\times G_0\big[\br_c; t, \xi_1|E_c \big] G_0\big[{\br}_{\bar{c}};t,\xi_1|E_{\bar{c}}\big]  G_0\big[\bar{\br}_g;\xi_1,t|E_g\big] \nonumber \\
  & \qquad\times \frac{-1}{2\, N_c}\, \left( e^{-i\bq \cdot[ \br_c(\xi_1) - \br_{\bar{c}}(\xi_1) ]} - 1\right)\, \nonumber \\
   &\qquad\times G_0\big[\br_c; \xi_1,\xi_2|E_c \big] G_0\big[{\br}_{\bar{c}};\xi_1,\xi_2|E_{\bar{c}}\big]  G_0\big[\bar{\br}_g;\xi_2,\xi_1|E_g\big] \text{Tr}\, \left[ T^{d} U^{d} \right]\, .
   \label{eq77.57}
\end{align}
In close analogy to \eqref{eq77.26}, the first of these diagrams comes with a colour factor   
$\text{Tr}\, \left[ T^b T^a T^b U^{a} \right]   = \frac{-1}{2\, N_c} \text{Tr}\, \left[ T^{d} U^{d} \right] $. The second diagram in \eqref{eq77.57} corrects 
for the subleading terms that we have dropped in writing $2 \times (-\tfrac{1}{2})C_F = -\frac{1}{2}N_c 
\times \left( 1 + \mathcal{O}\left(\tfrac{1}{N_c^2}\right) \right)$ in \eqref{eq77.27}. Adding \eqref{eq77.57} to \eqref{eq77.27} amounts to replacing in 
\eqref{eq77.32} 
\begin{equation}
\sigma_3\left(\br_r,\hat{\bm{\uprho}},z \right) \longrightarrow 
 \frac{N_c}{2} \sigma((1-z)\br_r + \hat{\bm{\uprho}}) 
   + \frac{N_c}{2} \sigma( -z\br_r + \hat{\bm{\uprho}})  -  \frac{1}{2\, N_c} \sigma(\br_r) \, .
   \label{eq77.58}
\end{equation}
With this replacement, the target average \eqref{eq77.34} \emph{for intermediate times} is correct to all $\mathcal{O}\left(\tfrac{1}{N_c^2}\right)$.
\item \emph{Target averages at times $\bar{t}< \xi$, section~\ref{sec7.4}}: \\
Including $\mathcal{O}\left(\tfrac{1}{N_c^2}\right)$-corrections in late-time target averages leads to a result that is significantly more complicated than \eqref{eq77.51}. The reason is two-fold. First, the subleading $\mathcal{O}\left(\tfrac{1}{N_c^2}\right)$-contribution \eqref{eq77.48} 
to the target average depends  on ${\bf R}$ and $\bar{\bf R}$. Therefore, the corresponding derivatives in \eqref{eq77.16} act non-trivially on the 
target averages and the corresponding vertex function becomes more complicated. Second, the colour average for $m$-fold scattering to  subleading $\mathcal{O}\left(\tfrac{1}{N_c^2}\right)$ does not factorise into a factor $-\tfrac{N_c}{2}$  times a contribution of $m-1$-fold scattering, and 
contributions from multiple scattering therefore cannot be obtained from exponentiating the 
one-scattering contribution \eqref{eq77.48}. 
\newline
The target average at times $\bar{t}< \xi$ is a target average over four fundamental Wilson lines with evolving transverse positions. Explicit expressions
for such so-called quadrupole terms have been given first in~\cite{Jalilian-Marian:2004vhw} in the eikonal limit, and they appear in several modern
applications of the BDMPS-Z formalism, see ~\cite{Apolinario:2014csa,Dominguez:2019ges} and Refs. therein. Numerical techniques exist for their evaluation beyond large-$N_c$~\cite{Isaksen:2020npj}. 
\end{enumerate}
Since \eqref{eq77.56} does not depend on target averages at late times $\xi > \bar{t}$, 
we conclude that the total medium-modified yield \eqref{eq77.56} of $c\bar{c}$-pairs can be written correctly to all $\mathcal{O}\left(\tfrac{1}{N_c^2}\right)$ 
by replacing $\sigma_3$ in that equation with \eqref{eq77.58} for $ \hat{\bm{\uprho}}=0$. With this replacement, the yield~\eqref{eq77.56} 
becomes consistent with the sub-leading $\mathcal{O}\left(\tfrac{1}{N_c^2}\right)$-terms in the collision kernel of QCD effective kinetic theory~\cite{Caron-Huot:2010qjx,Arnold:2008iy}. 

In contrast, correcting the more differential expressions \eqref{eq77.51} and \eqref{eq77.52} to all $\mathcal{O}\left(\tfrac{1}{N_c^2}\right)$ would yield
a significantly more complicated and more lengthy expression as it would include contributions with more complicated vertex functions and quadrupole terms. While all techniques for deriving and analysing such a parametrically more precise calculation are documented, we have not employed them here for reasons of technical simplicity that facilitate the physics discussion. 

In all previous sections of this work, we have therefore started from ~\eqref{eq77.52}  supplemented by \eqref{eq77.58}. These results are correct to all 
$\mathcal{O}\left(\tfrac{1}{N_c^2}\right)$ in the yield of $c\bar{c}$-pairs, and they are correct to leading 
$\mathcal{O}\left(\tfrac{1}{N_c^2}\right)$ in the $ \bm{\upkappa} $-differential distribution of this yield.

\section{Conclusions and Outlook}

 We have analysed the medium modification of the QCD leading order gluon splitting function into a massive quark--anti-quark pair. Our main result \eqref{eq2.2} is derived 
 to leading $\mathcal{O}\left(1/N_c^2\right)$ in a close-to-eikonal approximation, which was previously used to calculate medium-induced gluon radiation off quarks and gluons. The main emphasis of our work is on exhibiting the physics contained in \eqref{eq2.2} by analytical and numerical studies in certain limiting cases. Here, we recall some of our findings, and we discuss how they may contribute to further developments.

 In the opacity expansion of section~\ref{sec3}, we explicitly demonstrated that the modification of the $g\to{c\bar{c}}$ splitting function is consistent with a probabilistic interpretation in terms of two mechanisms: a probability-conserving transverse momentum broadening of the relative quark- anti-quark pair momentum and an enhanced splitting probability. We also observed for both static and expanding media the emergence of a formation time $\tau_{g\to c \bar{c}}=2 E_g/Q^2$, which sets the scale for the transition from the coherent to the incoherent limit of the splitting function. 
Though the notions of short- and long-distance physics are common in high-energy physics, 
 they are not testable in elementary collisions 
 where the splitting of partons remains unperturbed by interactions.
 This is different in nucleus-nucleus collisions where the parton-medium interaction depends on the location of a particular parton splitting and thus can cause observable outcomes. 
Proposals for direct experimental tests of the distances related to medium-modified parton branching processes are scarce. One noteworthy exception is the proposal to utilise the reconstructed $W$-boson mass in boosted semi-leptonic top quark decay topologies~\cite{Apolinario:2017sob}, but this will remain luminosity-limited for many years to come. Since the mass threshold for $g\to c\bar{c}$ sets a finite scale in units of which formation time can be measured, and since special techniques exist to access the $g\to c\bar{c}$ splitting experimentally, we hope that further studies can use the present work to identify experimental signatures of $\tau_{g\to c \bar{c}}$.

In section~\ref{sec4}, we have studied $P_{g\to c\bar{c}}^\text{med}$ in the multiple soft scattering approximation in which the absolute scale and the kinematic dependence of medium-modifications is fully governed by the quenching parameter $\hat{q}(\xi)$. Technically, we have managed to express the medium-modified splitting in terms of one-dimensional integrals suitable for numerical implementation for a static medium, which we expect is also possible for all other known splitting functions.
We illustrated the characteristic behaviour of the splitting function in the $({\bm{\upkappa}}^2,z)$ plane and found sizeable enhancements compared to the vacuum splitting within a broad phase space region of transverse momentum ${\bm{\upkappa}}^2 \sim \hat{q}L$ (see, e.g., Fig.~\ref{figMSratio}). 
We also derived formulas for media undergoing arbitrary expansion and we quantified, for not too large gluon energy, the difference between time-dependent $\hat{q}$ profiles and an equivalent static brick. 

In principle, knowledge of medium-modified splitting functions is the basis for formulating a medium-modified parton shower and testing it against jet and hadron measurements in heavy ion collisions. In practice, this requires significant work that lies outside the scope of the present manuscript. In section~\ref{sec6}, we have found that 
a simple reweighting of a vacuum-generated distribution of $g\to c\bar{c}$ by a factor 
$\left(1 + P_{g\to c\bar{c}}^\text{med}/P_{g\to c\bar{c}}^\text{vac}\right)(z,\bm{\upkappa}^2,E_g)$ can be used to estimate the medium-modified $g\to c\bar{c}$ branching probability. 
In the absence of a full medium-modified parton shower including $P_{g\to c\bar{c}}^\text{med}$, 
this reweighting prescription will allow us to explore some experimental signatures of enhanced 
$c\bar{c}$ radiation in a companion paper~\cite{Attems:2022otp}.

\acknowledgments We thank Fabio Dom\'\i{}nguez for pointing out a mistake that affected the finite $z$-dependence of \eqref{eq2.2} in the first preprint version. 
We thank F. Dom\'\i{}nguez, A. Huss, J.G. Milhano, P.F. Monni, K. Rajagopal, C.A. Salgado,  and K. Tywoniuk for useful discussions. MA acknowledges support 
through H2020-MSCA-IF-2019 ExHolo 898223.

\begin{appendix}
\section{The splitting function at order \texorpdfstring{$N=1$}{N=1} in opacity.  }
\label{appb}
In this appendix, we provide details of the derivation of \eqref{eq3.6}. Expansion of the integrand of  \eqref{eq2.2} to first
order in $n\, \sigma$ involves two contributions. 
The first arises from expansion of the absorption factor $\exp\left[ -\tfrac{1}{2} \int_{\bar t}^\infty d\xi\, n(\xi)\, \sigma_3({\bf r}_\text{out},z) \right]$, and it places a scattering centre at late times $\xi > \bar{t}$ at which the $c\bar{c}$-pair is \emph{fully formed (ff)}. 
The second arises from expansion of the propagator ${\cal K}$, and it
places a single scattering centre at times $t < \xi < \bar{t}$ at which the $c\bar{c}$-pair is \emph{not fully formed (nff)} (in the sense that it appears in the amplitude but not yet in the complex conjugate).
In the following, we evaluate these contributions separately.

The \emph{fully formed} $N=1$ opacity contribution to \eqref{eq2.2} reads
\begin{align}
&	\left(\frac{1}{Q^2}\, P_{g \to c\, \bar{c}} \right)^{\text{med}}_{\text{ff}} = -\, \mathfrak{Re}\, \frac{1}{8\, E_g^2}\, \int_{0}^{\infty} dt \int_t^{\infty} d\bar{t}\,\int_{\bar{t}}^\infty d\xi\, n(\xi) 
		\exp\left[ i \frac{m_\text{c}^2}{2 \mu} (t-\bar{t}) - \epsilon |t| - \epsilon |\bar{t}| \right]\, 
		\nonumber \\
		 & \int d{\bf r}_\text{out}
		   \, \sigma_3({\bf r}_\text{out},z)  
		  e^{-i\, {\bm{\upkappa}}\cdot{\bf r}_\text{out}} \left[ \left( m_\text{c}^2 + \frac{\partial}{\partial {\bf r}_\text{in}} \cdot \frac{\partial}{\partial {\bf r}_\text{out}}
		  \right) \frac{z^2 + (1-z)^2}{z(1-z)}  + 2 m_c^2  \right] \, {\mathcal K}_0\big[{\bf r}_\text{in}=0,t;{\bf r}_\text{out},\bar{t}| \mu \big]\nonumber\\
     &= \int \frac{d{\bf q}}{(2\pi)^2}\, \vert a_3({\bf q},z) \vert^2 \int_{0}^{\infty} d\xi n(\xi) \Bigg\{(1-\cos[\Gamma_1\xi])
     \left(\frac{1}{Q^2_1}\, P_{g \to c\, \bar{c}} \right)^{\text{vac}} \nonumber\\
     &\hspace{5cm}- (1-\cos[\Gamma_0\xi])
\left(\frac{1}{Q^2_0}\, P_{g \to c\, \bar{c}} \right)^{\text{vac}} \Bigg\}\, .
		  \label{eqb.2}
\end{align}
Here, to get to \eqref{eqb.2}, one inserts \eqref{eq2.5} for the dipole cross section and one does the ${\bf r}_\text{out}$
integration according to \eqref{eq2.9}.
Noting that 
\begin{equation}
 \int_{0}^{\infty} dt \int_t^{\infty} d\bar{t}\,\int_{\bar{t}}^\infty d\xi = \int_{0}^\infty d\xi \int_{0}^\xi dt  \int_{t}^\xi d\bar{t}\, ,
\end{equation}
we can perform the longitudinal integrals for density profile $n(\xi)$ with finite support 
\begin{align}
  &\lim_{\epsilon\to 0}\frac{1}{4 E_g^2} \int_{0}^\xi dt  \int_{t}^\xi d\bar{t}\cos[\Gamma_0(\bar{t}-t)] e^{- \epsilon (t+\bar{t})}= \frac{1-\cos[\Gamma_0 \xi]}{Q^4}\, .
\end{align}
Analogously, one integrates over the phase factor set by the shifted transverse energy $\Gamma_1$ in \eqref{eq3.7}. 

The \emph{not fully formed} $N=1$ opacity contribution to \eqref{eq2.2} is found with the help of \eqref{eq3.3}, 
\begin{align}
&	\left(\frac{1}{Q^2}\, P_{g \to c\, \bar{c}} \right)^{\text{med}}_\text{nff}= -\, \mathfrak{Re}\, \frac{1}{8\, E_g^2}\, \int_{0}^{\infty} dt \int_t^{\infty} d\bar{t}\, \int^{\bar{t}}_t d\xi\, n(\xi)
		\exp\left[ i \frac{m_\text{c}^2}{2 \mu} (t-\bar{t}) - \epsilon |t| - \epsilon |\bar{t}| \right]\, 		\nonumber \\
		 &\qquad\qquad\times \int d{\bf r}_\text{out}
		  e^{-i\, {\bm{\upkappa}}\cdot {\bf r}_\text{out}}
		\left[ \left( m_\text{c}^2 + \frac{\partial}{\partial {\bf r}_\text{in}}\cdot \frac{\partial}{\partial {\bf r}_\text{out}}
		  \right) \frac{z^2 + (1-z)^2}{z(1-z)}  + 2 m_c^2  \right] \, \nonumber \\
     &\qquad\qquad\times \int d{\bm{\uprho}}	\, 
	{\cal K}_0\big[{\bf r}_\text{in},t;{\bm{\uprho}},\xi  \vert \mu \big] \, \sigma_3({\bm{\uprho}},z)\, {\cal K}_0\big[{\bm{\uprho}},\xi ;{\bf r}_\text{out},\bar{t}  \vert \mu \big]\nonumber\\
     &=\int \frac{d{\bf q}}{(2\pi)^2}\, \vert a_3({\bf q},z) \vert^2 \int_{0}^{\infty} d\xi n(\xi) \Bigg\{ (1-\cos[\Gamma_0 \xi])
\left(\frac{1}{Q^2_0}\, P_{g \to c\, \bar{c}} \right)^{\text{vac}} \nonumber\\
     & \qquad \qquad - (1-\cos[\Gamma_1 \xi])
\frac{1}{2Q^2 Q_1^2} \left[ \left( m_\text{c}^2 + {\bm{\upkappa}}\cdot ({\bm{\upkappa}}+{\bf q})
   \right) \frac{z^2 + (1-z)^2}{z(1-z)}  + 2 m_c^2  \right]\Bigg\}\, .
\label{eqb.4}
\end{align}
As the scattering centre is placed in this contribution at a position $\xi$ with $t < \xi < \bar{t}$, the derivative 
$ \frac{\partial}{\partial {\bf r}_\text{out}}$ that acts at time $\bar{t}$ picks up a transverse momentum $\bm{\upkappa}$,
while the derivative $ \frac{\partial}{\partial {\bf r}_\text{in}}$ at earlier time $t$ picks up a shifted transverse momentum 
$\bm{\upkappa} + {\bf q}$. Compared to the fully formed contribution \eqref{eqb.2}, this complicates the transverse momentum 
dependence slightly. 
 
The full $N=1$ opacity contribution to \eqref{eq2.2} is the sum of \eqref{eqb.2} and \eqref{eqb.4}. In this sum, 
contributions proportional to the interference factor  $(1-\cos[\Gamma_0 \xi])$
cancel, and we are left with an integrand proportional to $(1-\cos[\Gamma_1 \xi])$.
To reorganise this integrand, we use
\begin{align}
  \frac{-1}{Q^2 Q_1^2} &=\frac{1}{2}\left(\frac{1}{Q_1^2} - \frac{1}{Q^2}\right)^2 - \frac{1}{2} \frac{1}{Q_1^4} - \frac{1}{2}\frac{1}{Q^4}\, ,\label{eqb.5}\\
  \frac{-{\bm{\upkappa}}\cdot\left({\bm{\upkappa}} + {\bf q}  \right) }{Q^2 Q_1^2} &=\frac{1}{2}\left(\frac{{\bm{\upkappa}}+{\bf q}}{Q_1^2} - \frac{{\bm{\upkappa}}}{Q^2}\right)^2 - \frac{1}{2} \frac{({\bm{\upkappa}}+{\bf q})^2}{Q_1^4} - \frac{1}{2}\frac{{\bm{\upkappa}}^2}{Q^4}\, .\label{eqb.6}
\end{align}
With the help of \eqref{eqb.5} and \eqref{eqb.6}, the sum of \eqref{eqb.2} and \eqref{eqb.4} takes the form
\begin{align}
	\left(\frac{1}{Q^2}\, P_{g \to c\, \bar{c}} \right)^{\text{med}}=&\frac{1}{2}\int \frac{d{\bf q}}{(2\pi)^2}\, \vert a_3({\bf q},z) \vert^2 \int_{0}^{\infty} d\xi n(\xi)\left(1-\cos\left[\Gamma_1 \xi\right]\right)
\nonumber\\
		\times\Bigg[&\left(\frac{1}{Q^2}\, P_{g \to c\, \bar{c}} \right)^{\text{vac}}_{{\bm{\upkappa}} \to {\bm{\upkappa}}+{\bf q}} - \left(\frac{1}{Q^2}\, P_{g \to c\, \bar{c}} \right)^{\text{vac}} \nonumber\\
&
    +\frac{m_c^2}{2z(1-z)} \left( \frac{1}{Q_1^2} - \frac{1}{Q^2} \right)^2 + \frac{z^2 + (1-z)^2}{2z(1-z)}\left(\frac{{\bm{\upkappa}}+{\bf q}}{Q_1^2} - \frac{{\bm{\upkappa}}}{Q^2}\right)^2\Bigg].
\end{align}
This completes the derivation of  \eqref{eq3.11}. For static medium with $n(\xi) = n_0 \theta(\xi-L)\theta(\xi)$ the $\xi$ integral is trivial and we obtain \eqref{eq3.6}.

 \section{Derivation of the splitting functions \texorpdfstring{\eqref{eq4.11}}{eq4.11} and \texorpdfstring{\eqref{eq4.12}}{eq4.12} in the  multiple soft scattering approximation for static medium\label{app:Isoftderiv}}
 
 To derive \eqref{eq4.11} and \eqref{eq4.12}, we focus on two contributions of the full splitting function \eqref{eq2.2} defined in terms of the integrals in \eqref{eq4.8}:  $I_4$ and $I_5$. In the saddle-point approximation \eqref{eq4.3} and for constant $\hat{q}(\xi)=\hat{q}$ in the region $0<\xi<L$ the term $I_4$ receives $\hat q$ contributions from  both the absorption factor and the path integral of the two dimensional harmonic oscillator~\eqref{eq4.4}. For $I_5$ the absorption factor is unity, but the propagator is a convolution of the harmonic oscillator with the free propagator. Our goal is to perform the remaining transverse and one longitudinal integral.
 
\subsection{\texorpdfstring{$I_4$}{I4}}
We first substitute the propagator expression for  the harmonic oscillator~\eqref{eq4.4} in \eqref{eq2.2}. Then we take the $\partial_{{\bf r}_\text{in}}$ derivative,
integrate over ${\bf r}_\text{out}$ by parts
and set ${\bf r}_\text{in}=0$. We obtain
\begin{align}
	  		 I_4&=2\, \mathfrak{Re}\, \frac{1}{8\, E_g^2}\, \int_{0}^{L} dt \int_t^{L} d\bar{t}\, 
		\exp\left[ i \frac{m_\text{c}^2}{2 \mu} (t-\bar{t}) \right]\, \int d{\bf r}_\text{out}
		  e^{ - \frac{1}{4} \hat{q}\,c(z)(L-\bar{t}) {\bf r}_\text{out}^2 
		 -i\, {\bm{\upkappa}}\cdot {\bf r}_\text{out}}
		  \nonumber \\		
     & \times \left[  \mu \Omega\frac{{\bf r}_\text{out} \cdot {\bm{\upkappa}} -i \frac{1}{2} \hat{q}\,c(z)(L-\bar{t}) {\bf r}_\text{out}^2   }{\sin \Omega (\bar{t}-t)}		\frac{z^2 + (1-z)^2}{z(1-z)}  + \frac{ m_c^2}{z(1-z)}  \right]  \frac{\mu\Omega}{2 \pi i \sin\Omega (\bar{t}-t)} e^{- \frac{-i 2\mu\Omega \cot\Omega (\bar{t}-t)}{4}{\bf r}_\text{out}^2}
\end{align}
Noticing that the trigonometric functions depend only on the difference of longitudinal coordinates, we make the change of variables to $x= \bar{t}-t$ and $u=L-\bar{t}$. The integration bounds are changed using the following equivalence of inequalities
\begin{equation}
  0 < t < L\, , \quad t < \bar{t} < L\, , \quad\leftrightarrow \quad
  0< x < L\, , \quad 0 < u <L-x \, .
\end{equation}
Expressing the transverse integral in polar coordinates, we write
\begin{align}
	  		 I_4&= 2\, \mathfrak{Re}\, \frac{1}{8\, E_g^2}\, \int_0^{L} dx \int_{0}^{L-x} du\, 
		\exp\left[- i \frac{m_\text{c}^2}{2 \mu} x \right]\, \int \frac{1}{2}d{r}^2_\text{out} d\phi
 e^{ - \frac{1}{4} \hat{q}\, c(z) u { r}_\text{out}^2 
		 -i\, \upkappa { r}_\text{out}\cos\phi}
		  \nonumber \\		
     & \times \left[  \mu \Omega\frac{{ r}_\text{out} {\upkappa}\cos\phi -i \frac{1}{2} \hat{q}\,c(z) u {r}_\text{out}^2   }{\sin \Omega x}		\frac{z^2 + (1-z)^2}{z(1-z)}  + \frac{ m_c^2}{z(1-z)}  \right]  \frac{\mu\Omega}{2 \pi i \sin\Omega x} e^{- \frac{-i 2\mu\Omega \cot\Omega x}{4}{ r}_\text{out}^2}\, .
\end{align}
Here, the $u$-integral is done easily and the azimuthal integral over $\phi$ leads to 
 Bessel functions. Using $r_\text{out} J_1(\upkappa r_\text{out}) = - \frac{\partial}{\partial \upkappa} J_0(\upkappa r_\text{out})$, we find
\begin{align}
	  		 I_4&=2\, \mathfrak{Re}\, \frac{1}{8\, E_g^2}\, \int_0^{L} dx \, 
		\exp\left[- i \frac{m_\text{c}^2}{2 \mu} x \right]\, \int_0^\infty {r}_\text{out}d{r}_\text{out} 
 	e^{ - \frac{1}{8} \hat{q}\,c(z)(L-x) { r}_\text{out}^2}	  \nonumber \\		
     & \times \Bigg\{ \frac{\sinh[\frac{1}{8} \hat{q}\,c(z)(L-x) { r}_\text{out}^2]}{\frac{1}{8} \hat{q}\,c(z) { r}_\text{out}^2}\left[  \mu \Omega\frac{i { \upkappa}\partial_{\upkappa} -2i }{\sin \Omega x}		\frac{z^2 + (1-z)^2}{z(1-z)}  + \frac{ m_c^2}{z(1-z)}  \right] \nonumber\\
     &+ i2\mu \Omega\frac{ (L-x)  e^{ - \frac{1}{8} \hat{q}\,c(z)(L-x) { r}_\text{out}^2}}{\sin \Omega x}		\frac{z^2 + (1-z)^2}{z(1-z)} \Bigg\} \frac{\mu\Omega J_0(\upkappa r_\text{out})}{i \sin\Omega x} e^{- \frac{-i 2\mu\Omega \cot\Omega x}{4}{ r}_\text{out}^2}\, .
\end{align}
The remaining radial integral can be done, using the identities
\begin{align}
   &      \int_0^\infty dx e^{- \beta x} J_0(\upkappa \sqrt{x}) = \frac{1}{\beta} e^{-\frac{\upkappa^2}{4\beta}}\, ,\\
  &      \int_0^\infty dx e^{- \beta x} \frac{\sinh [\gamma x]}{x}J_0(\upkappa \sqrt{x})
  = \frac{1}{2} \text{Ei}\left[- \frac{\upkappa^2}{4\left( \beta-\gamma \right) } \right] - \frac{1}{2} \text{Ei}\left[- \frac{\upkappa^2}{4\left( \beta+\gamma \right) } \right] \, 
,\end{align}
where  $4(\beta - \gamma) = -2i\mu\Omega  \cot\Omega x$ and $4(\beta + \gamma) = \hat{q}\,c(z) (L-x)-i2\mu\Omega  \cot\Omega x$. Here $\text{Ei}$ is the exponential integral function.
We obtain
\begin{align}
	  			I_4&= \frac{2\mu}{\hat{q}\,c(z) z(1-z)}2\, \mathfrak{Re}\, \frac{1}{8\, E_g^2}\, \int_0^{L} dx \frac{-i\Omega}{  \sin\Omega x}
e^{-i \frac{m_\text{c}^2}{2 \mu} x }	 		 \Bigg\{  \frac{-i2\mu \Omega}{\sin \Omega x}	(z^2 + (1-z)^2)   \Bigg( \text{Ei}\left[ \frac{\upkappa^2 \tan\Omega x}{i 2\mu\Omega   } \right] \nonumber\\
&- \text{Ei}\left[ \frac{\upkappa^2 \tan \Omega x}{  i 2\mu\Omega   - \hat{q}\,c(z) (L-x) \tan\Omega x} \right]-\exp\left[\frac{\upkappa^2\tan \Omega x}{ i 2\mu\Omega} \right] +\frac{\exp\left[- \frac{\upkappa^2\tan \Omega x}{ i 2\mu\Omega -\hat{q}\,c(z) (L-x) \tan\Omega x} \right]}{1- \frac{\hat q \,c(z) (L-x)}{i2 \mu \Omega}\tan \Omega x} \Bigg)\nonumber\\
&+ m_c^2\left( \text{Ei}\left[ \frac{\upkappa^2\tan\Omega x}{i 2\mu\Omega   } \right] - \text{Ei}\left[ \frac{\upkappa^2\tan \Omega x}{  i 2\mu\Omega   - \hat{q}\,c(z) (L-x) \tan\Omega x} \right] \right)\Bigg\}\, .
\end{align}
Expressing this equation in the dimensionless variables \eqref{eq4.13}, we obtain \eqref{eq4.11}.

\subsection{\texorpdfstring{$I_5$}{I5}}
 
For the case $0<t<L$ and $L< \bar{t}$ with constant $\hat{q}(\xi)=\hat{q}$ in the region $0<\xi<L$, the solution for the path integral is given by the convolution of the harmonic oscillator propagator \eqref{eq4.4} and the free propagator \eqref{eq2.8}
\begin{align}
 {\cal K}\big({\bf r}_\text{in}=0,t;{\bf r}_\text{out},\bar t \vert \mu \big)= &\int d {\bf x}\, {\cal K}_\text{osc}\big[{\bf r}_\text{in}=0,t;{\bf x},L \vert \mu \big]{\cal K}_0\big[{\bf x},L;{\bf r}_\text{out},\bar{t} \vert \mu \big].\label{eq:convol}
 \end{align}
We substitute this expression in \eqref{eq2.2} and we integrate over ${\bf r}_\text{out}$ by parts. 
We then integrate over ${\bf r}_\text{out}$ using \eqref{eq2.9}, we take the $\partial_{{\bf r}_\text{in}}$ derivative and we set ${\bf r}_\text{in}=0$. In this way, we obtain
\begin{align}
		I_5&= 2\, \mathfrak{Re}\, \frac{1}{8\, E_g^2}\, \int_{0}^{L} dt \int_L^{t_\infty} d\bar{t}\, 
		\exp\left[ i \frac{m_\text{c}^2}{2 \mu} (t-\bar{t})- \epsilon \bar{t} \right]\, \int d {\bf x}
		e^{ - i (\bar{t}-L) \frac{\bm{\upkappa}^2}{2 \mu}-i\, {\bm{\upkappa}}\cdot {\bf x}}
		\nonumber \\
		 & \times  \frac{\mu}{z(1-z)}
		  \left[ m_\text{c}^2+ \mu\Omega \frac{{\bf x}\cdot {\bm{\upkappa}}}{\sin\Omega (L-t)} ( z^2 + (1-z)^2 )\right]\frac{-i\Omega}{2 \pi \sin\Omega (L-t)} e^{-\frac{-i2\mu\Omega {\bf x}^2 \cot\Omega (L-t)}{4}}\, .
\end{align}
Doing the remaining Gaussian transverse integral, we find
\begin{align}
		I_5&= 2\, \mathfrak{Re}\, \frac{1}{8\, E_g^2}\, \int_{0}^{L} dt \int_L^{t_\infty} d\bar{t}\, 
		\exp\left[ i \frac{m_\text{c}^2}{2 \mu} (y-\bar{y}) -  \epsilon \bar{t} \right]\, 		\nonumber \\
		 &\times  \frac{1}{\cos \Omega (L-t)} e^{ - i (\bar{y}-L) \frac{{\bm{\upkappa}}^2}{2 \mu}}
     \frac{1}{z(1-z)}e^{- \frac{{\bm{\upkappa}}^2}{-2i \mu \Omega \cot \Omega x}}
		 \left[ m_\text{c}^2+ \frac{{\bm{\upkappa}}^2}{\cos\Omega (L-t)} ( z^2 + (1-z)^2 )\right]\, .
\end{align}
We again notice that the trigonometric functions only depend on the difference $L-y$. Therefore we make the change of variables $x= L-t$ and $u=\bar{t}-L$. The integration bounds are determined by the equivalence of
\begin{equation}
  0 < t < L, \quad L < \bar{t} < \infty \quad\leftrightarrow \quad
  0< x < L, \quad 0 < u < \infty 
.\end{equation}
The $u$ integral is simple and yields
\begin{align}
		I_5&=   \frac{2\mu}{z(1-z)} 2\, \mathfrak{Re}\, \frac{1}{8\, E_g^2}\, \int_{0}^{L} dx\,\frac{-i }{\cos \Omega x} 
    \exp\left[- i \frac{m_\text{c}^2}{2 \mu}x + \frac{ {\bm{\upkappa}}^2\tan \Omega x}{2i \mu \Omega }\right]
    \nonumber \\
		 &\qquad \times 
   \frac{1}{m_c^2+{\bm{\upkappa}}^2}
     \left[ m_\text{c}^2+ \frac{{\bm{\upkappa}}^2}{\cos\Omega x } ( z^2 + (1-z)^2 )\right]\, .
\end{align}
Expressing this equation in the dimensionless variables \eqref{eq4.13},  we obtain \eqref{eq4.12}.
 
  \section{Derivation of the splitting functions \texorpdfstring{\eqref{eq4.16}}{eq4.16} and \texorpdfstring{\eqref{eq4.17}}{eq4.17} in the multiple soft scattering approximation for an expanding medium\label{app:Igensoftderiv}}
 
For a general time-dependent quenching parameter $\hat{q}(\xi)$ with finite support for $0<\xi<L$, the two-dimensional path integral in \eqref{eq2.6} is given by~\cite{Baier:1998yf,schulman2012techniques}\footnote{See also the lecture notes on the single-dimensional harmonic oscillator by Andreas Wipf at \url{https://www.tpi.uni-jena.de/~wipf/lectures/pfad/pfad3}}
\begin{equation}
  {\cal K}\big[{\bf r}_\text{in}=0,t;{\bf r}_\text{out},\bar{t} \vert \mu \big]  = \frac{\mu}{2 \pi i D(\bar{t},t)} e^{i \frac{\mu}{2D(\bar{t},t)}\left( {\bf r}^2_\text{out} \frac{\partial}{\partial \bar{t}}D(\bar{t},t)-  {\bf r}^2_\text{in} \frac{\partial}{\partial {t}}D(\bar{t},t) -2 {\bf r}_\text{out}\cdot{\bf r}_\text{in}  \right) }
\label{eq:Kgeneral}
,\end{equation}
where the auxiliary function $D(\bar{t},t)$ solves the second order differential equation
\begin{equation}
  \frac{\partial^2}{\partial^2 \xi}D(\xi,t) = - \Omega^2(\xi) D(\xi,t),\quad D(t,t)=0,\quad \left.\frac{\partial D(\xi,t)}{\partial \xi}\right|_{\xi=t}=1,
\end{equation}
with $\Omega^2(\xi) = c(z)\frac{\hat{q}(\xi)}{2i\mu}$. For constant $\hat q$, the solution is $D(\bar{t},t) = \frac{1}{\Omega} \sin \Omega(\bar{t}-t)$ and we recover \eqref{eq4.4}.
In the following, we repeat the steps of Appendix \ref{app:Isoftderiv} for this
more general time-dependent case. 

 \subsection{\texorpdfstring{$I_4$}{I4}}
We first substitute the propagator expression for  generalised harmonic oscillator~\eqref{eq:Kgeneral} into 
\eqref{eq2.2}. Then we take the $\partial_{{\bf r}_\text{in}}$ derivative, integrate over ${\bf r}_\text{out}$ by parts 
and set ${\bf r}_\text{in}=0$. We obtain
\begin{align}
	 I_4	&= 2\, \mathfrak{Re}\, \frac{1}{8\, E_g^2}\, \int_{0}^{L} dt \int_t^{L} d\bar{t}\, 
    \exp\left[- i \frac{m_\text{c}^2}{2 \mu} (\bar{t}-t) \right]\, \int d{\bf r}_\text{out}
     e^{ - \frac{1}{4} \int_{\bar{t}}^L d\xi\, \hat{q} (\xi) c(z) {\bf r}^2_\text{out} } e^{-i\, {\bm{\upkappa}}\cdot {\bf r}_\text{out}}
		  \nonumber \\		  		
     &\qquad\qquad \times \left[ \frac{\mu }{D(\bar{t},t)}\left({\bf r}_\text{out}\cdot{\bm{\upkappa}} -i \frac{1}{2} \int_{\bar{t}}^L d\xi\, \hat{q}(\xi) c(z) {\bf r}^2_\text{out} \right) \frac{z^2 + (1-z)^2}{z(1-z)}  + \frac{ m_c^2}{z(1-z)}  \right]\nonumber\\
     &\qquad\qquad\times\frac{\mu}{2 \pi i D(\bar{t},t)} \exp\left[-\frac{1}{4}\{-2i\mu\partial_{\bar{t}}\log D(\bar{t},t)\} {\bf r}^2_\text{out}\right] \, .
\end{align}

Performing the remaining Gaussian integration yields
\begin{align}
	I_4&= 2\, \mathfrak{Re}\, \frac{1}{8\, E_g^2}\, \int_{0}^{L} dt \int_t^{L} d\bar{t}\, 
    \exp\left[- i \frac{m_\text{c}^2}{2 \mu} (\bar{t}-t)\right]\,
\Bigg[ \frac{m_c^2}{z(1-z)} +\frac{z^2 + (1-z)^2}{z(1-z)}  		\nonumber \\
    & \times\left(\frac{ \partial_{\bar{t}} D(\bar{t},t){\bm{\upkappa}}^2}{\{\partial_{\bar{t}}D(\bar{t},t) - D(\bar{t},t)\int_{\bar{t}}^L d\xi\, \Omega^2(\xi)\}^2} +\frac{2i\mu\int_{\bar{t}}^L d\xi\, \Omega^2(\xi)}{\partial_{\bar{t}}D(\bar{t},t) - D(\bar{t},t)\int_{\bar{t}}^L d\xi\, \Omega^2(\xi)}  \right)   \Bigg]\nonumber\\
     &\times\frac{1}{  \partial_{\bar{t}}D(\bar{t},t) - D(\bar{t},t)\int_{\bar{t}}^L d\xi\, \Omega^2(\xi) } \exp\left[-i \frac{{\bm{\upkappa}}^2}{2\mu} \{\partial_{\bar{t}}\log D(\bar{t},t) - \int_{\bar{t}}^L d\xi\, \Omega^2(\xi) \}^{-1}\right]\, .
\end{align}
Switching to dimensionless variables in \eqref{eq4.15} we obtain \eqref{eq4.16}.

\subsection{\texorpdfstring{$I_5$}{I5}}

We substitute generalised harmonic oscillator path integral \eqref{eq:Kgeneral} in  \eqref{eq2.2} (using \eqref{eq:convol}) and  integrate over ${\bf r}_\text{out}$ by parts. 
Then we can perform 
the ${\bf r}_\text{out}$ integral using \eqref{eq2.9}. Finally we take $\partial_{{\bf r}_\text{in}}$ derivative and set ${\bf r}_\text{in}=0$. We obtain
\begin{align}
	I_5= 2\, \mathfrak{Re}\, \frac{1}{8\, E_g^2}&\, \int_{0}^{L} dt \int_L^{t_\infty} d\bar{t}\, 
    \exp\left[- i \frac{m_\text{c}^2}{2 \mu} (\bar{t}-t) - \epsilon \bar{t} \right]\, \int d{\bf x}
		\nonumber \\
       &\times e^{-i\, {\bf k}_\text{c}\cdot {\bf x}- i (\bar{t}-L) \frac{{\bm{\upkappa}}^2}{2 \mu}  }\left[ \left( m_\text{c}^2 + \frac{\mu {\bf x}\cdot {\bm{\upkappa}}}{D(L, t)}		  \right) \frac{z^2 + (1-z)^2}{z(1-z)}  + 2 m_c^2  \right]\nonumber\\
       &\times\frac{-i\mu}{2 \pi  D(\bar{t},t)} \exp\left[- \frac{1}{4} (-2i\mu \left.\partial_{\bar{t}}\log D(\bar{t},t)\right|_{\bar{t}=L}) {\bf r}^2_\text{out} \right] \, .
\end{align}
We then perform the radial integral 
\begin{align}
		I_5&= 2\, \mathfrak{Re}\, \frac{1}{8\, E_g^2}\, \int_{0}^{L} dt \int_L^{t_\infty} d\bar{t}\, 
    \exp\left[- i \frac{m_\text{c}^2}{2 \mu} (\bar{t}-t) - \epsilon \bar{t} - i\frac{{\bm{\upkappa}}^2}{2 \mu}  (\{\left.\partial_{\bar{t}}\log D(\bar{t},t)\right|_{\bar{t}=L}\}^{-1}-L +\bar{t})  \right] 
		  \nonumber \\		  		
       &\qquad\qquad \times \left[ \frac{ {\bm{\upkappa}}^2}{\left.\partial_{\bar{t}}D(\bar{t},t)\right|_{\bar{t}=L}} \frac{z^2 + (1-z)^2}{z(1-z)}  +  \frac{m_c^2}{z(1-z)}  \right]\frac{1}{\left.\partial_{\bar{t}} D(\bar{t},t)\right|_{\bar{t}=L}} \, .
\end{align}
In this term, also the $\bar{t}$ integral can be done analytically,
\begin{align}
		I_5&=2\, \mathfrak{Re}\, \frac{1}{8\, E_g^2}\, \int_{0}^{L} dt \frac{-i2\mu}{m_c^2+{\bm{\upkappa}}^2}\, 
    \exp\left[- i \frac{m_\text{c}^2}{2 \mu} (L-t) - \epsilon \bar{t} - i\frac{{\bm{\upkappa}}^2}{2 \mu} \{\left.\partial_{\bar{t}}\log D(\bar{t},t)\right|_{\bar{t}=L}\}^{-1}  \right] 
		  \nonumber \\		  		
       &\qquad\qquad \times \left[ \frac{ {\bm{\upkappa}}^2}{\left.\partial_{\bar{t}}D(\bar{t},t)\right|_{\bar{t}=L}} \frac{z^2 + (1-z)^2}{z(1-z)}  +  \frac{m_c^2}{z(1-z)}  \right]\frac{1}{\left.\partial_{\bar{t}} D(\bar{t},t)\right|_{\bar{t}=L}} \, .
\end{align}
 Switching to dimensionless variables in \eqref{eq4.15} we obtain \eqref{eq4.17}.
 \section{Comments on the relation between \texorpdfstring{$N=1$}{N=1} hard and multiple soft scattering}
 \label{appc}
 
Our numerical study of the $N=1$ opacity expansion in section ~\ref{sec3.5}
was based on a medium parameterised by Yukawa-type scattering centres \eqref{eq3.20}.
We refer to it as $N=1$ \emph{hard} since these scattering centres display power-law tails that one expects to arise in scatterings with sufficiently high (\emph{hard}) momentum transfer. In a multiple scattering picture, this accounts for Moli\`ere-type~\cite{Moliere:1948zz,ParticleDataGroup:2020ssz} rare large-angle scattering. 
For such a Yukawa potential, the dipole cross section \eqref{eq2.4} takes the form
\begin{equation}
	\sigma({\bf r}) =\sigma_\text{el} \left( 1 - \mu_Dr K_1(\mu_Dr) \right)  \xrightarrow[ ]{r\mu_D\ll 1 } \sigma_\text{el}
  \frac{\mu_D^2}{2} \left(\frac{1}{2}-\gamma_E + \log\left[ \frac{2}{\mu_Dr}\right] \right) {\bf r}^2\, ,
	\label{eq6.3}
\end{equation}
where  $K_1$ denotes a modified Bessel function and $\gamma_E = 0.577 ....$ is Euler's constant.
Here, the logarithmic correction $\log\left[ \frac{2}{\mu_Dr}\right]$ prevents us from relating the prefactor of ${\bf r}^2$ to the quenching parameter 
 used in the saddle point approximation \eqref{eq4.3}. The problem is that \eqref{eq4.3} assumes the existence of the second moment
 $\int \tfrac{d{\bf q}^2}{(2\pi)^2}  \vert a({\bf q})\vert^2\, {\bf q}^2$ whereas this moment is logarithmically UV divergent for a Yukawa-type potential.
 The term $\log \left[ \tfrac{2}{\mu_Dr}\right] $ is the tell-tale sign of this logarithmic divergence in configuration space.
The saddle point approximation \eqref{eq4.3} amounts to cutting off the ${\bf q}^{-4}$ tail of $\vert a({\bf q})\vert^2$ at large ${\bf q}$ by regulating
the $\log r$-dependence at small $r$. 

In contrast,  by construction, a Gaussian ansatz for $\vert a({\bf q})\vert^2$ is free of UV problems and it is consistent with the dipole approximation  \eqref{eq4.3}
since 
  \begin{equation}
  	\vert a({\bf q})\vert^2 =  \sigma_\text{el}\frac{4\pi}{\langle {\bf q}^2 \rangle_\text{med}} e^{- {\bf q}^2 / \langle {\bf q}^2 \rangle_\text{med} }
		\Longrightarrow  \sigma({\bf r}^2) = \sigma_\text{el} \left(1 - e^{-\langle {\bf q}^2 \rangle_\text{med} {\bf r}^2/4} \right)
		= \frac{1}{4}\sigma_\text{el}\langle {\bf q}^2\rangle_\text{med} {\bf r}^2 + \mathcal{O}({\bf r}^4)\,.
		\label{eq6.1} 
  \end{equation}
We consider this Gaussian ansatz with a normalisation such that 
\begin{equation}
  \frac{1}{4} n_0\, L \vert a({\bf q})\vert^2
 \equiv \frac{c_\text{opaque}}{\langle {\bf q}^2 \rangle_\text{med}} \vert \tilde a({\tilde{\bf q}})\vert^2, \quad \vert \tilde a(\tilde{{\bf q}})\vert^2 \equiv 4\pi \exp[-\tilde{\bf{q}}^2] \label{eq6.2gaus}
  \, .
\end{equation}
We now ask to what extent the resulting $\left( \tfrac{1}{Q^2} P_{g\to c\bar{c}} \right)^{\text{med}}$ differs from the medium-modification calculated for Yukawa-type potentials with hard power-low tails.

\begin{figure}[t]
    \centering
\subfig{a}{\includegraphics[width=.53\textwidth]{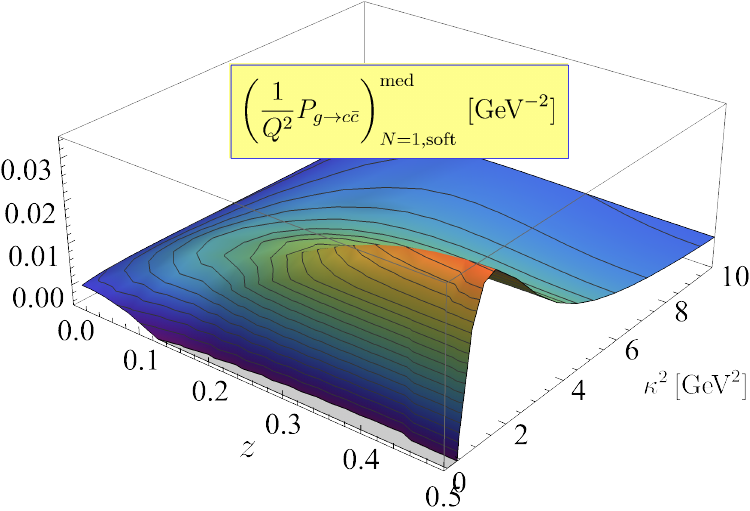}}%
\subfig{b}{\includegraphics[width=.46\textwidth]{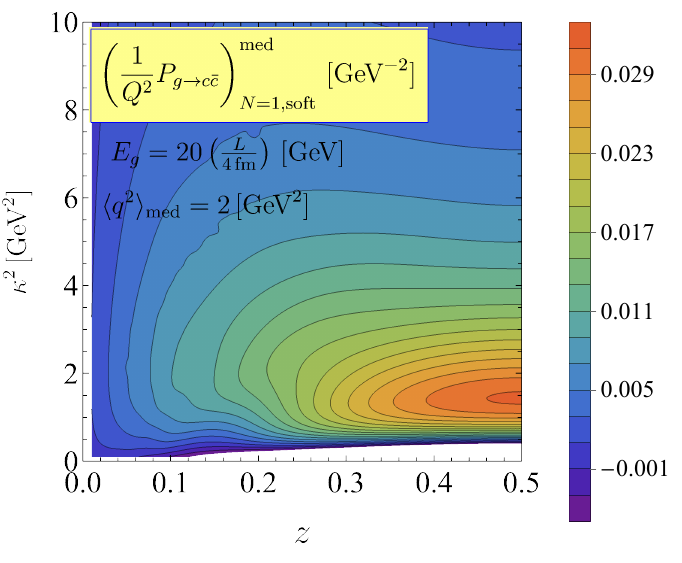}}
    \caption{The medium-modification \eqref{eq3.25} of the $g\to c\bar{c}$ splitting in the $N=1$ opacity expansion for `soft' Gaussian-type scattering centres~\eqref{eq6.2gaus}, plotted as a function of $z$ and final transverse momentum $\bm{\upkappa}^2$. A similar figure for Yukawa-type scatterings was shown in Fig.~\ref{figN1}.
    }
    \label{fig2}
\end{figure}
 
Using \eqref{eq6.1} 
amounts to replacing in the expression  \eqref{eq3.22}
 \begin{equation}
\frac{1}{2\mu_D^2}  \frac{2\pi A^2 }{ \left( \tfrac{1}{2}A^2 + \tilde{\bf q}^2\right)^2 }  \quad \longrightarrow \quad
 \frac{1}{{\langle {\bf q}^2 \rangle_\text{med}}} \frac{4 \pi}{A^2} e^{-\frac{\tilde{\bf q}^2}{A^2}} \,,\quad \text{where } A = 1, z, 1-z\,.
	\label{eq6.4}
\end{equation}
 The dimensionless variables in \eqref{eq3.25}  are then understood as measuring masses and momenta in units of 
 ${\langle {\bf q}^2 \rangle_\text{med}}$,
 \begin{equation}
 \tilde{m}_\text{c}^2 = \frac{{m}_\text{c}^2}{\langle {\bf q}^2 \rangle_\text{med} }\, ,\qquad  \tilde{\bm \upkappa}^2 = \frac{{\bm \upkappa}^2}{\langle {\bf q}^2 \rangle_\text{med}},\qquad  
 \tilde{\bf q}^2 = \frac{{\bf q}^2}{\langle {\bf q}^2 \rangle_\text{med}}\, ,\qquad  
 E_g \equiv e_g \frac{1}{2} {\langle {\bf q}^2 \rangle_\text{med}} L\,  .
 \end{equation}
 
Numerical results for this $N=1$ opacity expansion with Gaussian momentum distribution \eqref{eq6.1} 
are displayed in Fig.~\ref{fig2}. Close comparison with the results for $N=1$ hard in Fig.~\ref{figN1} shows that the hard Yukawa-type tails of single scattering centres yield only to a 
very mild if not negligible change in the $\bm{\upkappa}^2$-distribution. This can be understood on general grounds: 2/3rd (8/9th) of the Yukawa cross section in \eqref{eq6.1} lies in the region of soft momentum transfers $\tilde{q} < 1$  ($\tilde{q} <2$). Therefore, also the $N=1$ opacity approximation is dominated by small-angle scattering and differences to a Gaussian distribution may be expected to be only gradual.

There has been repeated interest in isolating Moli\`ere-type large angle scattering contributions to jet quenching as these may inform us about the nature of the constituents in the 
QCD plasma. For a recent summary of the phenomenological motivation, see Ref.~\cite{DEramo:2018eoy}.
That Moli\`ere-type large-angle scatterings can be included in jet quenching formulations via a logarithmic dependence of the dipole cross section has been known since~\cite{Baier:1996kr,Zakharov:1996fv}. Indeed, in describing $e^+e^-$ QED pair radiation off extended targets, Yukawa-type potentials of the form \eqref{eq6.3} 
had been used to describe the screening of nuclei in QED matter by their electron clouds; the observation of a logarithmic term in the corresponding dipole cross section had been made in this context much earlier~\cite{Zakharov:1987wy,Zakharov:1996fv}. Conceptually the same logarithmic dependence appears also in the configuration-space description of QCD saturation physics via Glauber-Mueller multiple scattering effects~\cite{Kovchegov:1996ty,Jalilian-Marian:1996mkd}. In applying this formalism to the longitudinal dynamics
of incoming partons that undergo multiple scattering in a nuclear target, it was shown that the rare Moli\`ere-type large-angle scattering encoded in an $\log r$-term could result in a Cronin-type enhancement at a characteristic intermediate transverse momentum scale~\cite{Baier:2003hr}. In quenched jet fragmentation patterns, however,
parametric reasoning indicates that Moli\`ere-type large-angle contributions can become numerically relevant only in a very narrow region of the entire phase space~\cite{Kurkela:2014tla}. We are not aware of any proposal that would allow for a clear identification of such a large-angle scattering contribution within realistic hadronic distributions. Consistent with these findings, 
the comparison between $N=1$ hard and $N=1$ Gaussian presented here indicates that also the medium-modification
of $g\to c\bar{c}$ shows only a very small if not negligible dependence on the presence of a large-angle scattering contribution.

\end{appendix}

\bibliographystyle{JHEP}
\bibliography{jhepccbar}
  \end{document}